\documentclass[twocolumn,author-year]{aastex701}

\usepackage{float}
\usepackage{pgf,pgffor}
\usepackage{textcomp}
\usepackage{threeparttable} 
\usepackage{indentfirst}
\usepackage{academicons}
\usepackage{pifont}
\usepackage{longtable}
\usepackage {amsmath}
\IfFileExists{url.sty}{%
  \usepackage{url}%
}{%
  \providecommand{\url}[1]{\texttt{#1}}%
}
\definecolor{orcidlogocol}{HTML}{A6CE39}
\usepackage{subcaption}
\usepackage{multirow}
\usepackage{supertabular}
\usepackage{booktabs}
\usepackage{siunitx}
\usepackage{mathrsfs}

\hypersetup{
    pdfborder={0 0 0},
    colorlinks=true,
    linkcolor=red,   
    citecolor=blue,      
    urlcolor=black
}

\DeclareCaptionLabelFormat{boldlarge}{\textbf{\Large #2}}

\providecommand{\apjs}{The Astrophysical Journal Supplement}
\providecommand{\apjl}{The Astrophysical Journal Letter}

\providecommand{\aap}{Astronomy \& Astrophysics}

\newcommand{\D}{\mathrm{d}}

\newcommand{\AFF}[1]{$^{\foreach\d[count=\ni]in{#1}{\ifnum\ni=1\ref{\d}\else,\ref{\d}\fi}}$}

\definecolor{dkblue}{RGB}{54, 86, 169}

\def\scititle{Detection of 2 MeV spectral bump and $\sim$40 MeV emission line from GRB 221009A}
\begin{document}

\title{\scititle}

%\correspondingauthor{Xinqiao Li, Zhuo Li, Zhenghua An, \newline Shaolin Xiong}
%\email{lixq@ihep.ac.cn, zhuo.li@pku.edu.cn,  anzh@ihep.ac.cn, \newline xiongsl@ihep.ac.cn}

\def\scisuppauthors{
Zhi-Qiang Ding,
Zhenxia Zhang,
Yanqiu Zhang,
Rui-Ze Shi,
Xinqiao Li,
Zhuo Li,
Zhenghua An,
Shaolin Xiong,
Roberto Battiston,
Shuangnan Zhang,
Hua Feng,
Dali Zhang,
Yanbing Xu,
Ping Wang,
Xiaoyun Zhao,
Yuqian Ma,
Hong Lu,
Xiangyang Wen,
Yongjie Zhang,
Xuhui Shen,
Shigeng Yuan,
Siquan Yang,
Rongxue Kang,
Chao Zheng,
Yan Huang,
Wen Su,
Shengli Liu,
Hui Wang,
Xiaohua Liang,
Jinzhou Wang,
Ke Gong,
Xiaojing Liu,
Yaqing Liu,
Sheng Yang,
Zeren Zhima,
Jianping Huang,
Wei Chu,
Na Zhou,
Roberto Iuppa,
Francesco Maria Follega,
Marco Cristoforetti,
Wangchen Xue,
Lu Wang,
Jiaqi Zheng,
Jiacong Liu,
Chenwei Wang,
Wenjun Tan,
Zhenghang Yu,
Shujie Li,
Hao Hong
}

\author[0009-0005-5459-3600]{Zhiqiang Ding}
\altaffiliation{These authors contributed equally to this work.}
\affiliation{State Key Laboratory of Particle Astrophysics, Institute of High Energy Physics, Chinese Academy of Sciences, Beijing 100049, China}
\affiliation{University of Chinese Academy of Sciences, Chinese Academy of Sciences, Beijing 100049, China}
\email{}

\author{Zhenxia Zhang}
\altaffiliation{These authors contributed equally to this work.}
\affiliation{National Institute of Natural Hazards, Ministry of Emergency Management of the People's Republic of China, Beijing 100085, China}
\email{}

\author{Yanqiu Zhang}
\altaffiliation{These authors contributed equally to this work.}
\affiliation{School of Physics and Electronic Science, Guizhou Normal University, Guiyang 550001, China}
\affiliation{Guizhou Provincial Key Laboratory of Radio Astronomy and Data Processing, Guizhou Normal University, Guiyang 550001, China}
\email{}

\author{Ruize Shi}
\altaffiliation{These authors contributed equally to this work.}
\affiliation{Department of Astronomy, School of Physics, Peking University, Beijing 100871, China}
\email{}

\author{Xinqiao Li}
\altaffiliation{Corresponding authors: Xinqiao Li, Zhuo Li, Zhenghua An, Shaolin Xiong; \\ 
lixq@ihep.ac.cn, zhuo.li@pku.edu.cn, anzh@ihep.ac.cn, xiongsl@ihep.ac.cn}
\affiliation{State Key Laboratory of Particle Astrophysics, Institute of High Energy Physics, Chinese Academy of Sciences, Beijing 100049, China}
\email{lixq@ihep.ac.cn}

\author{Zhuo Li}
\altaffiliation{Corresponding authors: Xinqiao Li, Zhuo Li, Zhenghua An, Shaolin Xiong; \\ 
lixq@ihep.ac.cn, zhuo.li@pku.edu.cn, anzh@ihep.ac.cn, xiongsl@ihep.ac.cn}
\affiliation{Department of Astronomy, School of Physics, Peking University, Beijing 100871, China}
\affiliation{Kavli Institute for Astronomy and Astrophysics, Peking University, Beijing 100871, China}
\email{zhuo.li@pku.edu.cn}

\author{Zhenghua An}
\altaffiliation{Corresponding authors: Xinqiao Li, Zhuo Li, Zhenghua An, Shaolin Xiong; \\ 
lixq@ihep.ac.cn, zhuo.li@pku.edu.cn, anzh@ihep.ac.cn, xiongsl@ihep.ac.cn}
\affiliation{State Key Laboratory of Particle Astrophysics, Institute of High Energy Physics, Chinese Academy of Sciences, Beijing 100049, China}
\email{anzh@ihep.ac.cn}

\author{Shaolin Xiong}
\altaffiliation{Corresponding authors: Xinqiao Li, Zhuo Li, Zhenghua An, Shaolin Xiong; \\ 
lixq@ihep.ac.cn, zhuo.li@pku.edu.cn, anzh@ihep.ac.cn, xiongsl@ihep.ac.cn}
\affiliation{State Key Laboratory of Particle Astrophysics, Institute of High Energy Physics, Chinese Academy of Sciences, Beijing 100049, China}
\email{xiongsl@ihep.ac.cn}

\author{Roberto Battiston}
\affiliation{University of Trento, Department of Physics, V. Sommarive 14, 38123 Trento, Italy}
\affiliation{INFN-TIFPA, V. Sommarive 14, 38123 Trento, Italy}
\email{}

\author{Shuangnan Zhang}
\affiliation{State Key Laboratory of Particle Astrophysics, Institute of High Energy Physics, Chinese Academy of Sciences, Beijing 100049, China}
\email{}

\author{Hua Feng}
\affiliation{State Key Laboratory of Particle Astrophysics, Institute of High Energy Physics, Chinese Academy of Sciences, Beijing 100049, China}
\email{}

\author{Dali Zhang}
\affiliation{State Key Laboratory of Particle Astrophysics, Institute of High Energy Physics, Chinese Academy of Sciences, Beijing 100049, China}
\email{}

\author{Yanbing Xu}
\affiliation{State Key Laboratory of Particle Astrophysics, Institute of High Energy Physics, Chinese Academy of Sciences, Beijing 100049, China}
\email{}

\author{Ping Wang}
\affiliation{State Key Laboratory of Particle Astrophysics, Institute of High Energy Physics, Chinese Academy of Sciences, Beijing 100049, China}
\email{}

\author{Xiaoyun Zhao}
\affiliation{State Key Laboratory of Particle Astrophysics, Institute of High Energy Physics, Chinese Academy of Sciences, Beijing 100049, China}
\email{}

\author{Yuqian Ma}
\affiliation{State Key Laboratory of Particle Astrophysics, Institute of High Energy Physics, Chinese Academy of Sciences, Beijing 100049, China}
\email{}

\author{Hong Lu}
\affiliation{State Key Laboratory of Particle Astrophysics, Institute of High Energy Physics, Chinese Academy of Sciences, Beijing 100049, China}
\email{}

\author{Xiangyang Wen}
\affiliation{State Key Laboratory of Particle Astrophysics, Institute of High Energy Physics, Chinese Academy of Sciences, Beijing 100049, China}
\email{}

\author{Yongjie Zhang}
\affiliation{State Key Laboratory of Particle Astrophysics, Institute of High Energy Physics, Chinese Academy of Sciences, Beijing 100049, China}
\email{}

\author{Xuhui Shen}
\affiliation{National Space Science Center, Chinese Academy of Sciences, Beijing 100190, China}
\email{}

\author{Shigeng Yuan}
\affiliation{Institute of Spacecraft System Engineering, Beijing 100094, China}
\email{}

\author{Siquan Yang}
\affiliation{National Institute of Natural Hazards, Ministry of Emergency Management of the People's Republic of China, Beijing 100085, China}
\email{}

\author{Rongxue Kang}
\affiliation{National Institute of Natural Hazards, Ministry of Emergency Management of the People's Republic of China, Beijing 100085, China}
\email{}

\author{Chao Zheng}
\affiliation{State Key Laboratory of Particle Astrophysics, Institute of High Energy Physics, Chinese Academy of Sciences, Beijing 100049, China}
\affiliation{TIANFU Cosmic Ray Research Center, Chengdu, Sichuan 610000, China}
\email{}

\author{Yan Huang}
\affiliation{School of Physics and Optoelectronics Engineering, Anhui University, Hefei 230601, China}
\email{}

\author{Wen Su}
\affiliation{DFH Satellite Co. Ltd., Beijing 100081, China}
\email{}

\author{Shengli Liu}
\affiliation{DFH Satellite Co. Ltd., Beijing 100081, China}
\email{}

\author{Hui Wang}
\affiliation{State Key Laboratory of Particle Astrophysics, Institute of High Energy Physics, Chinese Academy of Sciences, Beijing 100049, China}
\email{}

\author{Xiaohua Liang}
\affiliation{State Key Laboratory of Particle Astrophysics, Institute of High Energy Physics, Chinese Academy of Sciences, Beijing 100049, China}
\email{}

\author{Jinzhou Wang}
\affiliation{State Key Laboratory of Particle Astrophysics, Institute of High Energy Physics, Chinese Academy of Sciences, Beijing 100049, China}
\email{}

\author{Ke Gong}
\affiliation{State Key Laboratory of Particle Astrophysics, Institute of High Energy Physics, Chinese Academy of Sciences, Beijing 100049, China}
\email{}

\author{Xiaojing Liu}
\affiliation{State Key Laboratory of Particle Astrophysics, Institute of High Energy Physics, Chinese Academy of Sciences, Beijing 100049, China}
\email{}

\author{Yaqing Liu}
\affiliation{State Key Laboratory of Particle Astrophysics, Institute of High Energy Physics, Chinese Academy of Sciences, Beijing 100049, China}
\email{}

\author{Sheng Yang}
\affiliation{State Key Laboratory of Particle Astrophysics, Institute of High Energy Physics, Chinese Academy of Sciences, Beijing 100049, China}
\email{}

\author{Zeren Zhima}
\affiliation{National Institute of Natural Hazards, Ministry of Emergency Management of the People's Republic of China, Beijing 100085, China}
\email{}

\author{Jianping Huang}
\affiliation{National Institute of Natural Hazards, Ministry of Emergency Management of the People's Republic of China, Beijing 100085, China}
\email{}

\author{Wei Chu}
\affiliation{National Institute of Natural Hazards, Ministry of Emergency Management of the People's Republic of China, Beijing 100085, China}
\email{}

\author{Na Zhou}
\affiliation{National Institute of Natural Hazards, Ministry of Emergency Management of the People's Republic of China, Beijing 100085, China}
\email{}

\author{Roberto Iuppa}
\affiliation{University of Trento, Department of Physics, V. Sommarive 14, 38123 Trento, Italy}
\affiliation{INFN–TIFPA, V. Sommarive 14, 38123 Trento, Italy}
\email{}

\author{Francesco Maria Follega}
\affiliation{University of Trento, Department of Physics, V. Sommarive 14, 38123 Trento, Italy}
\affiliation{INFN-TIFPA, V. Sommarive 14, 38123 Trento, Italy}
\email{}

\author{Marco Cristoforetti}
\affiliation{University of Trento, Department of Physics, V. Sommarive 14, 38123 Trento, Italy}
\affiliation{INFN-TIFPA, V. Sommarive 14, 38123 Trento, Italy}
\affiliation{Fondazione Bruno Kessler, Via Sommarive 18, 38123 Trento, Italy}
\email{}

\author{Wangchen Xue}
\affiliation{State Key Laboratory of Particle Astrophysics, Institute of High Energy Physics, Chinese Academy of Sciences, Beijing 100049, China}
\affiliation{University of Chinese Academy of Sciences, Chinese Academy of Sciences, Beijing 100049, China}
\email{}

\author{Lu Wang}
\affiliation{State Key Laboratory of Particle Astrophysics, Institute of High Energy Physics, Chinese Academy of Sciences, Beijing 100049, China}
\affiliation{University of Chinese Academy of Sciences, Chinese Academy of Sciences, Beijing 100049, China}
\email{}

\author{Jiaqi Zheng}
\affiliation{University of Chinese Academy of Sciences, Chinese Academy of Sciences, Beijing 100049, China}
\affiliation{National Institute of Natural Hazards, Ministry of Emergency Management of the People's Republic of China, Beijing 100085, China}
\email{}

\author{Jiacong Liu}
\affiliation{State Key Laboratory of Particle Astrophysics, Institute of High Energy Physics, Chinese Academy of Sciences, Beijing 100049, China}
\affiliation{University of Chinese Academy of Sciences, Chinese Academy of Sciences, Beijing 100049, China}
\email{}

\author{Chenwei Wang}
\affiliation{State Key Laboratory of Particle Astrophysics, Institute of High Energy Physics, Chinese Academy of Sciences, Beijing 100049, China}
\affiliation{University of Chinese Academy of Sciences, Chinese Academy of Sciences, Beijing 100049, China}
\email{}

\author{Wenjun Tan}
\affiliation{State Key Laboratory of Particle Astrophysics, Institute of High Energy Physics, Chinese Academy of Sciences, Beijing 100049, China}
\affiliation{University of Chinese Academy of Sciences, Chinese Academy of Sciences, Beijing 100049, China}
\email{}

\author{Zhenghang Yu}
\affiliation{State Key Laboratory of Particle Astrophysics, Institute of High Energy Physics, Chinese Academy of Sciences, Beijing 100049, China}
\affiliation{University of Chinese Academy of Sciences, Chinese Academy of Sciences, Beijing 100049, China}
\email{}

\author{Shujie Li}
\affiliation{National Institute of Natural Hazards, Ministry of Emergency Management of the People's Republic of China, Beijing 100085, China}
\email{}

\author{Hao Hong}
\affiliation{National Institute of Natural Hazards, Ministry of Emergency Management of the People's Republic of China, Beijing 100085, China}
\email{}

\begin{abstract}

Gamma-ray bursts (GRBs) are the most energetic transients in the Universe, but prompt MeV--sub-GeV observations remain challenging with current gamma-ray observatories. Here, using unsaturated HEPP-H measurements on board CSES-01 and cross-calibrating with likewise unsaturated GECAM-C data, we reconstruct the continuous spectral evolution of GRB 221009A over 2--255 MeV. We identify an unusual spectral bump at $\sim2$ MeV in the first light-curve peak. We also confirm a narrow emission line whose energy decreases from $\sim$40 MeV following a $t^{-1}$ evolution, and whose flux shows a ``rise-then-fall'' profile, implying weak emission from the central jet. These findings hint that a magnetically dominated, central-empty jet exists in the BOAT.

\end{abstract}


\section{Introduction}

 Gamma-ray bursts (GRBs) are the most energetic and violent astrophysical phenomena in the universe. The study of their prompt emission in the MeV--sub-GeV range is crucial to investigate the poorly known physics of the GRB central engines and their released relativistic jets \citep{2019pgrb.book.....Z}. GRBs mainly emit photons in the MeV domain, with spectral peaks typically lying around a few hundred keV. The prompt emission beyond MeV is expected to bring key information on the central engine and relativistic jets. However, the precise observations of the prompt emission from GRBs are conventionally hindered by a long sought-after MeV--sub-GeV gamma-ray gap \citep{2004NewAR..48..193S}. This energy range represents the region where Compton and Pair events overlap \citep{2005NIMPA.541..310K}, resulting in an inherently worse detection sensitivity \citep{1993ApJS...86..657S}.
Bright GRBs are important in this matter, but they are generally rare.

An unprecedented opportunity %to overcome these observational challenges emerged on 9 October 2022, 
emerged when GRB~221009A shattered records of GRB brightness. The initial real-time trigger of GRB 221009A was provided by the \textit{Fermi} Gamma-ray Space Telescope's Gamma-ray Burst Monitor (GBM) \citep{2023ApJ...952L..42L}, but its exceptional brightness caused severe instrumental saturation effects for GBM during the brightest phase. In contrast, the \textit{Insight}-Hard X-ray Modulation Telescope (\textit{Insight}-HXMT) recorded high-resolution unsaturated data, thus we adopt its trigger time, $T_{\rm ref} =$ 13:17:00.050 UT \citep{B9_2024_09A_afterglow}, as the reference time for our analysis here. The Neil Gehrels \textit{Swift} Observatory (\textit{Swift}) provided the first accurate localization of this burst \citep{2023ApJ...946L..24W}, which enables a worldwide multi-wavelength observations and the redshift measurement ($z = 0.151$) \citep{2022GCN.32648....1D, 2025A&A...701A.134M}. Notably, the Gravitational Wave High-energy Electromagnetic Counterpart All-sky Monitor-C (GECAM-C, also known as HEBS) provided unsaturated measurements, which measures an isotropic equivalent energy of $E_{\text{iso}} \sim 1.5 \times 10^{55}$\,erg and confirms GRB~221009A as the brightest of all time (BOAT) %and most energetic GRB ever detected 
\citep{2023arXiv230301203A, 2023ApJ...946L..31B, 2026ApJ...998L..21Y}. 
The extreme brightness of this burst leads to several record breaking discoveries. The Large High Altitude Air Shower Observatory (LHAASO) detected more than 64,000 photons above 0.2\,TeV within the first 3,000\,seconds \citep{2023Sci...380.1390L}. Line emission of $\sim10$ MeV is historically discovered in the BOAT using  \textit{Fermi}/GBM data \citep{2024Sci...385..452R}. 
%把慧眼的删除，只有前两个的数据
Further analysis combining data of GECAM-C and \textit{Fermi}/GBM show a power-law temporal evolution of the line energy $E_{\mathrm{line}} \propto (t-t_0)^{-1.05}$ where $t_0 = T_{\rm ref} + 226$\,s, starting from a high energy $E_{\mathrm{line}}\simeq37$\,MeV in the early time \citep{2024SCPMA..6789511Z}. 

%Despite these extensive observations across multiple wavelengths, the study of the MeV--sub-GeV prompt phase remains hindered by severe instrumental limitations. For GRB~221009A, this crisis was exacerbated by unprecedented photon fluxes that caused catastrophic saturation of most conventional detectors \citep{2023ApJ...952L..42L, 2025ApJ...986..170Z}. 
However, the extremely high photon fluxes of the BOAT caused catastrophic saturation of most conventional detectors. \textit{Fermi}/GBM suffered from the aforementioned pulse pileup issues, leading to the loss of TTE telemetry packets and rendering much of the data unrecoverable \citep{2009ApJ...702..791M, 2023ApJ...952L..42L}, while the \textit{Fermi} Large Area Telescope (LAT) faced a reduction in effective area and increased background contamination \citep{2009ApJ...697.1071A}.

These instrumental limitations restrict the study of spectral features, e.g., the spectral measurement at the light curve peaks are missing; the MeV line emission is only detected at the low-flux phase; and the 37\,MeV emission line is missing the high-energy wing of the profile at $\gtrsim35$\,MeV due to the limited energy range of \textit{Fermi}/GBM \citep{2024SCPMA..6789511Z}.

%These instrumental limitations restrict the study of fine spectral features, which are of critical importance in revealing the physics of relativistic jets \citep{2024SCPMA..6789511Z}. Consequently, while emission lines were initially reported using \textit{Fermi}/GBM data \citep{2024arXiv241000286B}, these results were restricted to less-bright periods due to instrumental issues, including a highly significant ($>6\sigma$) narrow feature at $\sim$10\,MeV \citep{2024Sci...385..452R}. At the same time, a multi-instrument analyses using GECAM-C, \textit{Fermi}/GBM, and \textit{Insight}-HXMT identified these spectral excesses as line emission through the discovery of a power-law temporal evolution of the line's central energy and flux  \citep{2024SCPMA..6789511Z}. Although this study identified a line reaching approximately 37\,MeV, the highest energy detected so far, the 4--35\,MeV coverage of \textit{Fermi}/GBM limited observation of its trailing edge, rendering previous observations incomplete \citep{2024SCPMA..6789511Z}.

Here, these observational deficiencies are overcome by the HEPP instruments aboard the CSES-01 Satellite. Primarily designed to monitor the geophysical charged particle environment, HEPP incorporates particle-detection sub-detectors---specifically the high-energy (HEPP-H) and low-energy (HEPP-L) units (Table~\ref{tab:hepp_params}). Beyond charged particle monitoring, the HEPP instruments are sensitive to gamma rays by capturing secondary electrons generated within the materials surrounding the sensitive detection \citep{2019RDTM....3...22L, 2018RAA....18..154N, 2023ApJ...946L..29B, 2024ApJ...960...21P, 2024ApJ...976..239B}. In particular, the trigger logic of HEPP-H, based on the plastic scintillator units of the calorimeter, has a low efficiency for direct gamma-ray detection while remaining sensitive to secondary electrons. This specific configuration, while unconventional for gamma-ray astronomy, proved to be a decisive advantage for the BOAT measurement, which can effectively mitigate pulse pile-up and saturation under extremely high photon fluxes.

\section{Observations and Data Reduction}

\subsection{Instruments and Observations}

\begin{figure}[t]
\centering
  \includegraphics[width=1\columnwidth]
  {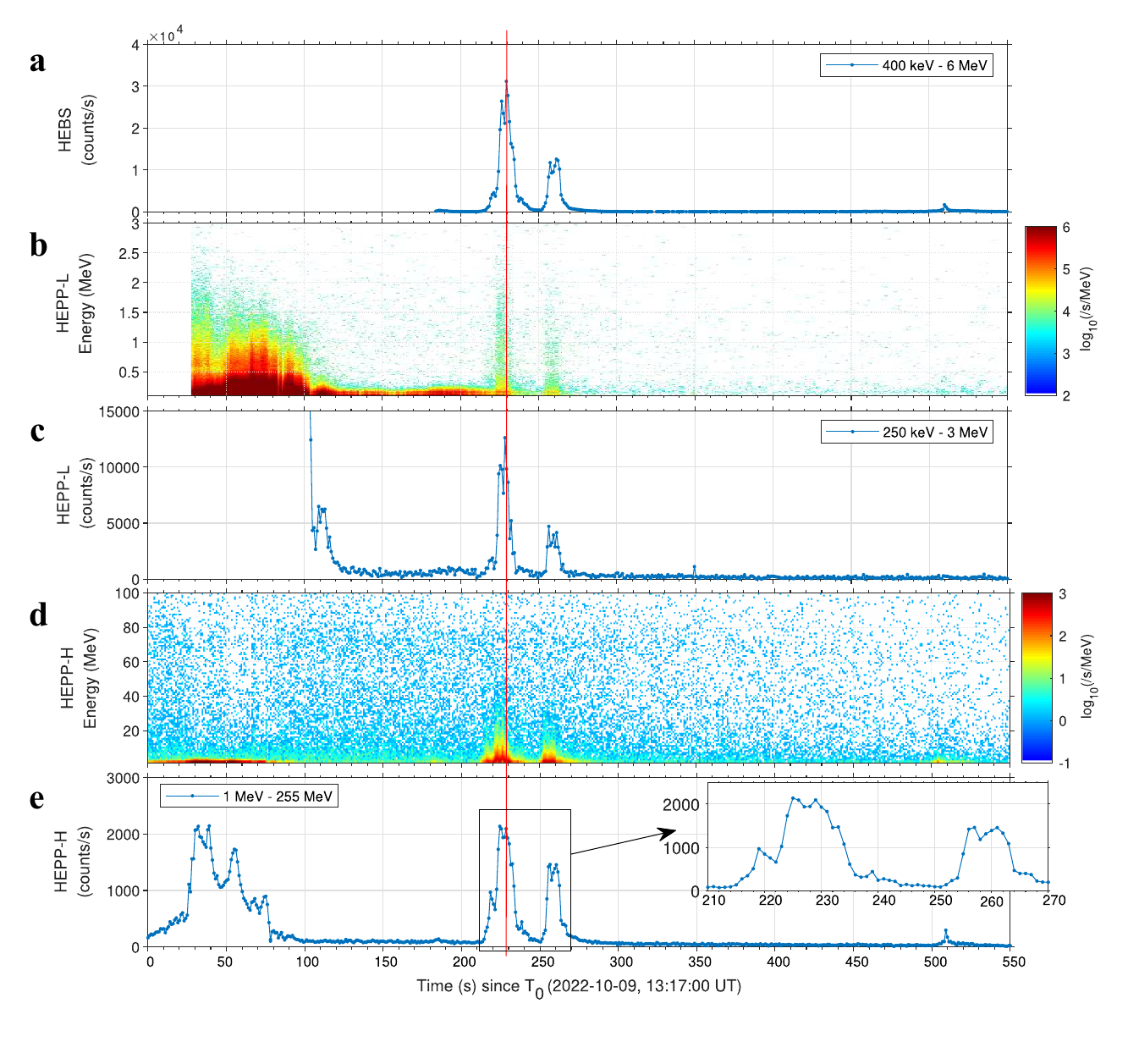}
  \caption{\textbf{Temporal evolution of GRB~221009A observed by GECAM-C(HEBS) and CSES-01.}
  \textbf{(a)} The light curve from GECAM-C in the 400\,keV--6\,MeV range, showing the primary gamma-ray emission. The GECAM-C data are shown starting from approximately $T_{\rm ref} + 180$\,s because the preceding interval was contaminated by high background due to the high-latitude effect \citep{2023arXiv230301203A}.
  \textbf{(a, c)} The energy-time spectrogram and integrated light curve (250\,keV--3\,MeV) of the secondary particles detected by HEPP-L. \textbf{(d, e)} The corresponding spectrogram and integrated light curve (1\,MeV--255\,MeV) from HEPP-H. The vertical red line marks the time of the main emission peak. %The striking temporal correlation between the primary gamma-ray peaks (\textbf{a}) and the secondary electron flux enhancements (\textbf{c, e}) provides unambiguous evidence that the particle events are induced by the GRB.
  }
  \label{fig:GE_CS_LC}
\end{figure}

% 原正文 Section 2 第一段
HEPP recorded the unattenuated signal from the primary burst phase across the 0.1--255\,MeV range. The temporal evolution of fluxes are shown in Fig.~\ref{fig:GE_CS_LC}. Detected within the stable low-background region of the inner radiation belt ($L$-shell = 2.2$-$2.7, where $L$ denotes the distance from the Earth's center to the magnetic equator in Earth radii), these signals represent the first direct measurement of GRB-induced secondary electrons spanning the MeV to sub-GeV region (Fig.~\ref{fig:GE_CS_LC}\textbf{d,e}). The two distinct enhancements in the fluxes recorded by HEPP (Fig.~\ref{fig:GE_CS_LC}\textbf{b-e}) are precisely synchronized with the primary peaks and subsequent flares observed from GECAM-C (Fig.~\ref{fig:GE_CS_LC}\textbf{a}) , confirming that the signal enhancements observed by HEPP originate from this GRB \citep{2023ApJ...946L..29B}. To correct instrumental effects driven by the unprecedented burst intensity, we developed a specialized calibration framework incorporating dynamic data acquisition (DAQ) modeling, detailed in Section~\ref{sec:DCHE}, which successfully recovered the intrinsic incident gamma-ray spectrum and effectively extended the reliable energy coverage up to $\sim$255\,MeV.

To ensure broadband spectral consistency, the HEPP measurements were anchored and validated against contemporaneous reference observations from GECAM-C and \textit{Fermi} (GBM and LAT). GECAM-C provides unsaturated low-energy coverage overlapping with HEPP-H, whereas \textit{Fermi}/GBM and LAT provide independent constraints across the MeV-to-GeV range. To anchor the low-energy analysis, we utilize contemporaneous observations from the GECAM-C instrument (HEBS) on board the SATech-01 satellite \citep{2025ExA....59...22L}. The GECAM-C payload is designed as a two-compartment structure to maximize its field of view and eliminate occultation from the satellite platform (Fig.~\ref{fig:GECAM_C}), providing a field of view of approximately 224$^{\circ}$ (excluding Earth occultation). The instrument comprises 12 individual Gamma-Ray Detector (GRD) probes, utilizing LaBr$_3$ or NaI(Tl) crystals coupled with Silicon Photomultipliers, which detect photons from approximately 6~keV up to 6~MeV \citep{2023NIMPA105668586Z}. Crucially for this work, when GRB~221009A occurred, GECAM-C was operating in a high particle radiation zone mode, with only one GRD probe and one CPD probe activated. Due to the fast electronic design of GRD and the fact that only one probe operates, it has not experienced signal saturation or count rate saturation (Fig.~\ref{fig:GRB221009A_LC}) \citep{2024SCPMA..6789511Z}. This study employs the GECAM-C dataset as detailed in previous publications to ensure consistency \citep{2023arXiv230301203A}.

To validate the high-energy portion of our spectra, we incorporate public data from the \textit{Fermi} Gamma-ray Space Telescope. The Gamma-ray Burst Monitor (GBM) \citep{2009ApJ...702..791M} provides complementary data in the 8 keV to 35 MeV range \citep{2023ApJ...952L..42L}. The Large Area Telescope (LAT) \citep{2009ApJ...697.1071A} is particularly vital, covering the high-energy gamma-ray band from 20~MeV to over 300~GeV \citep{2025ApJS..277...24A}. The LAT data serve as a high-energy benchmark, enabling us to verify that the spectrum recovered from HEPP-H smoothly connects to established measurements at higher energies.
% note

\subsection{Data Processing and Instrument Response}

A critical step in the data processing is the energy calibration, which corrects for the time-dependent gain attenuation of the calorimeter. Due to long-term exposure to cosmic rays and high-energy particles in the radiation belt, the gain of the scintillators and PMTs was affected, with a measured attenuation of 10--17\% (Table~\ref{tab:gain_degradation}) over approximately 56 months of in-orbit operation (February 2018 to October 2022). To correct for this, we performed a precise gain correction using cosmic-ray minimum ionizing particle (MIP) events. We collected MIPs that penetrated all calorimeter layers to obtain the spectra shown in Fig.~\ref{fig:HEPP_CAL_PSD}. We fitted these spectra with a Landau distribution, which describes the characteristic asymmetric energy-loss distribution of minimum-ionizing particles in detector materials, and adopted the most probable value (MPV) as the MIP peak position. The energy response is then precisely corrected by comparing the peak positions between the in-flight data and the pre-launch on-ground calibration results. We also assessed the instrument's response to the extreme brightness of GRB~221009A. The receiving area of the calorimeter is \SI{144}{cm^2}, and its detection sensitivity to gamma bursts is not high, implying that HEPP-H will not experience count saturation under such bright conditions.

For HEPP-H, which is more sensitive to high-energy cosmic rays and their secondaries, its background exhibited unpredictable variations between revisiting orbits, despite their nearly identical ground tracks. This suggests that these variations likely stem from fluctuations in the primary cosmic-ray flux or other non-local space weather phenomena \citep{2025SpWea..2304197Z}. Therefore, a simple background subtraction using a single orbit is unreliable for HEPP-H, as it introduces large, chaotic residuals that corrupt spectral analysis. To overcome this challenge of unstable HEPP-H background, we developed a robust background modeling method that effectively reduces the systematic error caused by these variations. To construct an average background model, we selected six revisiting orbits with clean background strictly prior to the burst, ensuring no contamination from the GRB's afterglow (Appendix B).

To establish the critical link between the observed count spectra and the incident gamma-ray spectrum, we generated the response matrix file (RMF) for HEPP-H through detailed Monte Carlo (MC) simulations using the GEANT4 toolkit \citep{agostinelli2003geant4, allison2016recent}. The detailed construction of the instrument mass model, the specific simulation configurations (e.g., secondary electron tracing), and the replication of on-orbit trigger conditions are comprehensively documented in Appendix B.

\subsection{Dynamic Correction for High-Flux Effects}
\label{sec:DCHE}
The unprecedented flux of GRB~221009A induced severe instrumental effects in HEPP-H, including dead time, pulse pile-up, and signal trailing, particularly during the peak emission phases. Consequently, the standard detector RMF, derived from simulations under more moderate flux conditions, is insufficient for an accurate spectral analysis. It is essential to develop a specific correction for these high-flux effects.

The primary instrumental effect is a non-negligible detector dead time of approximately $\tau = 240$~µs. In principle, the true event rate $N$ can be recovered from the measured count rate $n$ using the standard formula:
\begin{equation}
    N = \frac{n}{1 - n\tau}.
    \label{eq:dead_time}
\end{equation}

The effectiveness of this correction can be evaluated by examining the relationship between the energy spectrum counting rate (fully reconstructed events) and the total telemetry counting rate, which serves as a proxy for the incident flux. This total telemetry counting rate is acquired from a dedicated HEPP-H channel that functions as an independent counter. With a dead time of only 0.1~$\mu$s, the dead time effects are negligible in this GRB detection. For typical low-flux background orbits, the correction successfully restores a linear relationship between these two rates (Fig.~\ref{fig:Dead_Time_Correction}\textbf{a-f}). However, for the extreme flux of GRB~221009A, the correction partially linearizes the response and substantial dispersion remains, especially during the peak periods of highest telemetry rate (Fig.~\ref{fig:Dead_Time_Correction}\textbf{g}). This residual non-linearity indicates the presence of more complex, coupled instrumental effects. Therefore, the standard correction is not available.

To resolve these issues, we developed a comprehensive correction framework based on MC simulations of the full data-acquisition process \citep{2026ApJ...996....5D}. In each MC simulation, the energy and arrival time of every incident photon are sampled separately. The photon energy is drawn from the assumed input spectral model, while the arrival times are generated according to a Poisson process. For a trial incident photon rate $R_{\rm in}$, the rate parameter of the corresponding interarrival-time distribution is set to
\begin{equation}
\lambda=R_{\rm in}.
\label{eq:lambda}
\end{equation}
The time interval $\Delta t_n$ between the $n$th and $(n+1)$th incident photons follows an exponential distribution,
\begin{equation}
p(\Delta t_n)=\lambda\exp(-\lambda\Delta t_n).
\label{eq:deltatime_distribution}
\end{equation}
For each incident photon, a random number $u_n$ is drawn from a uniform distribution over $(0,1)$, and the corresponding time interval is generated using inverse-transform sampling:
\begin{equation}
\Delta t_n=-\frac{\ln(1-u_n)}{\lambda}.
\label{eq:deltatime_sampling}
\end{equation}
Starting from $t_0=0\,\mathrm{s}$, the arrival time of each subsequent photon is obtained recursively from
\begin{equation}
t_{n+1}=t_n+\Delta t_n,
\qquad
\Delta t_n=t_{n+1}-t_n,
\label{eq:arrival_time}
\end{equation}
where $t_n$ denotes the arrival time of the $n$th incident photon. Thus, a larger $\lambda$ produces a distribution concentrated toward shorter interarrival times and represents a higher incident photon rate.

Although $R_{\rm in}$, and hence $\lambda$, is controlled in each MC simulation, the appropriate value for a given observational interval is not known a priori. The telemetry rate cannot be used directly as $R_{\rm in}$ because it is a detector-level rate measured after photon--instrument interactions and trigger selection. We therefore scan the trial values of $\lambda$, simulate the complete detector and data-acquisition chain, and select the value for which the MC-corrected science-event rate agrees with the independently recorded telemetry rate. The selected simulation is then used to construct the dynamically corrected response matrix for that observational interval.

\section{Spectral Analysis and Results}

\subsection{Spectral Fitting Methodology}
To accurately characterize the broadband continuum and its evolution, we employed a set of standard astrophysical functions \citep{2019pgrb.book.....Z}, including a simple power-law (PL), a broken power-law (BKN), a cutoff power-law (CPL), and the Band function \citep{1993ApJ...413..281B}. Furthermore, to investigate the thermal emission from the photosphere and potential narrow spectral features, Blackbody (BB) and Gaussian components were also utilized. These models are mathematically described as follows:

\begin{equation}
A_{\rm PL}(E)=KE^{-\alpha},
\label{equ:pl_Model}
\end{equation}
where $K$ is the normalization amplitude constant ($\rm photons \cdot keV^{-1} \cdot cm^{-2} \cdot s^{-1}$), $\alpha$ is the photon index of power law.

\begin{equation}
A_{\rm CPL}(E) = K \left(\frac{E}{E_0}\right)^{-\alpha} \exp \left(-\frac{E}{E_{\rm c}}\right),
\label{equ:cpl_Model}
\end{equation}
where $E_0$ is the reference energy fixed at 1~keV, $\alpha$ is the power law photon index, $E_{\rm c}$ is the cutoff energy ($e$-folding energy) in keV, and $K$ is the amplitude ($\rm photons \cdot cm^{-2} \cdot s^{-1} \cdot keV^{-1}$).

\begin{equation}
    A_{\rm Band}(E)= \begin{cases} 
    K \bigg (\frac{E}{100}\bigg )^{\alpha} {\rm exp}\bigg(-\frac{E}{E_{\rm c}}\bigg ), \\
    \hfill (\alpha -\beta)E_{\rm c} \geq E, \\[2ex]
    K \bigg [\frac{(\alpha-\beta) E_{\rm c}} {100}\bigg ]^{\alpha-\beta} {\rm exp}(\beta-\alpha)\bigg (\frac{E}{100}\bigg )^{\beta}, \\
    \hfill (\alpha -\beta)E_{\rm c} \leq E. \\
    \end{cases}
    \label{equ:band_Model}
\end{equation}

where $K$ is the normalization amplitude constant ($\rm photons \cdot cm^{-2} \cdot s^{-1} \cdot keV^{-1}$), $\alpha$ is the first power law photon index, $\beta$ is the second power law photon index, and $E_{\rm c}$ is characteristic energy in keV.

\begin{equation}
A_{\rm BB}(E) = K \frac{C E^2}{(kT)^4 [\exp(E/kT) - 1]},
\label{equ:bb_Model}
\end{equation}
where $C \approx 8.0525$, $kT$ is the blackbody temperature in keV, and $K$ is the normalization constant defined as $L_{39}/D_{10}^2$, where $L_{39}$ is the source luminosity in units of $10^{39} \rm erg \, s^{-1}$ and $D_{10}$ is the distance to the source in units of 10 kpc.

\begin{equation}
A_{\rm Gauss}(E)=K\frac{1}{\sigma\sqrt{2\pi}}{\rm exp}(-\frac{(E-E_l)^2}{2\sigma^2}),
\label{equ:gauss_Model}
\end{equation}
where $K$ is the total $\rm photons \cdot cm^{-2} \cdot s^{-1} \cdot keV^{-1}$ in the line, $E_l$ is the line energy in keV and $\sigma$ is the line width in keV.

% \subsection{Thermal-like MeV emission}
\subsection{Additional MeV Component}
The interval $T_{\rm ref}+225$--$235$\,s represents the most intense phase of the burst and poses the most challenging observational conditions. The unprecedented photon flux caused severe saturation in most gamma-ray detectors, leaving the MeV spectral shape during this crucial phase poorly constrained. By HEPP-H's secondary-particle detection mechanism combined with aforementioned data reconstruction corrections \citep{2026ApJ...996....5D}, we can obtain an accurate measurement of the extreme fluxes of GRB, successfully overcoming the catastrophic saturation that is suffered by most of other instruments.

In this interval, GBM data were excluded due to severe saturation and pulse pile-up effects. Consequently, we utilized a joint dataset comprising GECAM-C (which remained unsaturated) and HEPP-H. To mitigate the influence of instrumental effects in HEPP-H caused by the extreme flux, we applied the dynamic correction framework detailed in Section~\ref{sec:DCHE}.

\begin{figure*}[htbp]
\centering

    \begin{subfigure}[t]{0.49\textwidth}
        \centering
        \includegraphics[width=\linewidth]{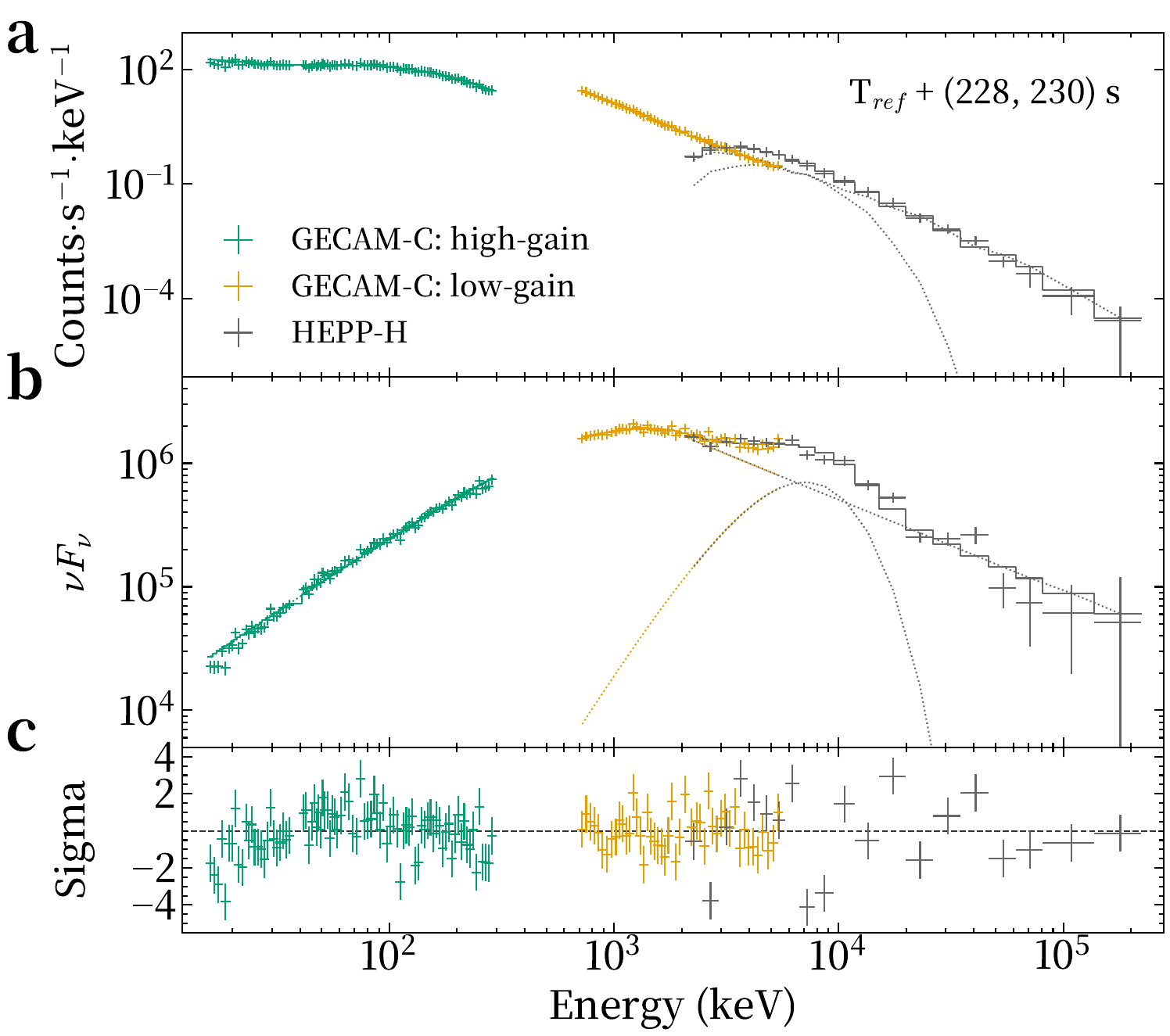}
    \end{subfigure}\hfill
    \begin{subfigure}[t]{0.49\textwidth}
        \centering
        \includegraphics[width=\linewidth]{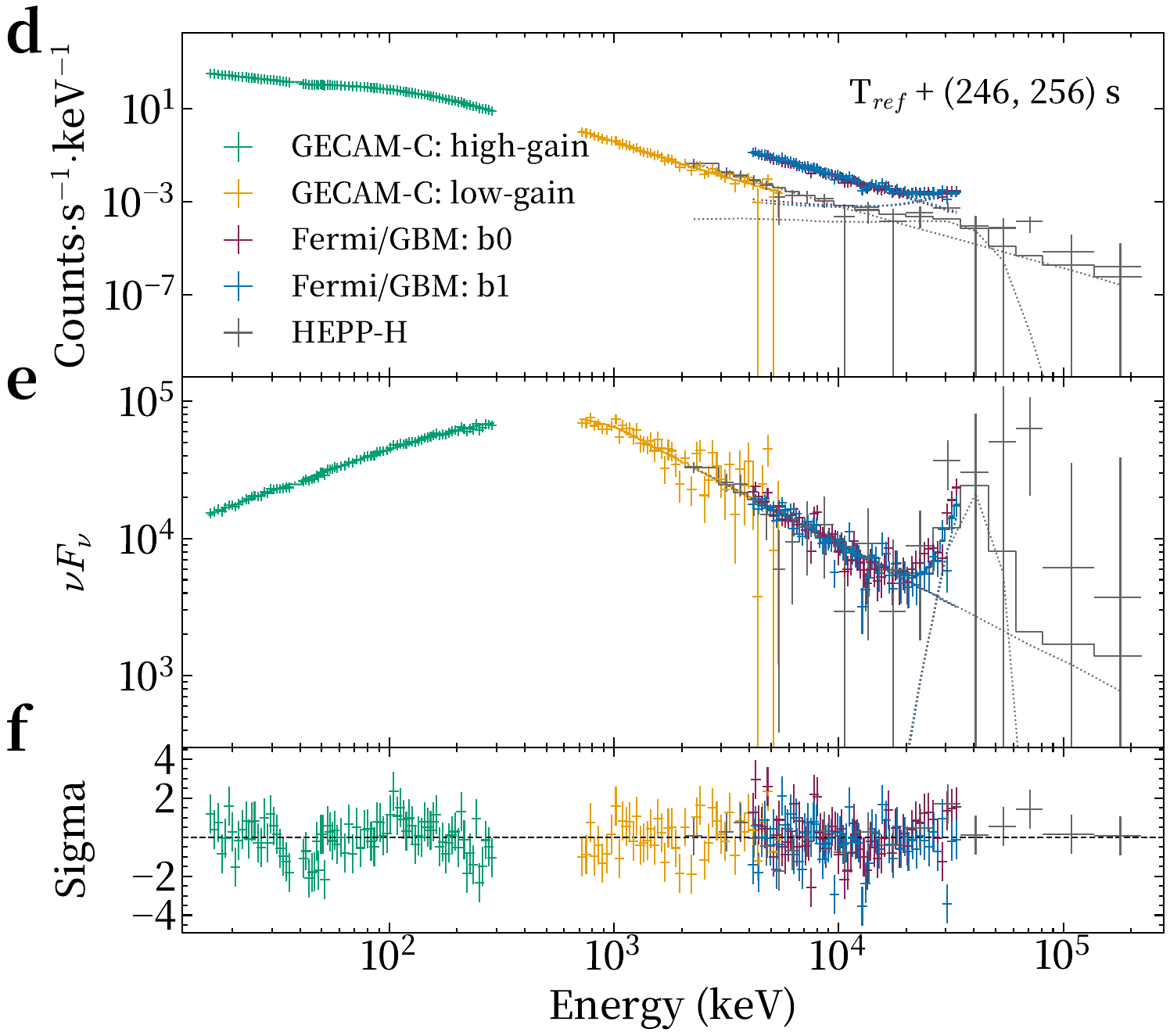}
    \end{subfigure}
    
  \caption{\textbf{Joint spectral characterization of GRB~221009A across multiple time periods.} 
  \textbf{(a--c)} Analysis for the high-flux peak phase ($T_{\rm ref} + 228$--$230$\,s) utilizing a Band + Blackbody model. 
  \textbf{(d--f)} Spectral analysis for the late decay phase ($T_{\rm ref} + 246$--$256$\,s) utilizing a multicomponent Band + Gaussian model. 
  \textbf{(a, d)} Observed count rate spectra compared with the folded best-fit models (solid lines). Data are from GECAM-C (high-gain, green; low-gain, orange), \textit{Fermi}/GBM BGO detectors  (b0, purple; b1, blue) and HEPP-H (grey). 
  \textbf{(b, e)} The $\nu F_\nu$ spectral distribution (units: $\mathrm{keV}^2 \cdot \mathrm{Photons} \cdot \mathrm{cm}^{-2} \cdot \mathrm{s}^{-1} \cdot \mathrm{keV}^{-1}$). For the peak phase (b), the displayed spectrum is fitted with the Band+Blackbody model. For the late phase \textbf{(e)}, the model includes a Band tail and a Gaussian line feature centered near 37\,MeV. In both time periods, HEPP-H data (grey) span the $2$--$255$\,MeV energy range. 
  \textbf{(c, f)} Residuals in units of standard deviation ($\sigma$) with $1\sigma$ error bars. }
  \label{fig:spectral_fit_results}
\end{figure*}

The joint spectral analysis yields an unprecedented isotropic-equivalent peak luminosity of $L_{\rm iso}\sim1.1\times10^{54}\,\mathrm{erg\,s^{-1}}$ during $T_{\rm ref}+230$--$231$\,s. Both $L_{\rm iso}$ and $E_{\rm iso}$ are calculated over the rest-frame energy range of 1\,keV--10\,MeV, corresponding to $1/(1+z)$--$10^{4}/(1+z)$\,keV in the observer frame. Over the $T_{\rm ref}+225$--$235$\,s interval, we obtain $E_{\rm iso}\sim4.9\times10^{54}$\,erg, consistent with the value of $\sim6.0\times10^{54}$\,erg independently derived from GECAM-C data \citep{2023arXiv230301203A}. To investigate the detailed spectral properties and their rapid evolution during the main peak, we divided this 10-second interval into six distinct time slices (detailed in Table~\ref{tab:methods_bb_summary}). Figure~\ref{fig:spectral_fit_results}\textbf{a-c} presents the representative spectral analysis for the high-flux interval of $T_{\rm ref} + 228$--$230$\,s.

The initial attempt to model the continuum using the standard Band function revealed a significant inadequacy: the fitting left a prominent, structured excess---a ``spectral bump''---specifically in the MeV energy range. To characterize the spectral curvature, we performed a comparative analysis of continuum models. A standard Band function ($H_0$) yielded unsatisfactory fits, leaving systematic residuals in the high-energy region. We therefore explored composite models incorporating a high-energy suppression or thermal component, specifically the ``Band + Cutoff Power-Law'' and ``Band + Blackbody'' models ($H_1$). The inclusion of these additional components resulted in a significant improvement in the goodness-of-fit.
As summarized in Table~\ref{tab:CPL_vs_BB_model}, both composite models substantially improve the fits relative to the Band-only model, with $\Delta\mathrm{AIC}\gg10$ in all time intervals. Notably, the ``Band + Cutoff Power-Law'' and ``Band + Blackbody'' models yielded comparable statistical performance, making them indistinguishable based on fit quality alone. Both models independently and consistently locate a distinctive spectral peak within the $\sim 2$\,MeV range (Fig.~\ref{fig:225_235_all_spectra}). This feature is consistently present across all time slices of the main peak, clearly demonstrating that during the brightest phase, the spectrum deviates from the typical high-energy dependence of the Band function.

\subsection{Evolution of the MeV line}

In the decay phase of $T_{\rm ref} + 246$--$256$\,s, by integrating data from HEPP-H, GECAM-C, and \textit{Fermi}/GBM, we resolved the continuous spectrum across the MeV window (Fig.~\ref{fig:spectral_fit_results}\textbf{d-f}), revealing that it cannot be adequately described by a simple Band function which leaves distinct structured residuals. The addition of a Gaussian emission line effectively eliminates these residuals and displays the optimal fitting (Fig.~\ref{fig:spectral_fit_results}\textbf{f}). In this joint fit, GECAM-C and \textit{Fermi}/GBM provide high-statistics constraints on the low energy wing of the line. Additionally, the HEPP-H data play an important role in describing the high energy wing, thereby resolving the complete spectral feature (Fig.~\ref{fig:spectral_fit_results}\textbf{d,e}). 

The inclusion of the Gaussian component contributed a substantial improvement in the goodness-of-fit compared to the baseline Band model. For the $T_{\rm ref} + (242, 251)$\,s interval, we obtained a $\Delta \text{cstat} = 44.06$ (defined as $\text{cstat}_{\rm Band} - \text{cstat}_{\rm Band+Gaussian}$), corresponding to a significance of $\approx 4.0\sigma$ (verified by $\sim 200,000$ MC simulations; Fig.~\ref{fig:line_significance}\textbf{a}). Furthermore, for the $T_{\rm ref} + (246, 256)$\,s interval, the improvement in the goodness of fit attains a higher value of $\Delta \text{cstat} = 190.22$, representing a statistical significance exceeding $4.5\sigma$ (Fig.~\ref{fig:line_significance}\textbf{b}). These level exceed the standard $3\sigma$ detection threshold, confirming the existence of such a feature.

Built upon the robust spectral line detection established in the late phase of the light curve ($T_{\rm ref} + 246$--$256$\,s), we confidently push our analysis backward into its earlier, more complex decay stage ($T_{\rm ref} + 237$--$255$\,s). Performing the same joint broadband analysis, we find that a distinct spectral excess emerges above the standard Band continuum in the tens of MeV range. By implementing a sliding-window analysis, we identify a significance of $\sim$4$\sigma$ in the $T_{\rm ref} + 242$--$251$\,s interval. Within this interval, the emission line can be robustly constrained with a free-fitted width, and the ratio of the line width to the line energy is consistent with the 10\% reported in previous study within uncertainties \citep{2024SCPMA..6789511Z}. Earlier intervals (e.g., $T_{\rm ref} + 237$--$246$\,s, Fig.~\ref{fig:line_fit}) also show consistent excesses, although its significance is relatively low (2.6$\sigma$). 

\begin{figure*}[p]
\centering
  \includegraphics[width=1.5\columnwidth]{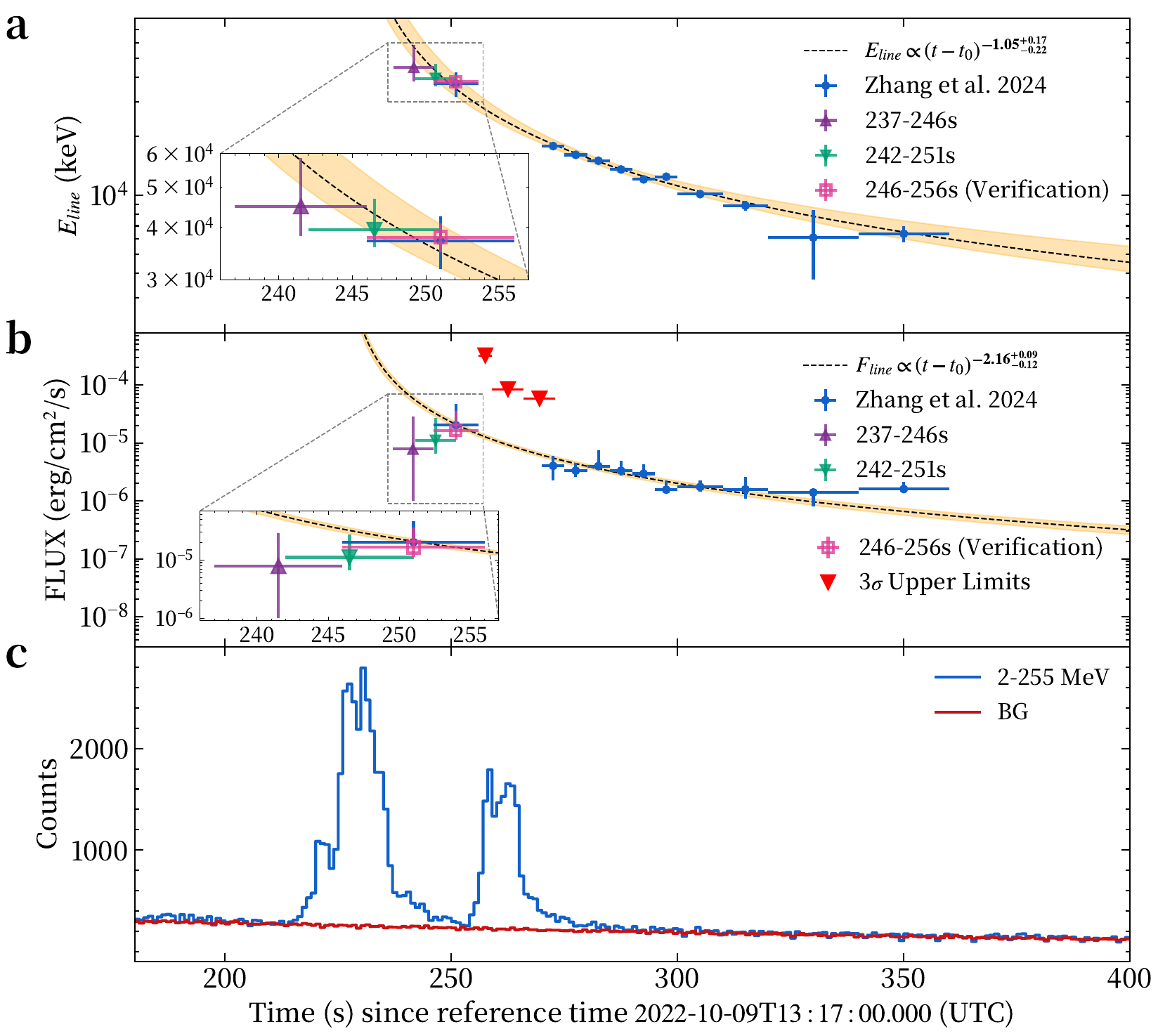}
  \caption{\textbf{Evolution of emission line parameters in GRB~221009A.}
  \textbf{(a)} Temporal evolution of the central line energy ($E_{\mathrm{line}}$). The black dashed line depicts the power-law decay model [$E_{\mathrm{line}} \propto (t-t_0)^{-1.05}$] constrained by data from later intervals. The purple, cyan, and pink crosses mark the measurement results for the $T_{\rm ref} + 237$--$246$\,s, $242$--$251$\,s, and $246$--$256$\,s intervals, respectively. Note that the blue crosses represent observations from previous study \citep{2024SCPMA..6789511Z}.
  \textbf{(b)}  Evolution of the emission line flux. The black dashed line indicates the power-law decay trend [$F_{\mathrm{line}} \propto (t-t_0)^{-2.16}$]. No significant emission line signals were detected in the 256--273\,s interval, providing only $3\sigma$ flux upper limits (red triangles), where $3\sigma$ represents the statistical uncertainty of the flux measurement (equivalent to a 99.7\% confidence level). \textbf{(c)} The HEPP-H count rate light curve (2--255\,MeV).}
  \label{fig:line_evolution}
\end{figure*}

To constrain the evolution of this spectral feature, we extended the analysis across three distinct time intervals, integrating the HEPP-H measurements with trends reported in contemporaneous studies \citep{2024SCPMA..6789511Z}. The fitted results are summarized as follows (see Fig.~\ref{fig:line_evolution}):

\begin{enumerate}
    \item \textbf{$T_{\rm ref}+(242, 251)$\,s:}  
    For this time interval, the fit obtains a central energy of $39.4_{-3.6}^{+7.4}$\,MeV (the asymmetric errors, where the upper bound exceeds the lower bound, are consistent with the power-law decay of the line energy), a Gaussian width ($\sigma_{\rm line}$) of $3.1 \pm 2.3$\,MeV, and a line flux of $1.11_{-0.45}^{+1.59} \times 10^{-5}$\,erg\,cm$^{-2}$\,s$^{-1}$. The results are consistent with the backward extrapolation of the central energy and the power-law evolution law [$E_{\mathrm{line}} \propto (t-t_0)^{-1.05}$] \citep{2024SCPMA..6789511Z, 2026arXiv260114257S}. However, the measured line flux remains lower than the value predicted by the extrapolation of the flux decay trend [$F_{\mathrm{line}} \propto (t-t_0)^{-2.16}$] \citep{2024SCPMA..6789511Z, 2026arXiv260114257S}.

    \item \textbf{$T_{\rm ref}+(246, 256)$\,s:}
    We performed a joint spectral fit incorporating HEPP-H, GECAM-C, and \textit{Fermi} (GBM and LAT). The resulting line parameters are consistent with those reported in the previous analysis work \citep{2024SCPMA..6789511Z}, and this successful reproduction serves as a validation of our data analysis pipeline. By extending the analysis to high energies ($>$200 MeV) with the help of \textit{Fermi}/LAT data, we reveal the complete broadband spectral shape (Fig.~\ref{fig:246_256s}). This global view unveils a clear bimodal structure: a Band$+$Gauss component is dominated for energy range of 10 keV-200 MeV, while the high-energy region exhibits a distinct spectral upturn following a power-law at $>200$\,MeV. Additionally, we explicitly determined the $3\sigma$ lower limit of the line flux for this interval, providing a constraint on the signal intensity profile during the main emission phase.

    \item \textbf{$T_{\rm ref}+(256, 273)$\,s:} No significant spectral line was detected in this phase. To constrain the fading behavior, we calculated the $3\sigma$ upper limits on the line flux. In this procedure, the central energy was fixed according to the evolutionary law $E \propto (t-t_0)^{-1.05}$ established by previous work \citep{2024SCPMA..6789511Z}, and the line width ($\sigma_{\rm line}$) was fixed to 10\% of the central energy. We then performed a spectral fit to determine the normalization of this fixed-shape Gaussian component. The $3\sigma$ upper limit was calculated as the best-fit flux value plus three times its $1\sigma$ statistical uncertainty. The resulting upper limits are plotted as red triangles in Fig.~\ref{fig:line_evolution}. 
\end{enumerate}

These time-resolved detections and upper limits bring two critical insights into the evolution of the emission line. Firstly, regarding the line energy, the observation extends the established evolutionary track \citep{2024SCPMA..6789511Z} earlier by approximately 5\,seconds. By integrating these early data with subsequent observational intervals (Fig.~\ref{fig:line_evolution}\textbf{a}), the measured central energies robustly confirm the overall temporal decay of the line energy \citep{2024SCPMA..6789511Z, 2026arXiv260114257S}. The results from the sliding-window analysis align well with the best fit of the line energy evolution from previous data \citep{2024SCPMA..6789511Z} within its uncertainty range, a power-law decay of $E_{\mathrm{line}} \propto (t-t_0)^{-1.05}$ where $t_0 = T_{\rm ref} + 226$\,s (Fig.~\ref{fig:line_evolution}\textbf{a}). The temporal index and $t_0$ were adopted from a previous analysis in
which both parameters were allowed to vary freely
\citep{2024SCPMA..6789511Z}. Given the correlated sliding windows and the relatively low significance of the earliest feature (2.6$\sigma$), we do not refit this relation here; our measurements are consistent with its backward extrapolation within their uncertainties. Secondly, within the $242$--$251$\,s window, the measured line flux is lower than what would be expected from an extrapolation of the later decay trend. Combined with the subsequent fading observed in later intervals, this early deficit represents a ``rise-then-fall'' flux profile (Fig.~\ref{fig:line_evolution}\textbf{b}).

\begin{table*}
    \centering
    \caption{Spectral fitting parameters obtained with the Band+Blackbody model}
    \label{tab:methods_bb_summary}
    \centering
    \rmfamily 
    \renewcommand{\arraystretch}{1.3}
    \resizebox{\textwidth}{!}{%
        \begin{tabular}{lcccccccc}
            \toprule
            \multirow{2}{*}{Time range (s)} & \multicolumn{3}{c}{Band Component} & \multicolumn{1}{c}{Blackbody} & \multicolumn{4}{c}{Statistics} \\
            \cmidrule(lr){2-4} \cmidrule(lr){5-5} \cmidrule(lr){6-9}
            & $\alpha$ & $\beta$ & $E_{\rm c}$ (keV) & $kT$ (keV) & chi/dof & AIC$_{\rm Band}$ & AIC$_{\rm Band+BB}$ & $\Delta\text{AIC}$\\
            \midrule
            (225, 228) & $-0.82 \pm 0.01$ & $-2.45 \pm 0.02$ & $1281 \pm 39$   & $2801 \pm 59$  & 526/150 & 1308 & 525 & 783\\
            (228, 230) & $-0.76 \pm 0.01$ & $-2.75 \pm 0.02$ & $1073 \pm 33$   & $1803 \pm 51$  & 267/150 & 705 & 277 & 428\\
            (230, 231) & $-0.77 \pm 0.02$ & $-2.72 \pm 0.02$ & $1560 \pm 63$  & $2105 \pm 71$  & 410/150 & 561 & 417 & 144\\
            (231, 232) & $-0.72 \pm 0.02$ & $-2.60 \pm 0.03$ & $970 \pm 42$   & $2269 \pm 89$  & 239/150 & 450 & 247 & 203\\
            (232, 233) & $-0.77 \pm 0.02$ & $-2.63 \pm 0.04$ & $863 \pm 44$   & $2123 \pm 112$  & 291/150 & 404 & 301 & 103\\
            (233, 235) & $-0.80 \pm 0.01$ & $-2.77 \pm 0.03$ & $761 \pm 38$   & $1019 \pm 62$  & 207/150 & 305 & 223 & 82\\
            \bottomrule
        \end{tabular}%
    }
    \vspace{1ex}
    \begin{flushleft}
    \footnotesize
    \textbf{Note:} Spectral parameters for the Band+Blackbody (BB) model. $\Delta\text{AIC}$ represents the improvement in the Akaike Information Criterion of the Band+BB model relative to the single Band model ($\Delta\text{AIC} = \text{AIC}_{\rm Band} - \text{AIC}_{\rm Band+BB}$). Positive values indicate a statistical preference for the multicomponent model.
    \end{flushleft}
\end{table*}

\section{Physical interpretation}

\subsection{Additional MeV component}
The additional spectral component at $\sim 2$\,MeV in $T_{\rm ref} + 225$--$235$\,s (Fig.~\ref{fig:spectral_fit_results}\textbf{a,b})  has not been reported in the other GRBs before. Such a high energy component beyond the GRB continuum cannot find a natural interpretation from existing GRB theories, except for the photosphere emission from a relativistic fireball \citep{1986ApJ...308L..43P, 2000ApJ...530..292M}. In the following, we consider it as photosphere origin and discuss the implication.

The GRB central engine is expected to form an initially hot ($\sim$MeV) and optically thick fireball with a huge radiation energy density. The fireball expands and produces the thermal photosphere emission when the optical depth drops to unity. Observations of thermal component in the GRB prompt emission have been reported often, but usually with temperatures of tens to hundreds of keV \citep{%2011AIPC.1358...33G, 2012ApJ...757L..31A, 
2011ApJ...730..141Z}, lower than the initial temperature because the photosphere emission occurs in the fireball coasting phase where the thermal radiation adiabatically cools significantly.

In contrast, the exceptionally high temperature ($\sim 2$\,MeV) observed here indicates that the thermal emission is emitted in the fireball acceleration phase, where the apparent temperature is equal to the initial temperature, because of the Lorentz boost compensating the adiabatic cooling. In this case the observed temperature is a measurement of the initial fireball temperature. 
Moreover, as in the acceleration phase the thermal radiation keeps a significant fraction of the initial fireball energy, not converted to kinetic energy of the fireball mass yet, the total energy of the observed thermal emission is roughly the initial fireball energy. 
However, the thermal component is only a small fraction $\sim6\%$ of the total emission energy; thus the bulk GRB emission is powered by the other larger energy component, which is very likely the magnetic field energy. 

In total, the observed MeV spectral bump hints that the thermal fireball is a sub-dominant component of a jet and that the bulk GRB prompt emission is powered by magnetic energy of a magnetically dominated jet released by the central engine (see Appendix \ref{sec:theory-bump} for more detailed discussion). 

In addition, the CPL model can fit comparably well as the blackbody model, but with a power-law index of $\sim-1$ (Table~\ref{tab:CPL_vs_BB_model}), significantly softer than the Planck distribution. This may be due to the fact that since the blackbody component is not significantly higher than the Band component, the low-energy power law of the Planck distribution cannot be fully displayed. Moreover, some facts may result in softer thermal emission from the photosphere of the GRB fireball, e.g. the high latitude effect \citep{2011ApJ...732...49P}, or the angular structure of jets \citep{2013MNRAS.428.2430L}. 

%Second, 
\subsection{MeV line emission}
The temporal evolution of the emission line, which exhibits a central energy of $\sim$40 MeV during its early stage, may provide hints for the jet physics. The line energy evolution $E_{\mathrm{line}} \propto (t-t_0)^{-1}$ (Fig.~\ref{fig:line_evolution}\textbf{a}), without fine tuning, naturally finds an interpretation of the curvature effect \citep{2000ApJ...541L..51K} of electron--positron annihilation line from a relativistically expanding sphere with a large radius of $r \approx 2.45 \times 10^{16} (\Gamma/500)$\,cm \citep{2024ApJ...973L..17Z}. Importantly, the narrow line width ($\sim10\%$) requires rapid radiative cooling of the electron--positron pairs before annihilation, which cannot be satisfied by inverse-Compton scatterings of prompt emission photons, thus may imply synchrotron/cyclotron cooling of pairs in a highly magnetized region, independently pointing to the existence of magnetically dominated jet in the BOAT \citep{2024ApJ...973L..17Z}. %This finding independently corroborates the conclusion drawn from the thermal emission analysis: the outflow must be a magnetically dominated jet. 

However, the flux evolution---specifically the ``rise-then-fall'' pattern (Fig.~\ref{fig:line_evolution}\textbf{b})---deviates significantly from the  scaling $F_{\mathrm{line}} \propto (t-t_0)^{-3}$ expected from the simple curvature effect. We attribute this early suppression to the angular structure of the jet, where emission is low near the jet axis but increases toward higher latitudes. The peak time of the line flux, $t_p$, corresponds to a characteristic viewing angle $\theta_p \sim 0.01 [(t_{p}-t_0)/20\,\mathrm{s}]^{1/2} (r/10^{16}\,\mathrm{cm})^{-1/2}$. This geometry implies a %``hollow-cone'' or 
ring-shaped magnetic structure, where the magnetic field transported from the central engine is relatively weak around the jet center but larger at the jet periphery (Fig.~\ref{fig:physical_process_picture}; see more detailed discussion in Appendix \ref{sec:theory-line}).

\begin{figure*}
    \centering
    \includegraphics[width=0.85\linewidth]{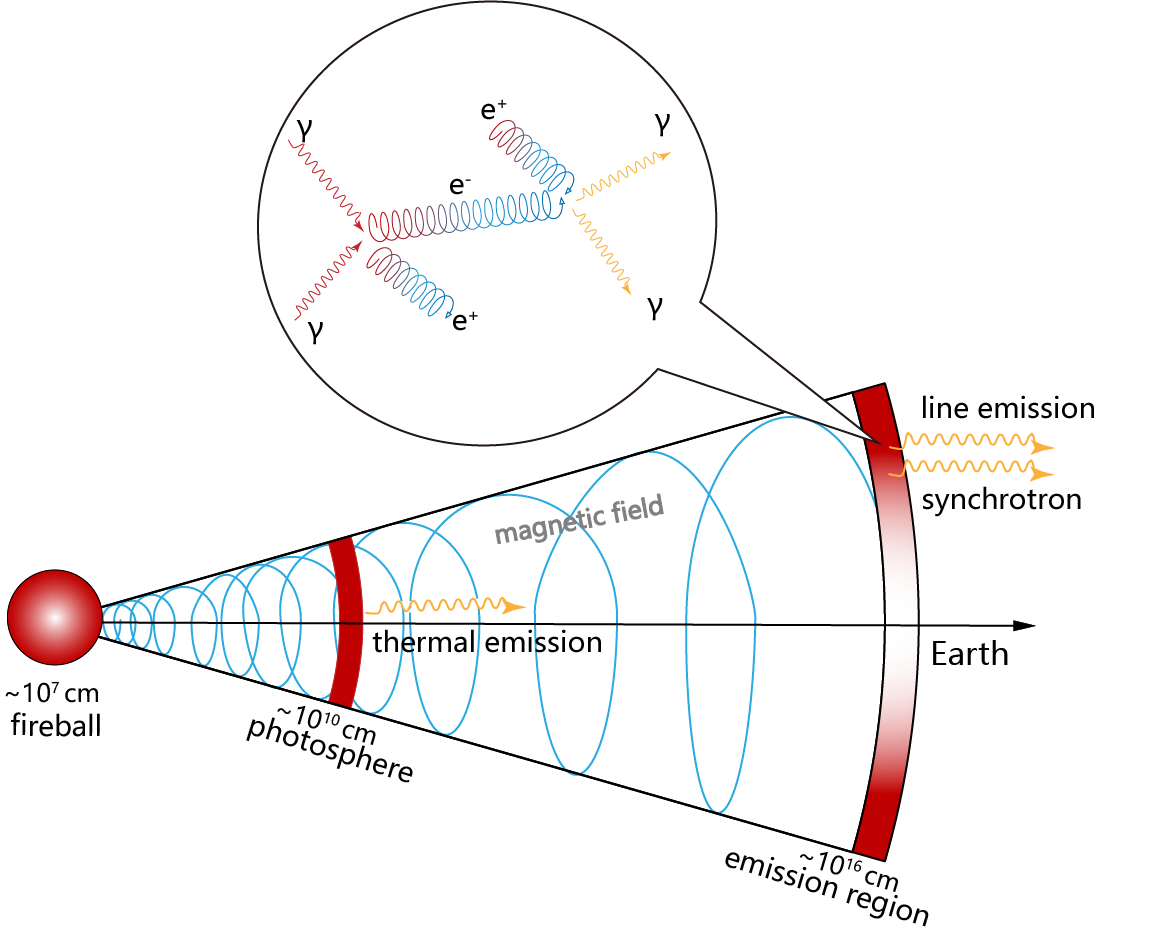}
    \caption{\textbf{Speculated schematics of the radiation processes in GRB~221009A.} The central engine generates both a hot thermal component (fireball) and a cold component (magnetic jet). The fireball, with initial radius $r_0\sim10^7$cm, expands and accelerates. A thermal emission with temperature of a few MeV, $T_{\rm ph}\simeq2$ MeV, and luminosity similar to the fireball power, $L_{\rm ph}\simeq L_w$, is released at the photosphere radius $\sim10^{10}$cm when the fireball become optically thin but before the acceleration phase ends. The energy dissipation of the magnetic jet occurs at much larger radii, around $\sim10^{16}$cm, where the prompt non-thermal emission at $\sim\rm keV-GeV$ is generated by synchrotron/inverse-Compton radiation. The thermal emission energy is only a fraction of the nonthermal emission energy, which implies that the GRB jet is not dominated by thermal fireball energy but by magnetic field. In the same emission region,  the $\gamma\gamma$ interactions lead to the production of $e^\pm$ pairs which quickly cool by synchrotron radiation before annihilating each other into emission line photons. However, at lower latitudes, due to the weaker magnetic field and thus slower electron cooling, the line emission is much lower, which leads to the "rise-then-fall" profile of temporal evolution of the line flux.
    }
    \label{fig:physical_process_picture}
\end{figure*}

\section{Summary}

In this study, we overcome the catastrophic instrumental saturation caused by the BOAT GRB 221009A using the unattenuated particle-detector measurements from CSES-01/HEPP-H. By dynamically correcting the dead-time effects under extreme fluxes and cross-calibrating with GECAM-C and \textit{Fermi}, we reconstruct the continuous prompt emission spectrum across the MeV--sub-GeV band. Our main findings and physical implications are summarized as follows:

\begin{enumerate}
    % \item We discover a prominent, unusually hot thermal-like component peaking at $\sim$2\,MeV during the most intense main emission phase ($T_{\rm ref} + 225$--$235$\,s). This component is consistent with photosphere emission originating from a highly pure fireball in its acceleration phase.
    \item We discover a prominent spectral component peaking at $\sim$2\,MeV during the most intense main emission phase ($T_{\rm ref} + 225$--$235$\,s). This component is consistent with photosphere emission originating from a highly pure fireball in its acceleration phase. %\textcolor{red}{However, we cannot rule out the possibility that the additional component originates from other emission mechanisms.}
    \item By time-integration, the thermal emission accounts for only $\sim$6\% of the total gamma-ray energy release. This indicates that the thermal fireball is a sub-dominant component, and the bulk of the GRB prompt emission is powered by the magnetic energy of a magnetically dominated jet.
    \item We robustly identify an emission line in the decay phase, whose central energy decays following a $E_{\rm line} \propto (t-t_0)^{-1.05}$ evolution, naturally explained by the curvature effect of pair annihilation. Furthermore, the early deficit in the line flux reveals a ``rise-then-fall'' temporal profile. This early suppression, coupled with the narrow line width ($\sim$10\%), implies rapid synchrotron cooling of pairs within a structured, magnetically dominated jet that is relatively weak around the jet axis but stronger at higher latitudes.
\end{enumerate}

% Overall, these unsaturated MeV--sub-GeV observations provide a rare and comprehensive glimpse into the central engine of the BOAT, pointing to a structured, magnetically dominated jet accompanying a sub-dominant hot fireball.
Overall, these unsaturated MeV--sub-GeV observations provide a rare and comprehensive glimpse into the central engine of the BOAT, pointing to a structured, magnetically dominated jet likely accompanied by a sub-dominant hot fireball.

\begin{acknowledgments}

We thank the CSES, GECAM, \textit{Fermi}, and LHAASO teams for their operations, and acknowledge the CSES Data Center, the China Centre for Resources Satellite Data and Application (CRESDA), the GECAM Science Data Center, the \textit{Fermi} Science Support Center, the LHAASO Data Center, and the National High Energy Physics Science Data Center for providing data access.

We thank the CSES, GECAM, \textit{Fermi}, and LHAASO teams for their operations, and acknowledge the CSES Data Center, the China Centre for Resources Satellite Data and Application (CRESDA), the GECAM Science Data Center, the \textit{Fermi} Science Support Center, the LHAASO Data Center, and the National High Energy Physics Science Data Center for providing data access.
\end{acknowledgments}

%\paragraph*{Funding:}
This work was supported by the National Natural Science Foundation of China (grant Nos. 12673120, 42474222, U2541210, 12173038, and 12575116) and the National Key R\&D Program of China (grant No. 2024YFA1611402). R.B. and R.I. acknowledge support from the Space It Up project, funded by the Italian Space Agency (ASI) and the Ministry of University and Research (MUR) under contract n. 2024-5-E.0 -- CUP n. I53D24000060005. This work was also supported by the Strategic Priority Research Program of the Chinese Academy of Sciences (grant Nos. XDA30050100 and XDA15360102).

\section*{Competing interests}
There are no competing interests to declare.

\section*{Data availability}
The data that support the findings of this study are available from the LHAASO, GECAM, and \textit{Fermi} (GBM and LAT) websites (\url{https://www.ihep.ac.cn/lhaaso/}, \url{http://gecam.ihep.ac.cn/}, and \url{https://Fermi.gsfc.nasa.gov/ssc/data/}). The CSES-01 data can be downloaded from the Data Service website (\url{https://www.leos.ac.cn/}), where scientists can access the data download service by registering online and filling the application form following the data policy. No physical materials were generated in this work.

\appendix
\restartappendixnumbering
\section{Instrument Configuration and Observational Geometry}

\subsection{CSES-01 Satellite Platform and Orbit}

\begin{figure*}[t]
\centering
  \includegraphics[width=0.8\textwidth]{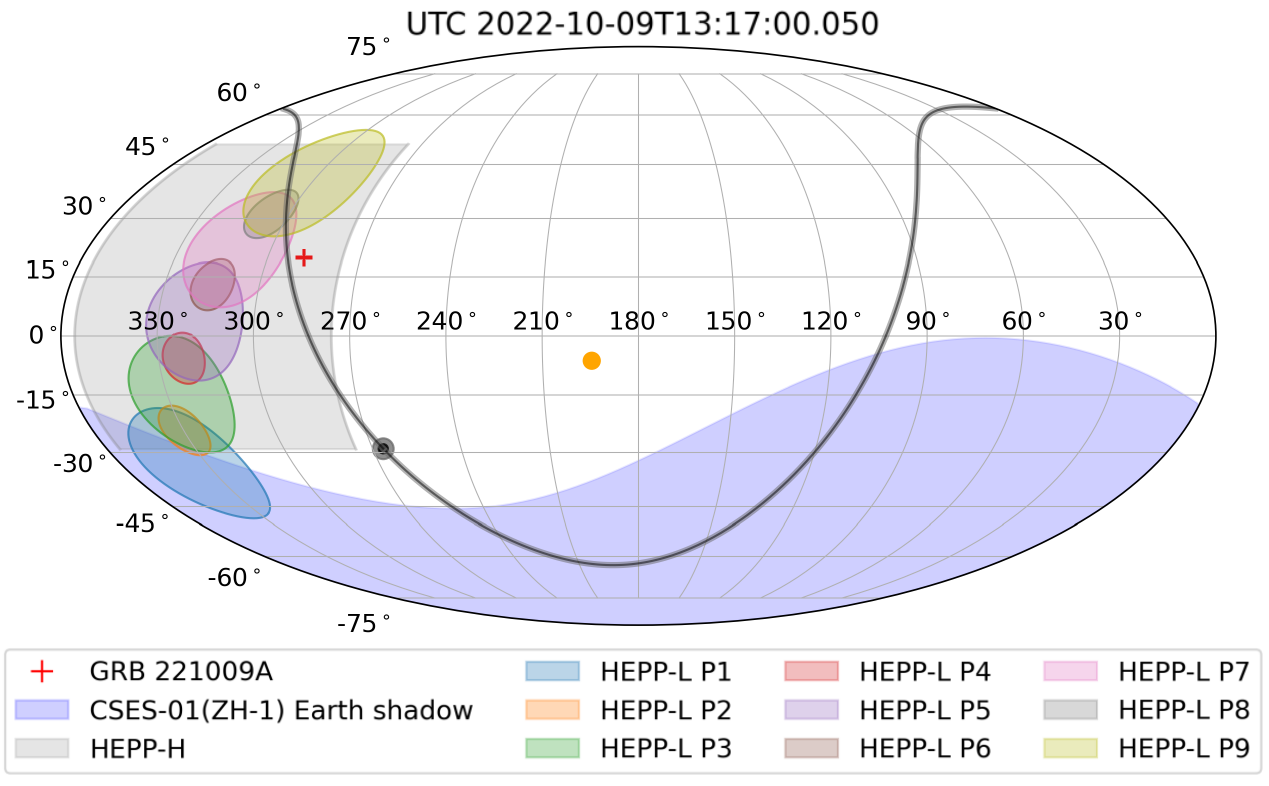}
  \caption{\textbf{Observational configuration of GRB~221009A relative to CSES-01.} Satellite-frame sky map at the reference trigger time ($T_{\rm ref}$), with the red cross indicating the GRB position. Shaded regions denote the fields of view (FoVs) for HEPP-H (gray) and the nine HEPP-L units (colored ellipses, P1--P9); blue shading represents Earth occultation. }
  \label{fig:coverage}
\end{figure*}

The primary observational data for this study were collected by the High-Energy Particle Package (HEPP) onboard the China Seismo-Electromagnetic Satellite (CSES-01). During GRB 221009A, CSES-01 was traversing a stable region of the inner radiation belt over Europe (Fig.~\ref{fig:coverage}), providing a low-background environment for detecting the burst-induced secondary-particle signals. Launched in February 2018, CSES-01 operates in a sun-synchronous orbit at an altitude of 507 km. The satellite performs science operations between latitudes of $\pm65^{\circ}$ along its 97$^{\circ}$ inclination orbit. The orbital period is 94.6 minutes, with a revisiting period of five days \citep{2019RDTM....3...22L}. The ascending node on the dayside flying from north to south is around 14:00 MLT, while the descending node on the nightside flying from south to north is around 2:00 MLT \citep{2018ScChE..61..634S}. At the time of the burst, the satellite was traversing a mid-latitude region over Europe within the inner radiation belt. This specific location is characterized by a stable particle background environment, which is addressed in the background modeling section.

\subsection{HEPP Detector Architecture}

\begin{figure*}[t]
\centering

  \begin{subfigure}[t]{0.4\textwidth}
      \caption{} % note
      \includegraphics[width=\linewidth]{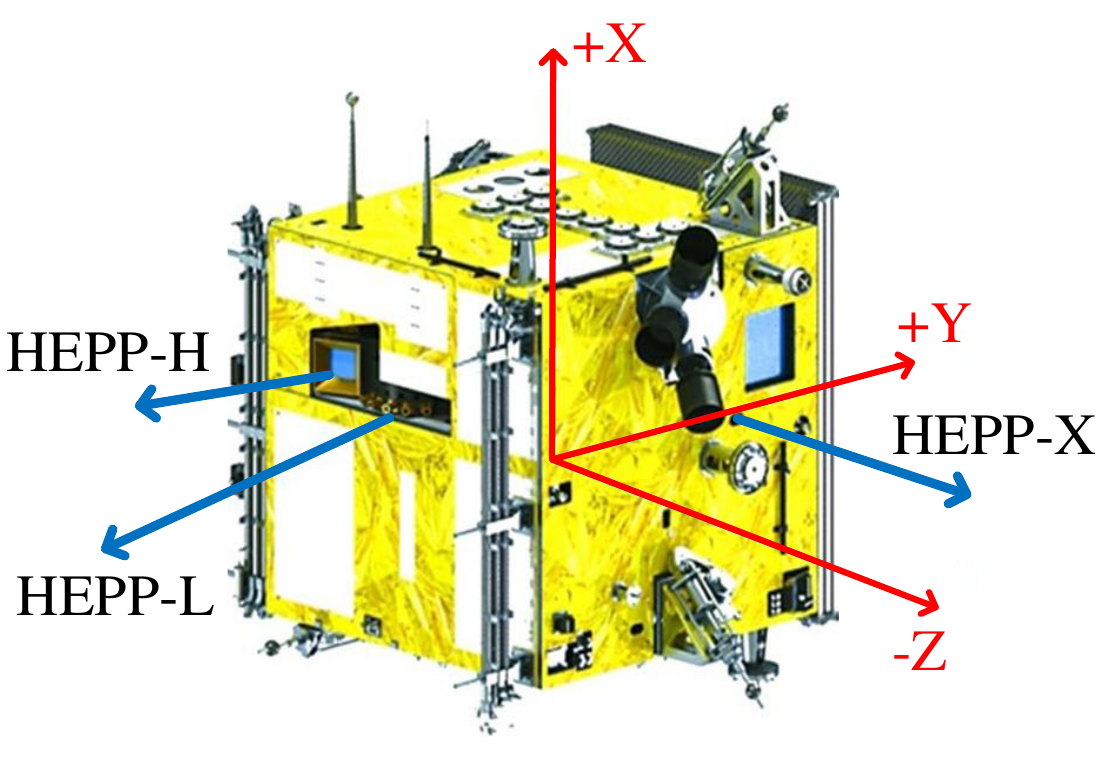}
      \label{subfig:cses_frame}
  \end{subfigure}
  \begin{subfigure}[t]{0.29\textwidth}
      \caption{} % note
      \includegraphics[width=\linewidth]{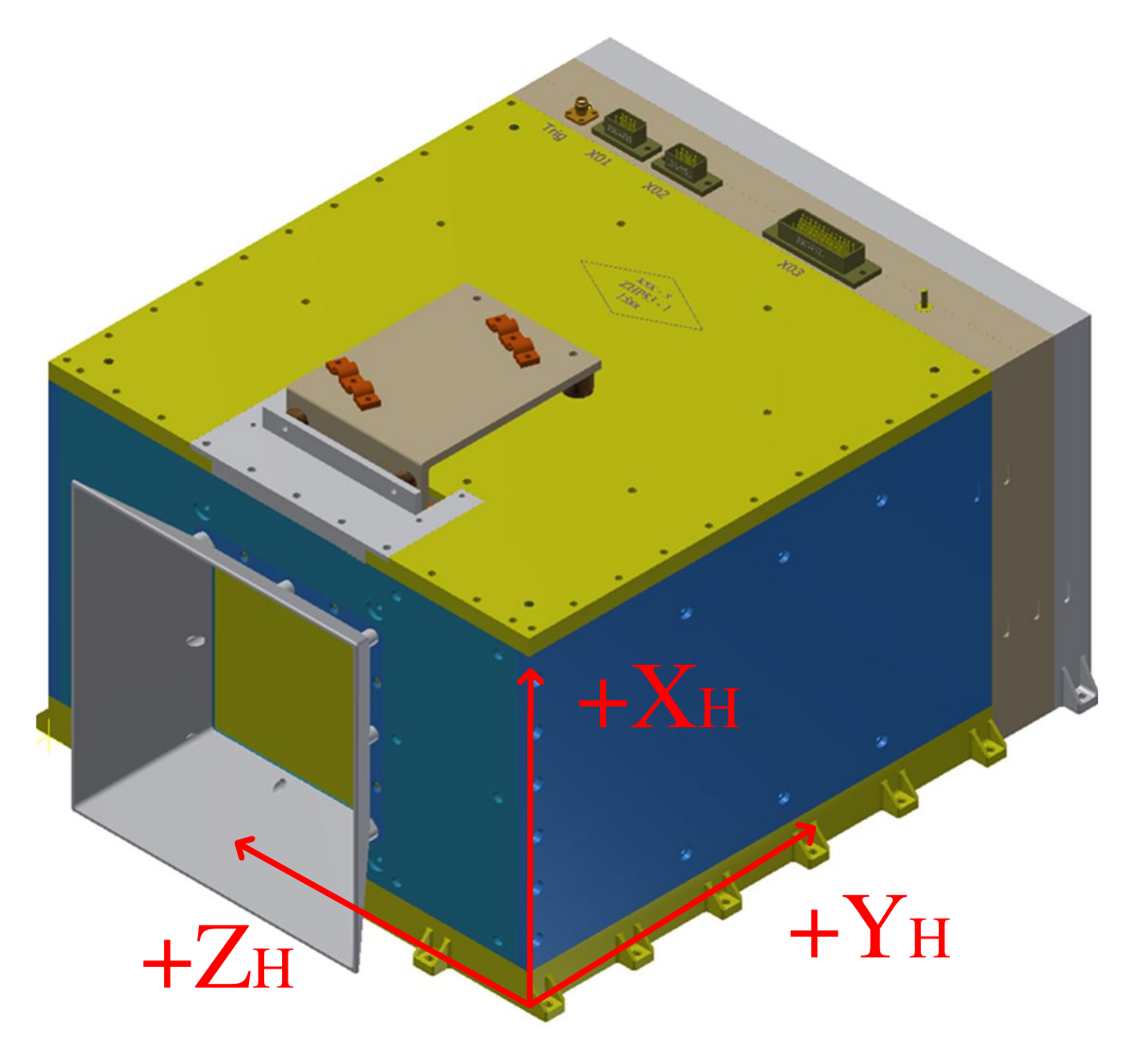}
      \label{subfig:hepp_h_frame}
  \end{subfigure}\hfill
  \begin{subfigure}[t]{0.29\textwidth}
      \caption{} % note
      \includegraphics[width=\linewidth]{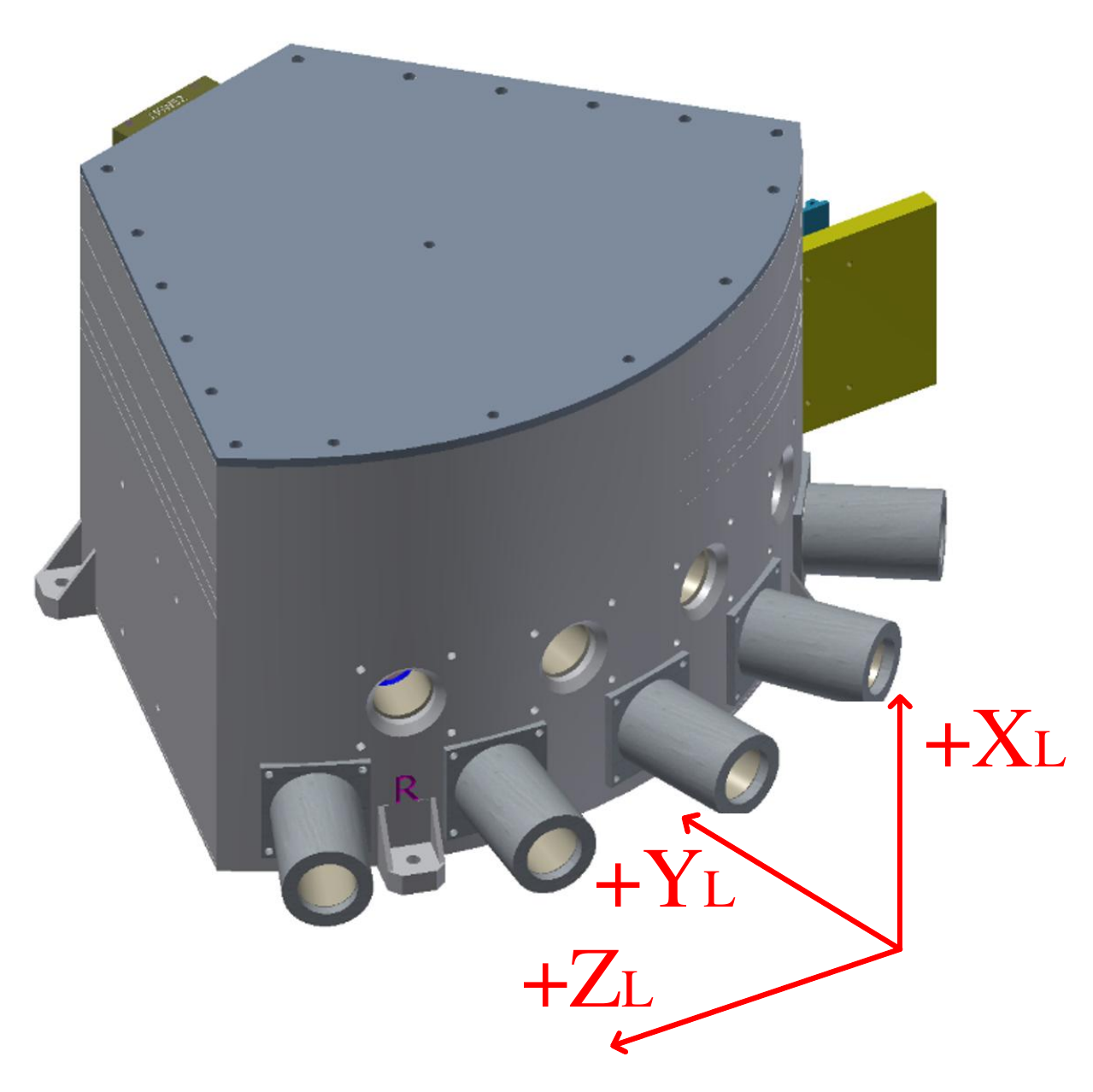}
      \label{subfig:hepp_l_frame}
  \end{subfigure}

  \vspace{0.4cm}

  \begin{subfigure}[t]{0.45\textwidth}
      \caption{} % note
      \includegraphics[width=\linewidth]{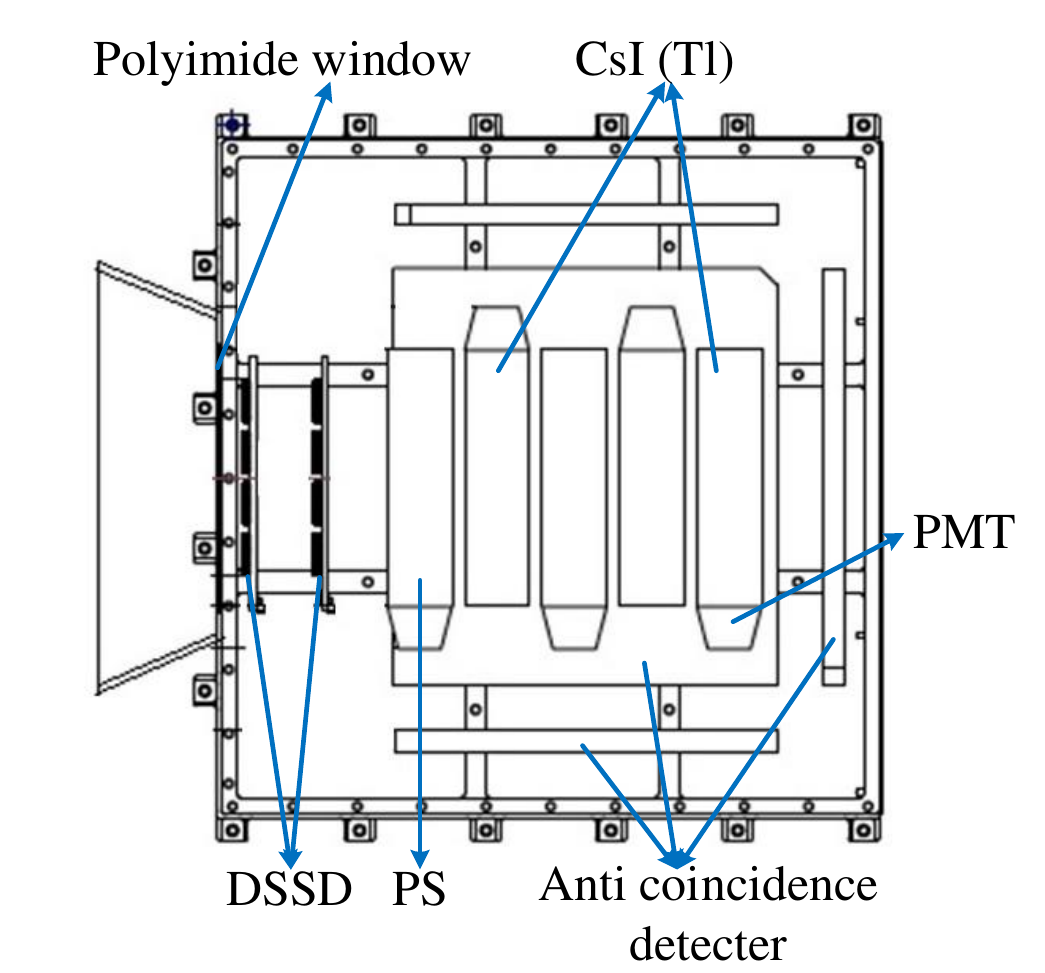}
      \label{subfig:hepp_h_diagram}
  \end{subfigure}\hfill
  \begin{subfigure}[t]{0.48\textwidth}
      \caption{} % note
      \includegraphics[width=\linewidth]{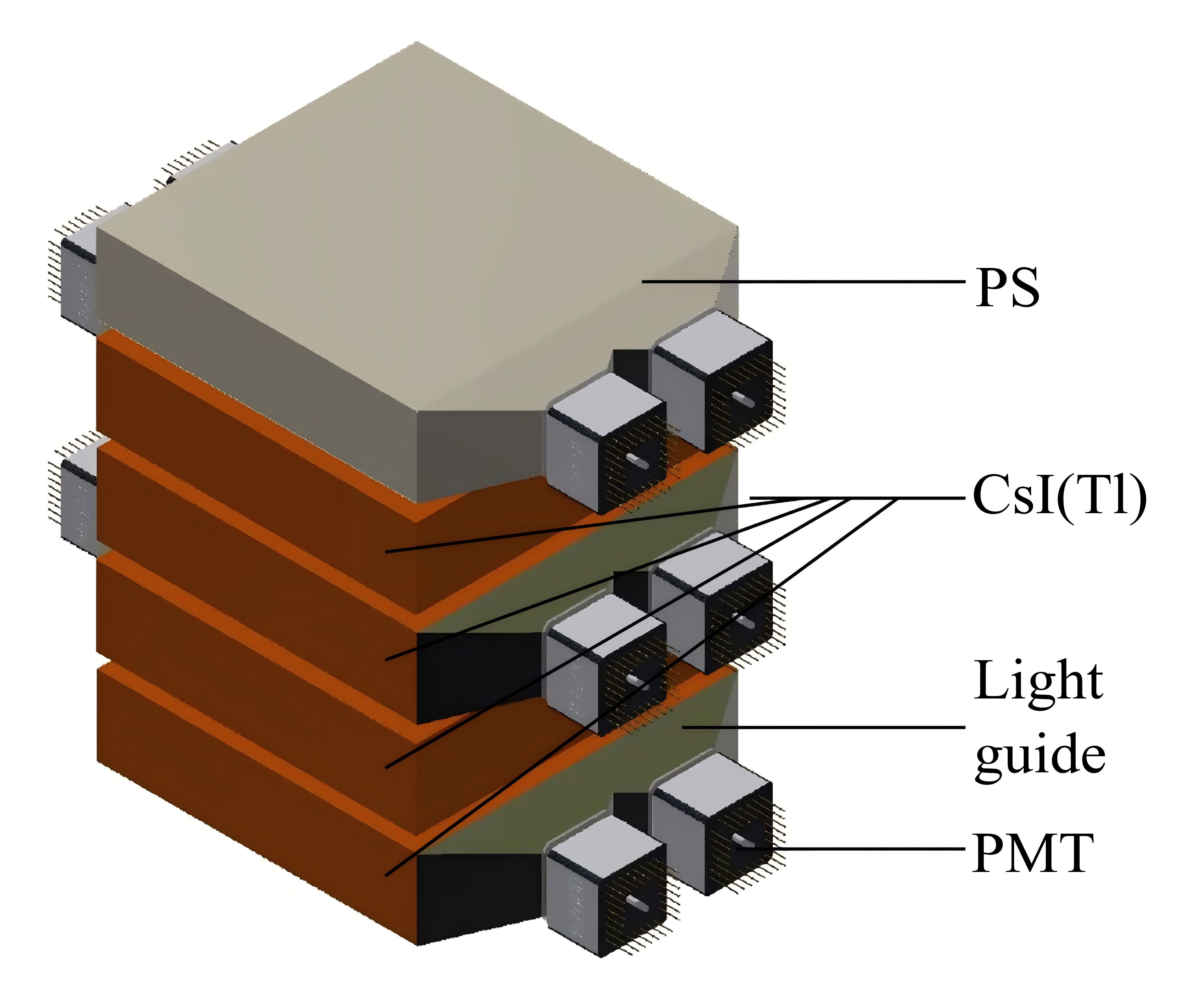}
      \label{subfig:hepp_h_model}
  \end{subfigure}
  
  \caption{\textbf{Coordinate reference frames and internal configuration of the CSES-01 satellite and HEPP instruments.} 
  \textbf{(a)} Schematic of the CSES-01 satellite platform and its primary reference frame ($X, Y, Z$), where the $+X$-axis aligns with the flight direction, the $-Z$-axis points towards the zenith, and the $+Y$-axis completes the right-handed orthogonal system (facing the Sun). 
  \textbf{(b, c)} Schematics of the HEPP-H \textbf{(b)} and HEPP-L \textbf{(c)} instruments; local coordinate systems are defined with $X_{H/L}$-axes parallel to the satellite $X$-axis frame and $Y_{H/L}/Z_{H/L}$ axes rotated within the $YZ$-plane by 20$^{\circ}$ and 12$^{\circ}$, respectively. 
  \textbf{(d)} Cross-sectional view of the HEPP-H assembly illustrating the arrangement of double-sided silicon strip detectors (DSSD), PS trigger layer, CsI(Tl) crystals, and peripheral anti-coincidence detectors. 
  \textbf{(e)} 3D structural diagram of the calorimeter, comprising one PS and four CsI(Tl) layers coupled to photomultiplier tubes (PMTs) for signal readout.}
  \label{fig:coordinate frames}
\end{figure*}

We use both the HEPP-H and HEPP-L detectors in this work \citep{2019RDTM....3...22L, 2018RAA....18..154N}; their key performance parameters are summarized in Table~\ref{tab:hepp_params}. The internal configuration and coordinate reference frames of these instruments are illustrated in Fig.~\ref{fig:coordinate frames}. A distinct advantage of HEPP is its large geometric factor, which significantly enhances its sensitivity to particle fluxes compared to other similar instruments \citep{2019RDTM....3...22L, 2022JGRA..12730550Z}.

\textbf{HEPP-H:} This detector serves as the primary instrument for spectral reconstruction. The front-end trigger is provided by the first 3-cm-thick plastic scintillator (PS) layer of the calorimeter stack. However, due to the low detection efficiency of the PS for gamma rays, consequently, the instrument remained unsaturated even under the extreme photon flux of the burst. The two-layer silicon microstrip detector (DSSD, 300 $\mu$m thick) is instead utilized for particle direction measurement and species identification (Fig.~\ref{fig:coordinate frames}\textbf{d}), while its contribution to the total energy measurement is negligible. The primary energy measurement is performed by the instrument's calorimeter stack, which consists of a 3-cm-thick plastic scintillator (PS) layer used for trigger and low-energy detection and four subsequent layers of 3-cm-thick CsI(Tl) crystals for higher-energy particle containment (Fig.~\ref{fig:coordinate frames}\textbf{e}). All calorimeter layers are matched with photomultiplier tubes (PMTs) for signal readout \citep{2019RDTM....3...22L}.

\textbf{HEPP-L:} This detector is used for independent temporal verification \citep{2023ApJ...946L..29B}. It consists of nine telescope detection units, collimation units, and a rear anti-coincidence detector. Each telescope unit utilizes two sensitive silicon detectors (300 $\mu$m and 2000 $\mu$m thick) based on the $\Delta E - E$ principle. However, due to the thinness of the silicon sensors, high-energy incident photons tend to be penetrating, depositing only a small fraction of their energy (predominantly below 2 MeV). Therefore, HEPP-L data are solely used to construct an independent light curve to confirm the temporal correlation of the observed signal \citep{2019RDTM....3...22L}.

\subsection{Observational Geometry and Pointing Stability}

\begin{figure*}[t]
\centering
  \includegraphics[width=1\textwidth]{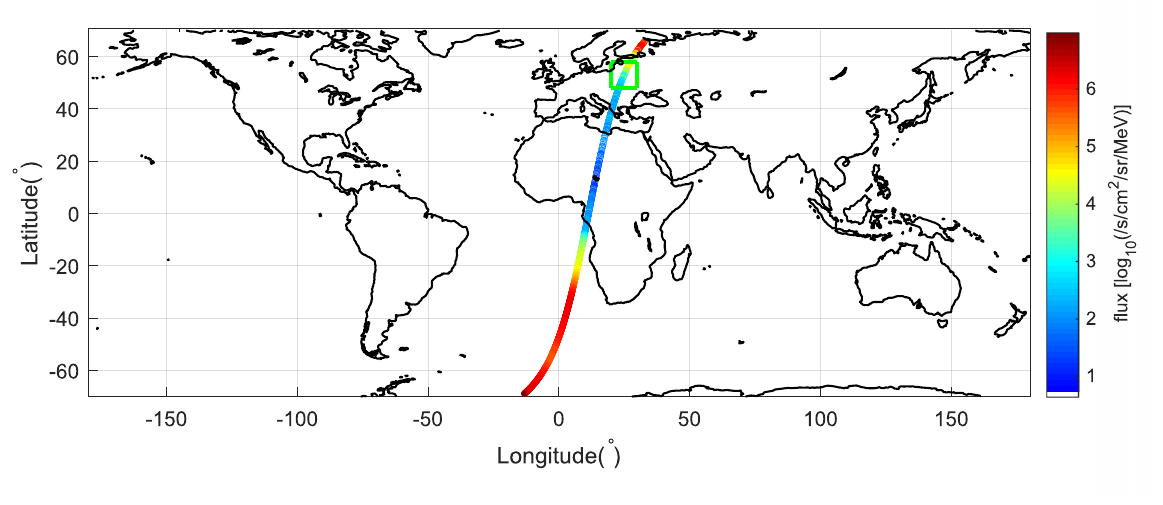}
  \caption{\textbf{Satellite orbit and flux distribution during gamma ray burst occurrence.} The figure shows the ground track of the CSES-01 satellite during orbit 259770 on October 9, 2022. The color scale along the track indicates the measured electron flux by HEPP-L. The green rectangular box highlights the segment of the orbit over Europe where electron pulses corresponding to the transient emission from GRB 221009A were detected. This specific location corresponds to a stable region within the inner radiation belt.}
  \label{fig:orbit_map}
\end{figure*}

Precise knowledge of the observational geometry during the GRB is critical for the analysis. The CSES-01 satellite maintains a three-axis stability relative to the ground, with a pointing accuracy of $\leq0.1^{\circ}$ ($\sigma$), a measurement accuracy of $\leq0.05^{\circ}$ (3$\sigma$) and an attitude stability of 0.005$^{\circ}$/s (3$\sigma$) \citep{yuan2018system}. The CSES-01 satellite's primary reference frame is defined with the $+X$-axis pointing along the flight direction and the $-Z$-axis pointing towards the zenith, while the $+Y$-axis is determined by the right-hand rule (facing the Sun). Relative to this primary frame, the local coordinate systems of the HEPP instruments share the same $+X$-axis but are generated by a right-handed rotation around it by 20$^{\circ}$ for HEPP-H and 12$^{\circ}$ for HEPP-L. This rotation aligns their local $Y$ and $Z$ axes within the satellite's $Y$-$Z$ plane. These alignments were measured post-assembly using a laser tracker with an angular error of less than 12$''$. Fig.~\ref{fig:coordinate frames}\textbf{a-c} provides schematics of the coordinate systems used in this analysis. The sky coverage and fields of view (FoVs) of the HEPP units at the burst time are shown in Fig.~\ref{fig:orbit_map}, providing the geometric context for the incident-angle calculation.

\begin{figure*}[t] 
    \centering 
    
    \begin{subfigure}[t]{0.7\textwidth}
        \caption{} 
        \includegraphics[width=\linewidth]{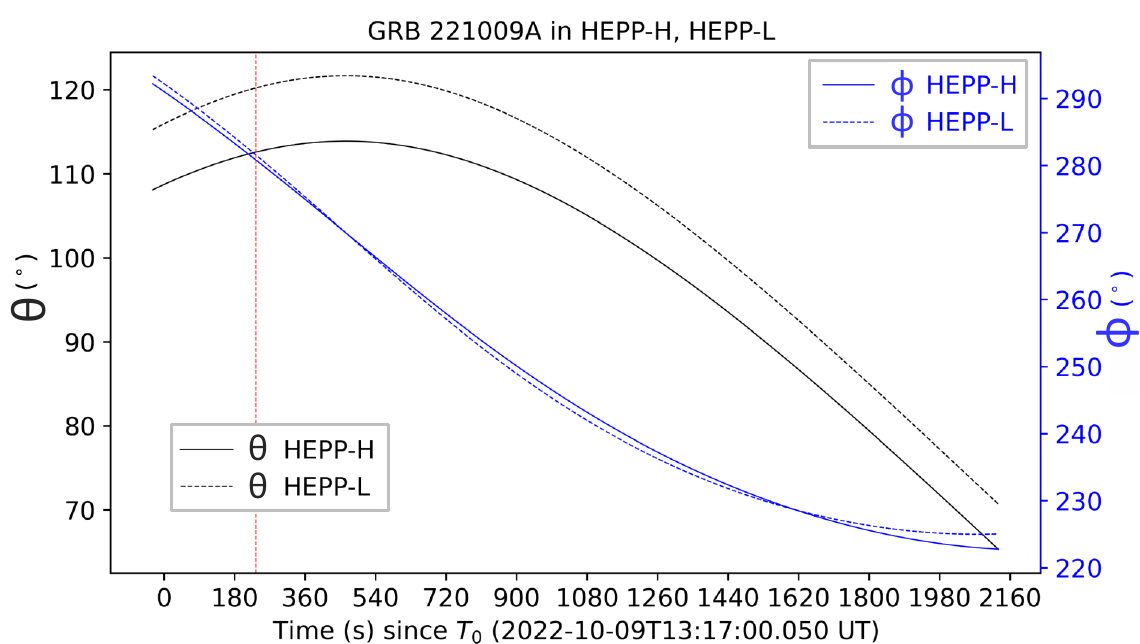}
        \label{subfig:evolution}
    \end{subfigure}

    %\vspace{0.1cm} 
    
    \begin{subfigure}[t]{0.86\textwidth}
        \caption{} 
        \includegraphics[width=\linewidth]{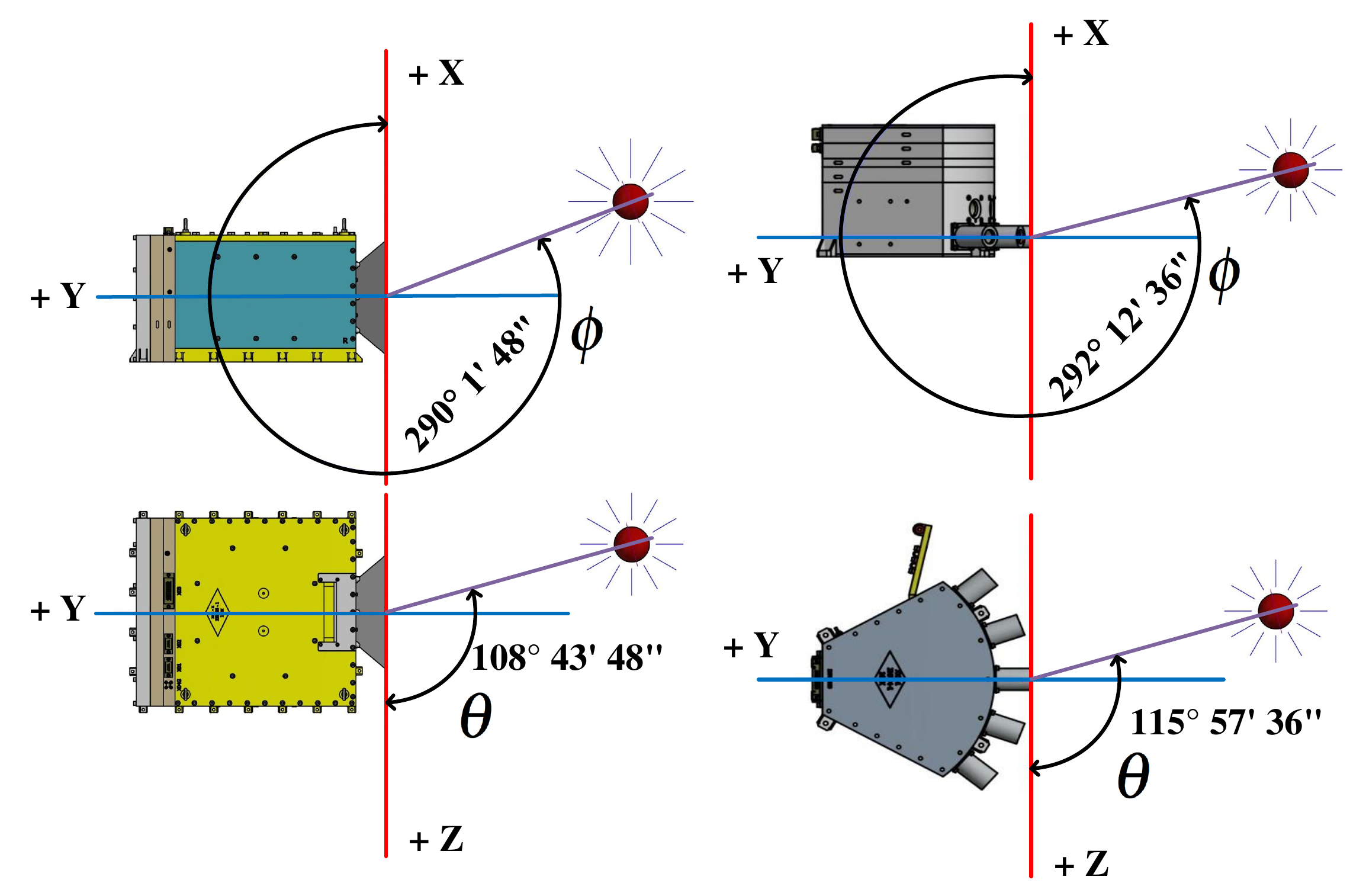}
        \label{subfig:angle}
    \end{subfigure}
    
    \caption{\textbf{Incident geometry of GRB~221009A relative to HEPP.} 
    \textbf{(a)} Temporal evolution of the polar ($\theta$; black, left axis) and azimuthal ($\phi$; blue, right axis) incident angles for HEPP-H (solid) and HEPP-L (dashed). The vertical red dotted line marks $T_{\rm ref}$. 
    \textbf{(b)} Detailed incident geometry of the burst source (red vector) within the detector coordinate systems ($+X, +Y, +Z$). Left and right panels display vector projections onto the $X$--$Y$ and $Y$--$Z$ planes for HEPP-H and HEPP-L, respectively.}
    \label{fig:incidentAng}
\end{figure*}

To determine the time-resolved incident direction of the GRB with respect to each detector (Fig.~\ref{fig:incidentAng}\textbf{b}), we used attitude quaternions along with the rotation matrix of the HEPP-L and the HEPP-H coordinate system relative to the satellite coordinate system. This calculation gives the incident polar ($\theta$) and azimuthal ($\phi$) angles of GRB 221009A in each local coordinate system. The temporal evolution of these angles is presented in Fig.~\ref{fig:incidentAng}\textbf{a}. The highly accurate localization of GRB 221009A from a multitude of space and ground-based observatories ensures that the intrinsic directional error of the burst is minimal \citep{2022GCN.32648....1D, 2023ApJ...946L..24W}. As observed in the analysis, the total variation of the incident angle during the main burst phase was small (approximately $\pm 5^{\circ}$). This stability justifies the approximation of using a single, fixed direction for the primary instrument response simulation, with the associated systematic uncertainty rigorously evaluated in the Error Analysis section.

\section{Data Processing and Calibration}

\subsection{Background Modeling and Sensitivity Analysis}

A final critical step in the data processing pipeline is the precise background subtraction. For HEPP-L, the background was found to be relatively stable across revisiting orbits in this specific region. Consequently, a standard background subtraction method, utilizing data from an adjacent revisiting orbit preceding the burst, proved sufficient and was effectively applied to its light curve analysis. 

\begin{figure*}[t]
\centering

\begin{subfigure}[t]{0.48\textwidth}
 \caption{}
    \centering
    \includegraphics[width=\linewidth]{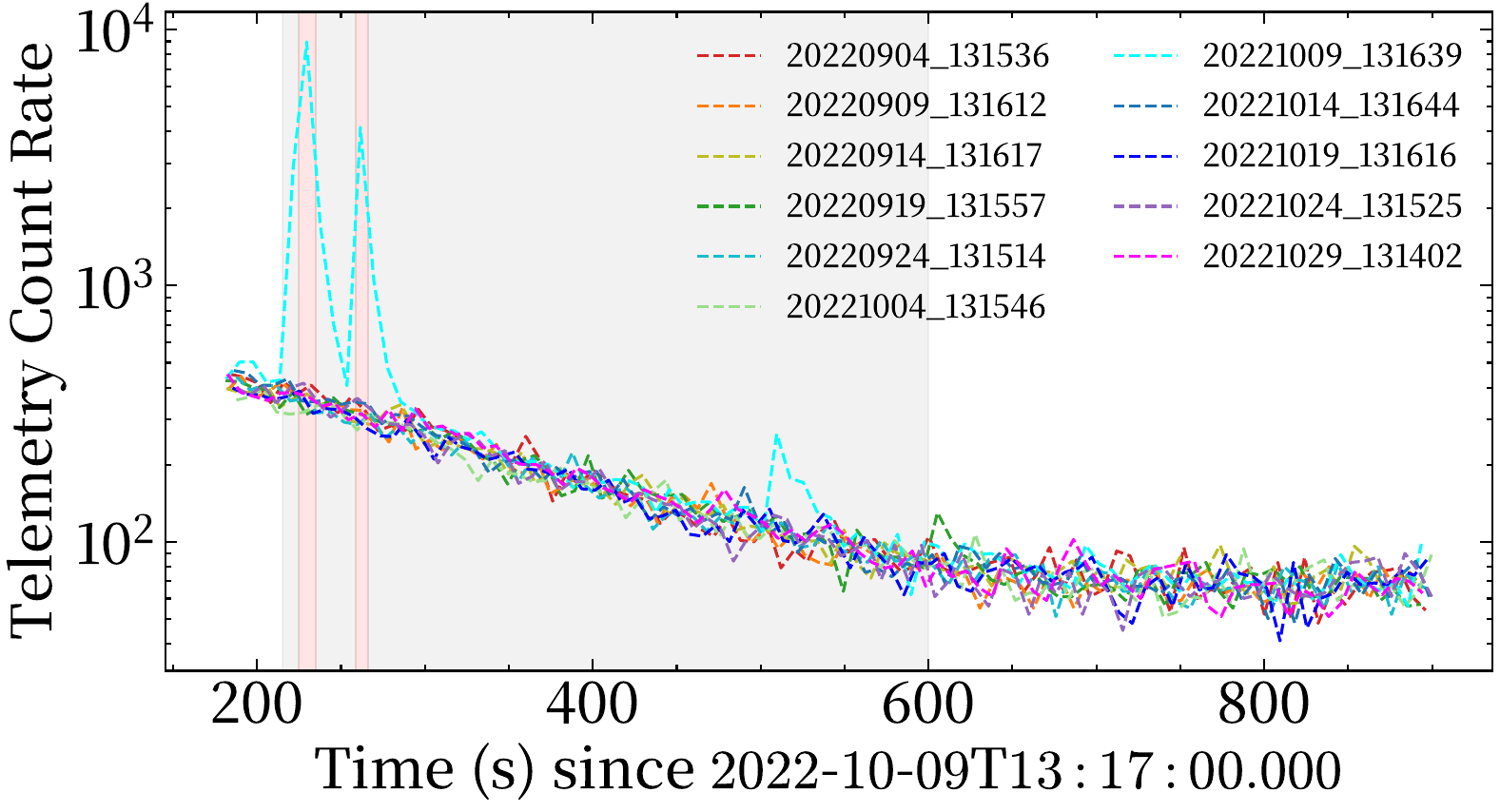}
    \label{subfig:telemetry_rate}
\end{subfigure}\hfill 
\begin{subfigure}[t]{0.48\textwidth}
 \caption{}
    \centering
    \includegraphics[width=\linewidth]{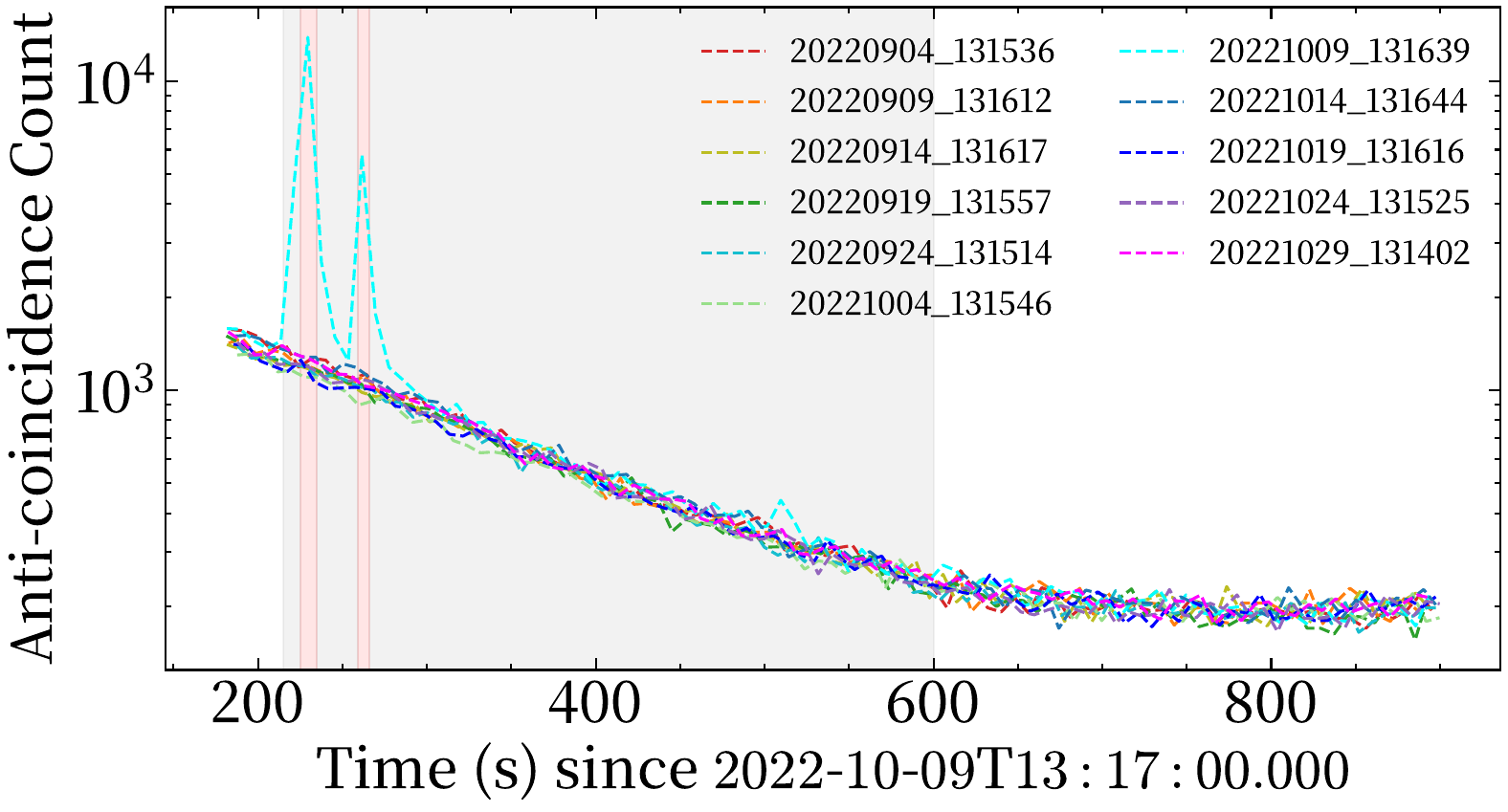}
    \label{subfig:anco_rate}
\end{subfigure} 

\vspace{0.4cm}

\begin{subfigure}[t]{0.48\textwidth}
 \caption{}
    \centering
    \includegraphics[width=\linewidth]{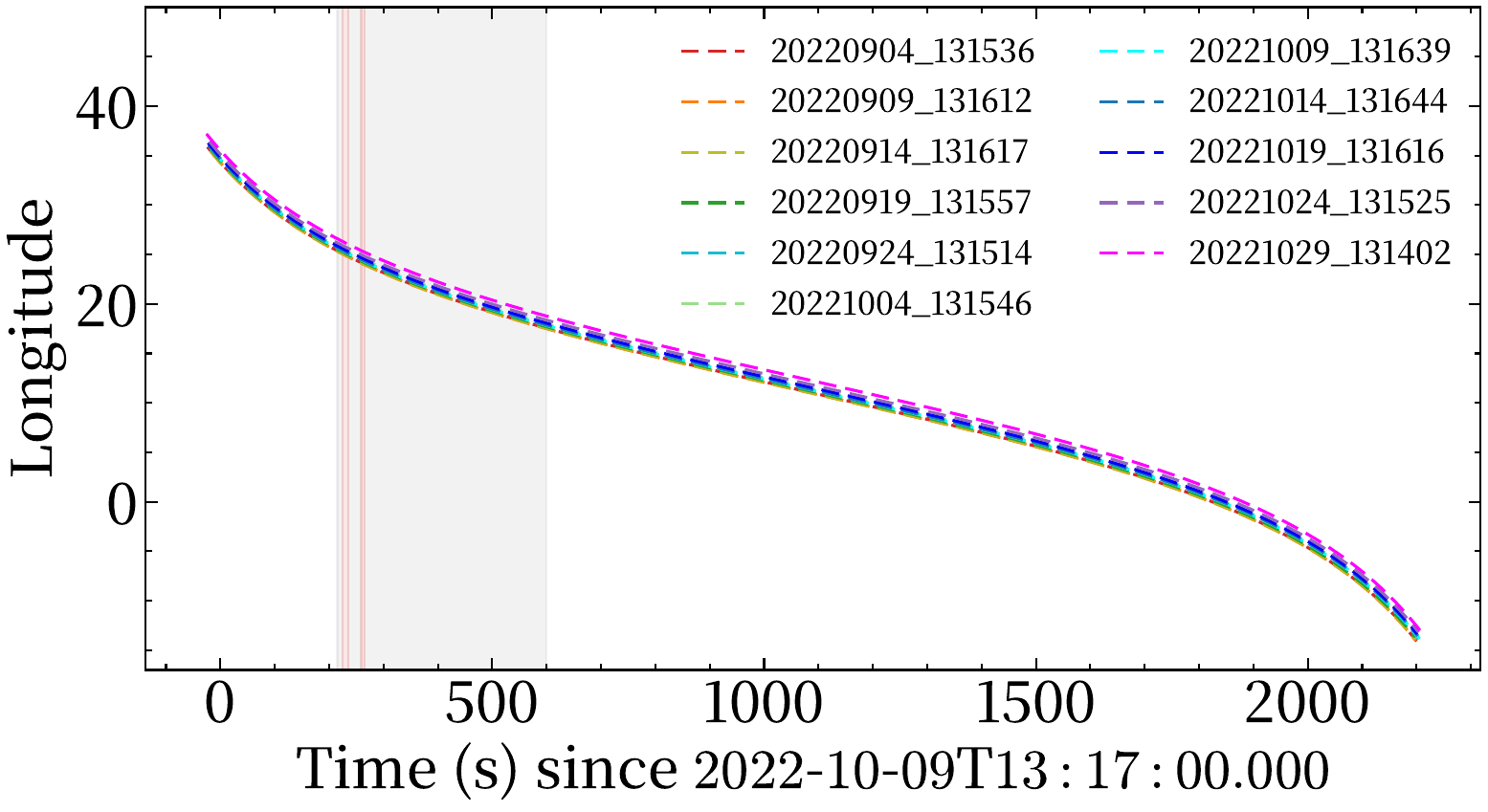}
    \label{subfig:longitude}
\end{subfigure}\hfill
\begin{subfigure}[t]{0.48\textwidth}
 \caption{}
    \centering
    \includegraphics[width=\linewidth]{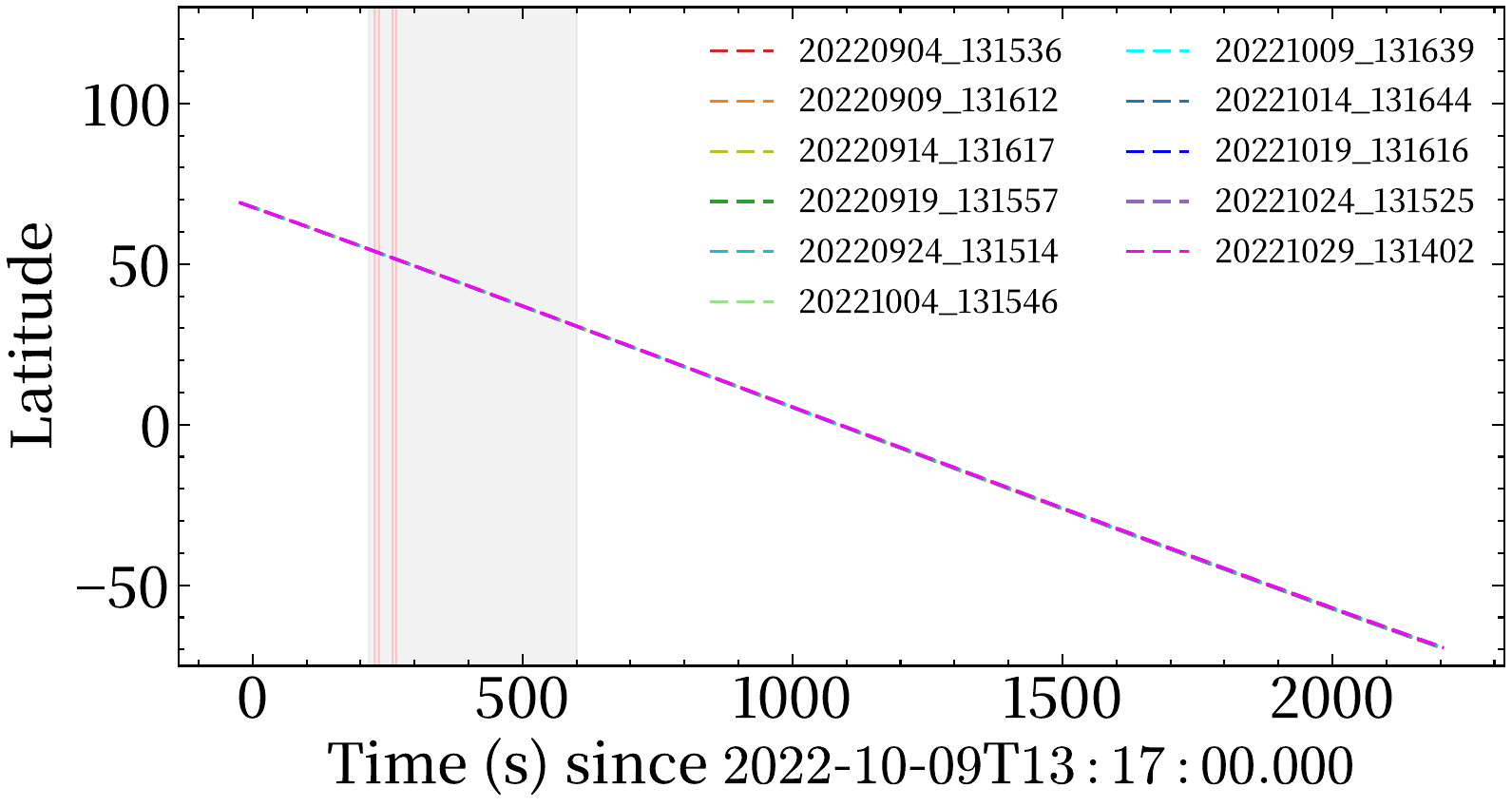}
    \label{subfig:latitude}
\end{subfigure}

\caption{\textbf{Comparison of count rates and ground tracks for revisiting orbits.} 
The orbit containing the GRB~221009A event is compared to six revisiting orbits with clean backgrounds strictly prior to the burst.
\textbf{(a)} Telemetry count rates. 
\textbf{(b)} Anti-coincidence detector count rates. 
\textbf{(c, d)} The corresponding longitude \textbf{(c)} and latitude \textbf{(d)} ground tracks for all selected orbits, which indicate a high degree of spatial overlap between the target orbit and the background orbits.}
\label{fig:orbital_path and count_rate}
\end{figure*}

In contrast, for HEPP-H, which is more sensitive to high-energy cosmic rays and their secondaries, its background exhibited unpredictable variations between revisiting orbits, despite their nearly identical ground tracks. To overcome this challenge of unstable HEPP-H background, we developed a robust background modeling method that effectively reduces the systematic error caused by these variations. To construct an average background model, we selected six revisiting orbits with clean background strictly prior to the burst, ensuring no contamination from the GRB's afterglow. The average background rate, $\bar{R}$, in each energy channel is calculated as:
\begin{equation}
    \bar{R} = \frac{1}{m} \sum_{i=1}^{m} \frac{N_i}{t_i},
    \label{eq:bg_mean}
\end{equation}
where $N_i$ and $t_i$ are the total counts and exposure time for the $i$-th background orbit, respectively, and $m$ is the total number of orbits used ($m=6$). To evaluate the stability of the background model, we performed a sensitivity analysis by systematically excluding each of the six background orbits one at a time. The spectral fitting results for key intervals show that the obtained parameters remain consistent within error ranges, regardless of which orbit is excluded (Fig.~\ref{fig:orbital_path and count_rate}). This demonstrates that our background estimation is robust and not affected by fluctuations in any single orbit.

\subsection{Detailed Instrument Response Simulation}

\begin{figure*}[t]
  \centering

  \begin{subfigure}[t]{0.46\textwidth}
      \caption{} % note
      \includegraphics[width=\linewidth]{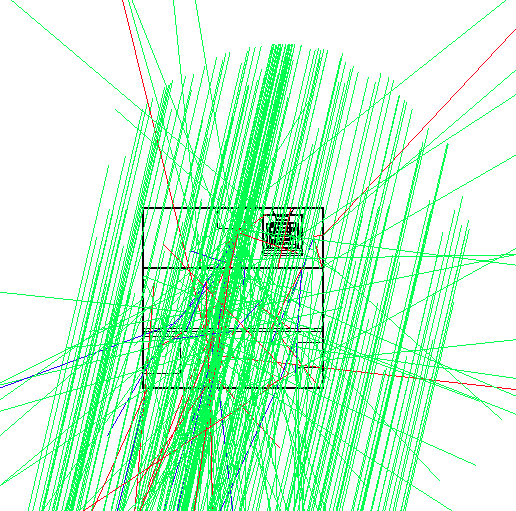}
      \label{subfig:hepp_h_geant4}
  \end{subfigure}\hfill
  \begin{subfigure}[t]{0.50\textwidth}
      \caption{} % note
      \includegraphics[width=\linewidth]{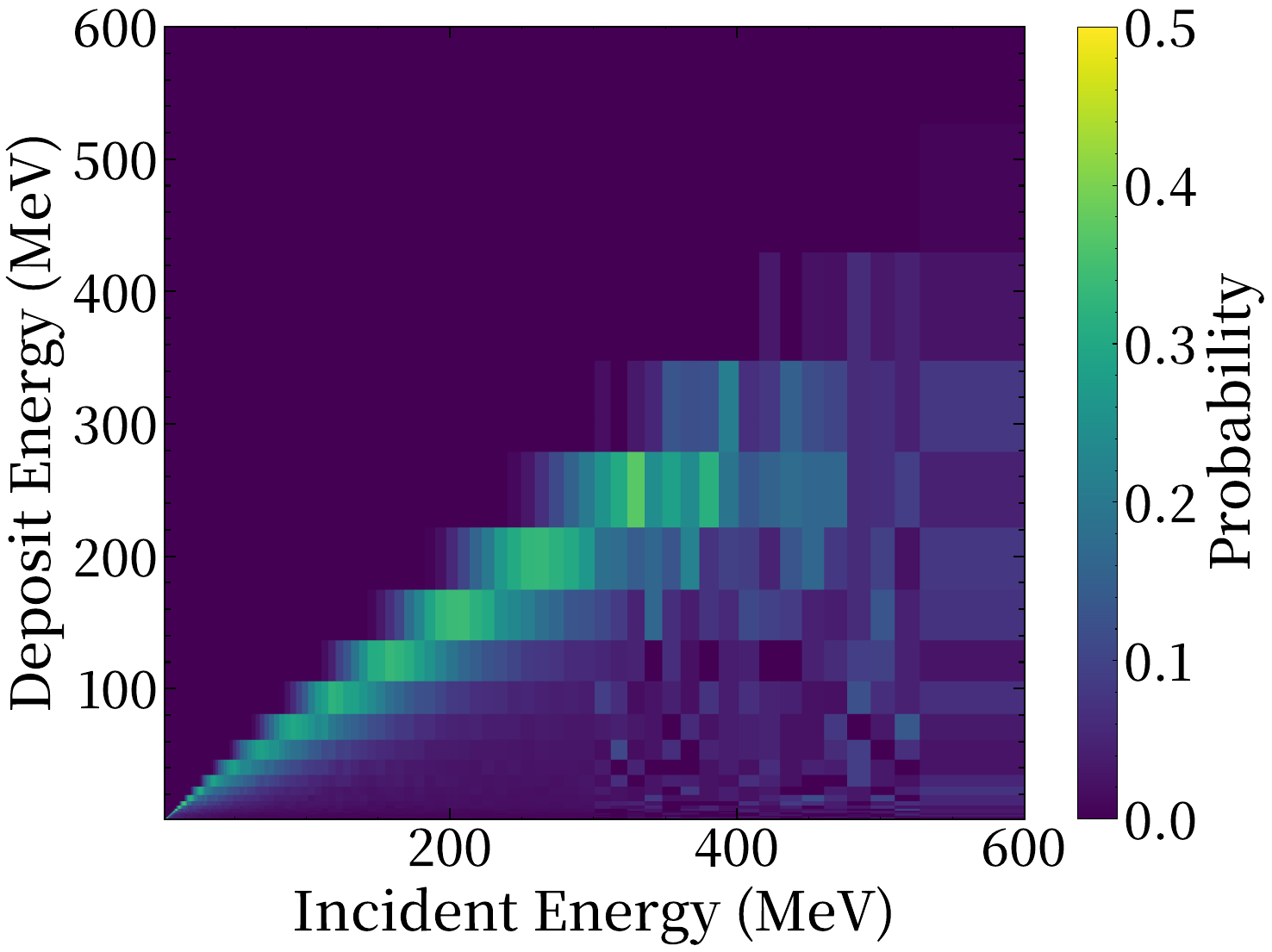}
      \label{subfig:hepp_h_RMF}
  \end{subfigure}

  \vspace{0.1cm} 

  \begin{subfigure}[t]{0.43\textwidth}
      \caption{} % note
      \includegraphics[width=\linewidth]{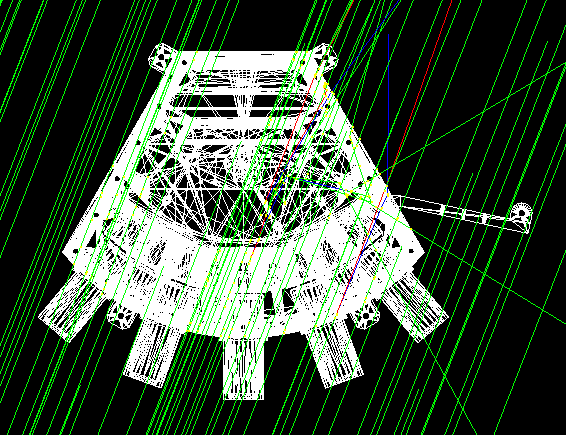}
      \label{subfig:hepp_l_geant4}
  \end{subfigure}\hfill
  \begin{subfigure}[t]{0.49\textwidth}
      \caption{} % note
      \includegraphics[width=\linewidth]{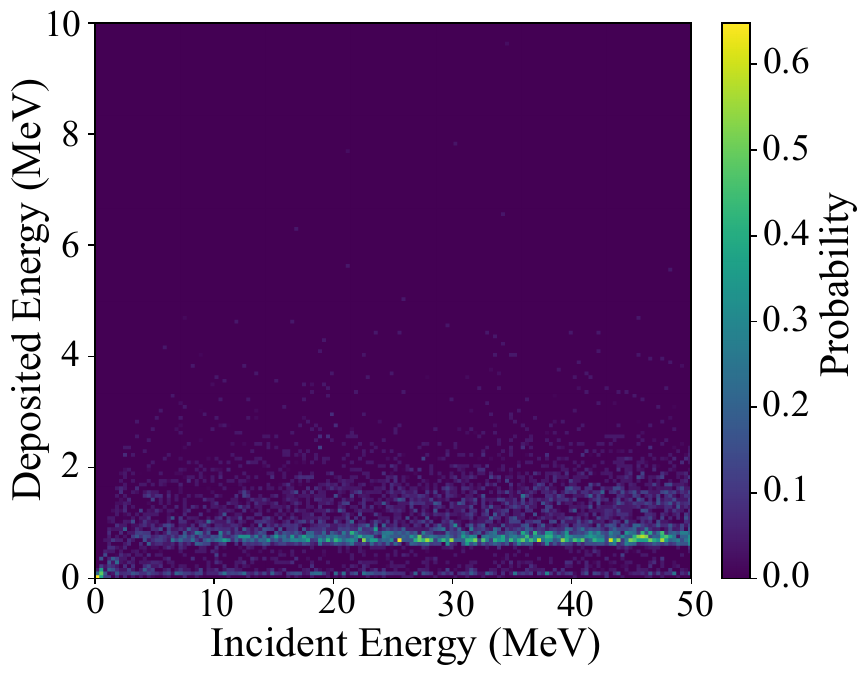}
      \label{subfig:hepp_l_RMF}
  \end{subfigure}

  \caption{\textbf{Geant4-simulated instrument response matrices for the HEPP-H and HEPP-L detectors.}
  \textbf{(a, c)} Visualization of the detector geometry and representative particle trajectories from the Geant4 simulations for HEPP-H \textbf{(a)} and HEPP-L \textbf{(c)}. Green lines illustrate incident gamma-rays, while red and blue tracks represent induced secondary electrons and positrons, respectively. 
  \textbf{(b)} Response Matrix File (RMF) for HEPP-H, correlating incident gamma-ray energy with the energy deposited in the calorimeter; the color scale represents the normalized probability density.  
  \textbf{(d)} Simulated energy response matrix for HEPP-L.}
  \label{fig:Simulation of HEPP-L and HEPP-H}
\end{figure*}

\textbf{HEPP-L:} The HEPP-L detector provides a valuable, independent temporal measurement of the GRB event. It is comprised of nine telescope detection units, collimation units, and a rear anti-coincidence detector. To robustly assess the instrument's capabilities for gamma-ray spectroscopy, we performed a dedicated Monte Carlo (MC) simulation using the GEANT4 toolkit. A model of HEPP-L was constructed from its computer-aided design (CAD) drawings. The simulation results, shown in Fig.~\ref{fig:Simulation of HEPP-L and HEPP-H}\textbf{c,d}, demonstrate the instrument's inherent limitations for high-energy gamma-ray spectroscopy. Due to the thinness of the silicon detectors, high-energy incident photons tend to be penetrating, depositing only a small and poorly-correlated fraction of their energy, predominantly below 2 MeV. This lack of a one-to-one correspondence between the incident and deposited photon energy makes a reliable inference of the incident gamma-ray spectrum unfeasible. Consequently, HEPP-L data are solely used to construct an independent light curve of the event to confirm the temporal correlation (Fig.~\ref{fig:GE_CS_LC}\textbf{b,c}).

\textbf{HEPP-H:} A mass model of the HEPP-H instrument was constructed by importing structural parameters from its CAD model. This model includes the dimensions, composition, and density of all sensitive components (silicon microstrips, PS, and CsI crystals) and surrounding passive materials (Fig.~\ref{fig:coordinate frames}\textbf{d,e}). Given that the GRB's incident direction was oblique to the detector's pointing axis, we also included a simplified mass model of the CSES-01 satellite platform (e.g., aluminum plates) to account for secondary particle production in surrounding structures.

In the simulation, we generated a beam of parallel gamma-ray photons originating from a radius\SI{150}{cm} semi-circular located directly above the instrument's collimator, mimicking a distant point source. The beam was directed toward the instrument at an angle matching the observed direction of GRB~221009A, with photon energies logarithmically spaced from 1~MeV to 300~MeV. A total of 7$\times$10$^8$ primary photons were simulated to ensure sufficient statistics for determining the geometric factor and the response matrix. The core of our analysis relies on the secondary electrons (including positrons) produced by photon interactions in the passive materials of the instrument and satellite. The simulation shows that these secondary electrons are the dominant source of recorded events, accounting for approximately 80\% of the total energy deposition that leads to a trigger (Fig.~\ref{fig:Simulation of HEPP-L and HEPP-H}\textbf{a}). They are the primary source of triggers for HEPP-H because the trigger efficiency for direct gamma-ray energy deposition in the first plastic scintillator layer is very low \citep{2024ApJ...960...21P, 2024ApJ...976..239B}. To replicate the on-orbit trigger conditions, a simulated event was recorded if the energy deposited in the plastic scintillator layer exceeded 1.5~MeV. The anti coincidence logic was not applied to the event selection algorithm in the simulation, mirroring the data analysis choice to maximize event statistics. Finally, to account for instrumental broadening, the simulated energy depositions were convolved with the in-orbit energy resolution model derived from the MIPs calibration, rather than performing a full optical photon transport simulation. This approach significantly reduces computation time while accurately reproducing the observed spectral features. The final output of this simulation pipeline is an instrument response matrix (Fig.~\ref{fig:Simulation of HEPP-L and HEPP-H}\textbf{b}), which is essential for all subsequent spectral fitting.

\section{Cross-calibration and Methodological Robustness}
\setcounter{equation}{0}
\subsection{Multi-instrument Cross-calibration}

\begin{figure*}[t]
  \centering

  \begin{subfigure}[t]{0.49\textwidth}
      \caption{} % note
      \includegraphics[width=\linewidth]{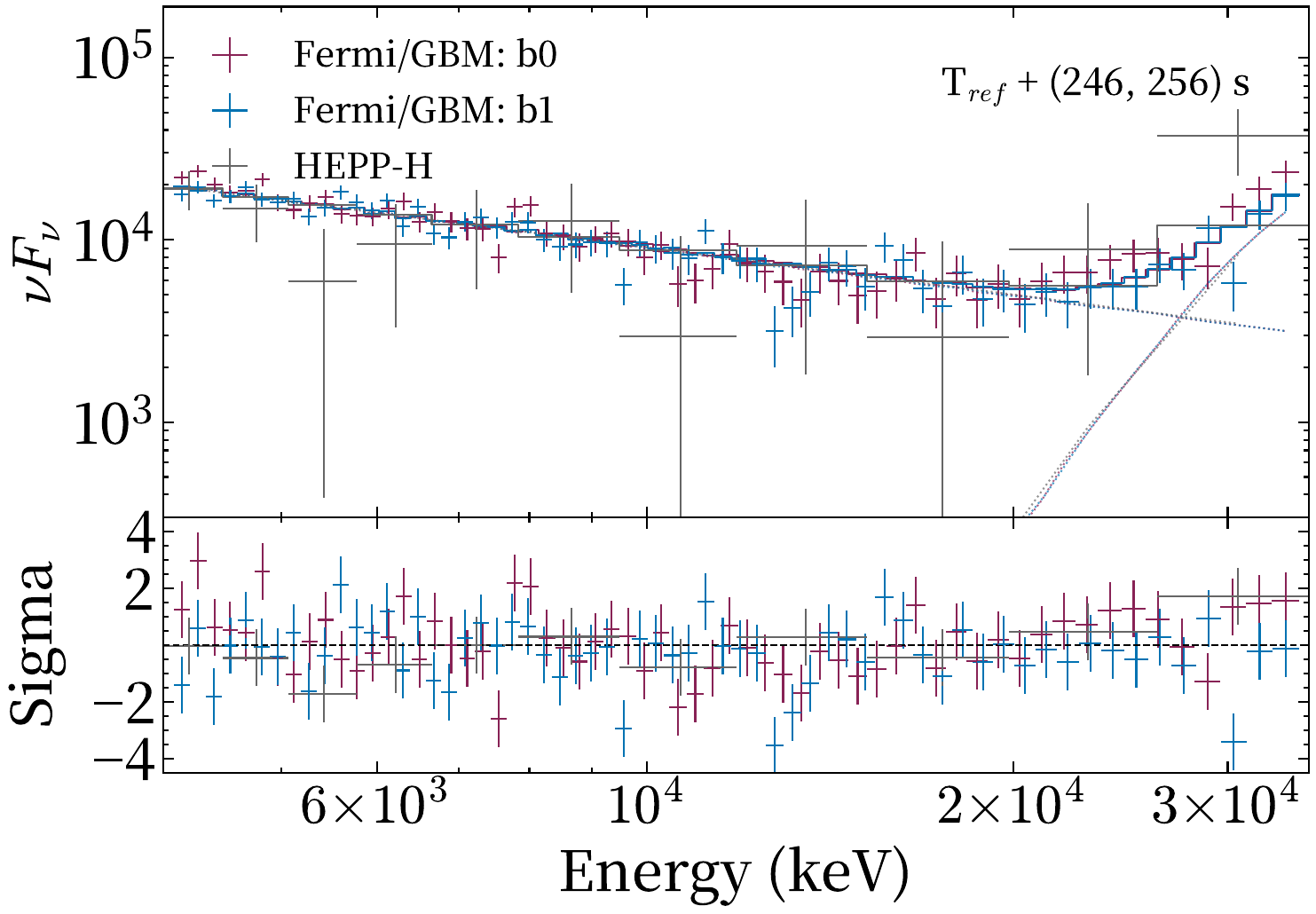}
      \label{subfig:246-256_GH}
  \end{subfigure}\hfill
  \begin{subfigure}[t]{0.49\textwidth}
      \caption{} % note
      \includegraphics[width=\linewidth]{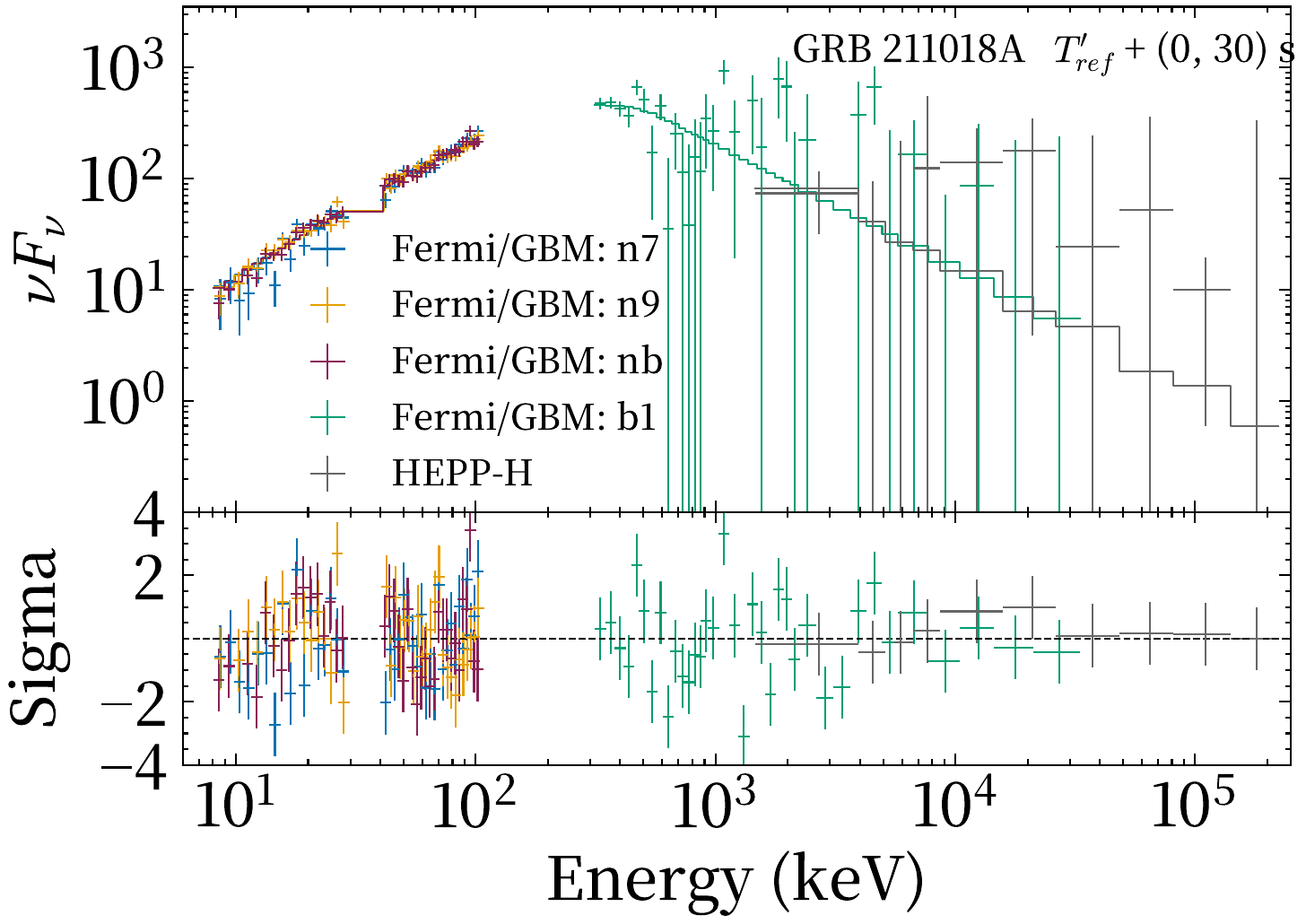}
      \label{subfig:18A_0-30_GH}
  \end{subfigure}

  \vspace{0.1cm} 

  \begin{subfigure}[t]{0.48\textwidth}
      \caption{} % note
      \includegraphics[width=\linewidth]{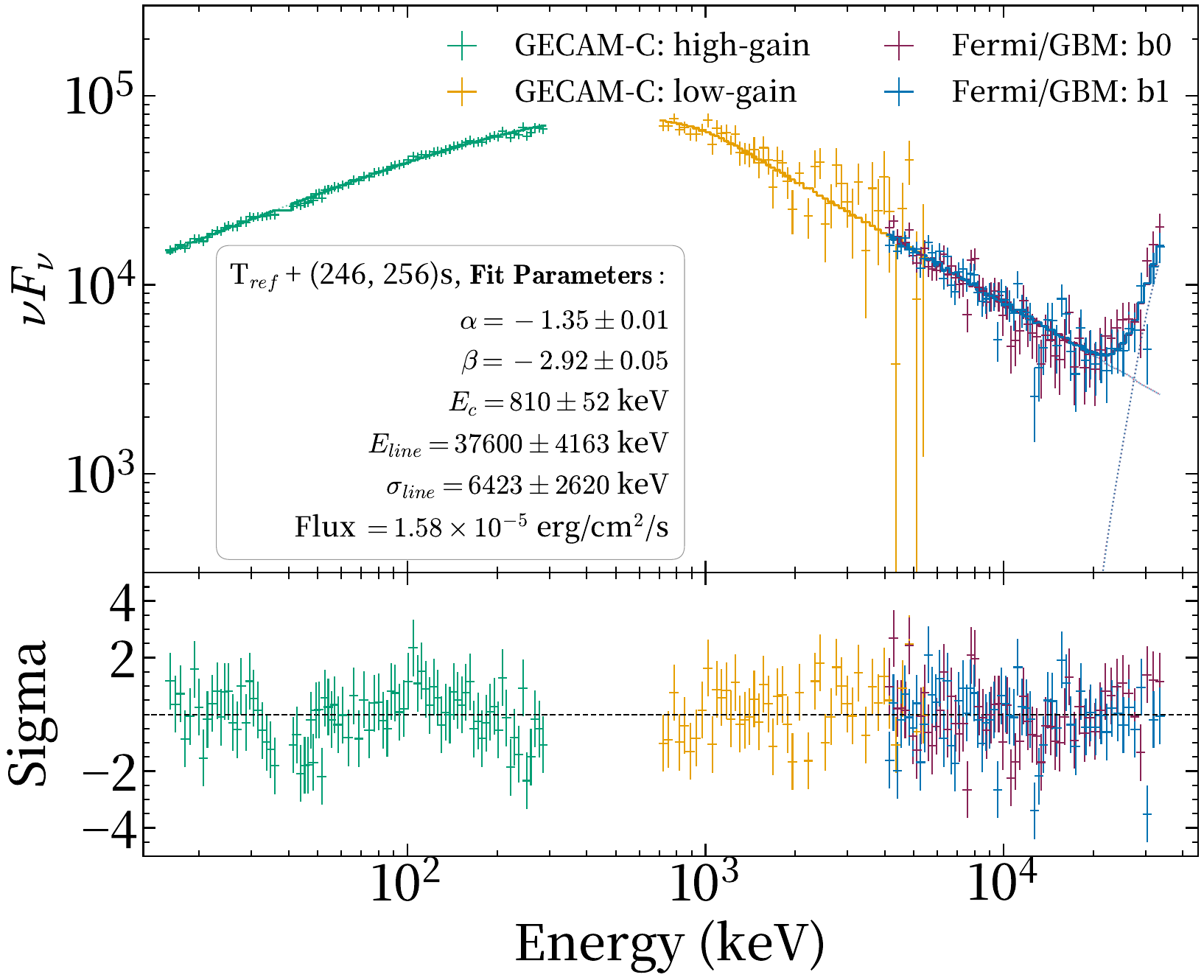}
      \label{subfig:NoHE_parameter}
  \end{subfigure}\hfill
  \begin{subfigure}[t]{0.49\textwidth}
      \caption{} % note
      \includegraphics[width=\linewidth]{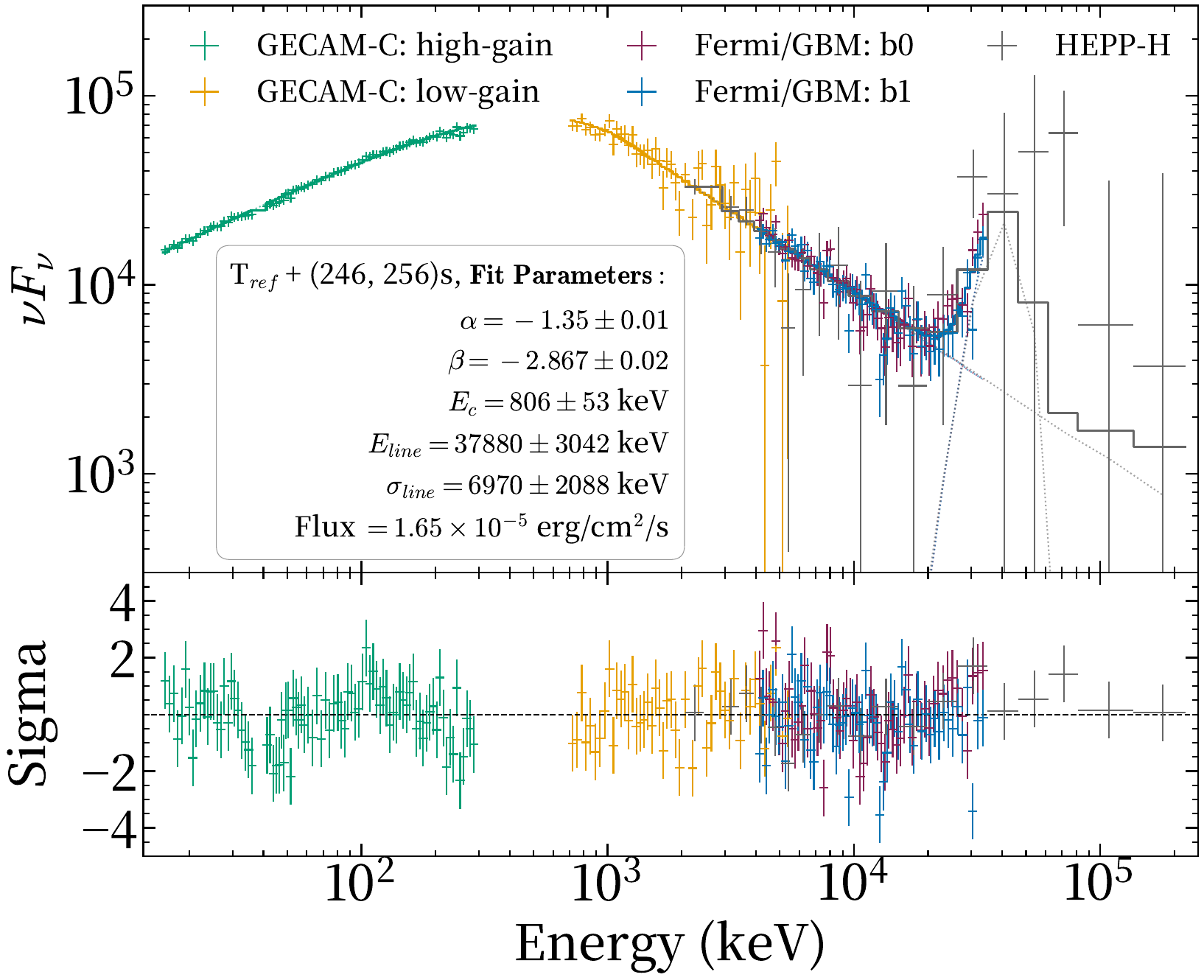}
      \label{subfig:YeHE_parameter}
  \end{subfigure}

  \caption{\textbf{Multi-instrument and multi-event verification of the HEPP-H data processing pipeline.}
  \textbf{(a)} Comparison of the $\nu F_\nu$ spectra (units: $\mathrm{keV}^2 \cdot \mathrm{Photons} \cdot \mathrm{cm}^{-2} \cdot \mathrm{s}^{-1} \cdot \mathrm{keV}^{-1}$) and residuals in the 4--35~MeV overlapping range of HEPP-H and \textit{Fermi}/GBM for the interval and $T_{ref} + (246, 256)$~s. 
  \textbf{(b)} Independent cross-event verification using GRB~211018A ($T'_{\rm ref} + 0$--$30$~s), comparing the $\nu F_\nu$ spectra of HEPP-H (grey) with \textit{Fermi}/GBM (blue, orange, purple, and green).
  \textbf{(c, d)} Joint spectral fit results for the $T_{ref} + (246, 256)$~s interval, showing fits using GECAM-C and \textit{Fermi}/GBM data only \textbf{(c)} and with the inclusion of HEPP-H data \textbf{(d)}. Solid lines represent the best-fit model (Band + Gaussian), with the corresponding spectral parameters listed in the legends.}
  \label{fig:HEPP-H_GBM_Calibration_Verufication}
\end{figure*}

To validate the non-saturation and the energy calibration accuracy of HEPP-H under extreme high-flux conditions, we performed a cross-comparison with the \textit{Fermi}/GBM. Within the overlapping energy range of 4--35 MeV, the $\nu F_\nu$ spectral shapes of HEPP-H and GBM exhibit excellent agreement (Fig.~\ref{fig:HEPP-H_GBM_Calibration_Verufication}\textbf{a}). This high degree of consistency directly validates that HEPP-H maintained a non-saturated state under extreme photon fluxes, ensuring that its recorded count rates accurately reflect the energy distribution of incident photons. Additionally, although the three line-analysis intervals $T_{\rm ref}+237$--$246$\,s, $242$--$251$\,s, and $246$--$256$\,s lie within the officially reported GBM bad-time interval, we used the 4--35\,MeV BGO CSPEC data following the data-selection and cross-validation procedure established for this burst \citep{2024SCPMA..6789511Z}. Taking GECAM-C as the reference, the GBM cross-normalization factors are $1.14\pm0.07$, $1.13\pm0.05$, and $1.12\pm0.05$ for these three intervals, respectively. These factors are mutually consistent and remain close to unity. In particular, our value of $1.12\pm0.05$ for the $T_{\rm ref}+246$--$256$\,s interval is consistent with the value of $1.05\pm0.08$ reported for the same interval \citep{2024SCPMA..6789511Z}. With this data selection, the best-fit Band-function parameters ($\alpha$, $\beta$, and $E_c$) and Gaussian-line parameters ($E_{\rm line}$ and $\sigma_{\rm line}$) remained stable and mutually consistent both before and after the inclusion of the HEPP-H data (Fig.~\ref{fig:HEPP-H_GBM_Calibration_Verufication}\textbf{c,d}). This stability demonstrates that the spectral shape of HEPP-H is consistent with that of GBM during this unsaturated period.

\subsection{Multi-event Cross-validation via GRB 211018A}

To further demonstrate the general applicability of our data processing pipeline, we performed an independent cross-validation using GRB 211018A during the interval $T'_{\rm ref} + (0, 30)$ s (where $T'_{\rm ref} = \text{2021-10-18T22:28:14}$; Fig.~\ref{fig:HEPP-H_GBM_Calibration_Verufication}\textbf{b}). This event was specifically selected as it represents a bright burst that, unlike the extreme case of GRB 221009A, exhibits no significant instrumental saturation in conventional gamma-ray detectors. Furthermore, at the time of the burst, HEPP-H was uncontaminated by the Earth radiation belts, providing an ideal benchmark for comparison. The HEPP-H data, processed using the same methodology, show good agreement with \textit{Fermi}/GBM observations in the MeV-to-sub-GeV energy range. This cross-event consistency further corroborates the robustness and reliability of our approach.

\subsection{Alternative Continuum Model Test}
\label{sec:double_break_test}

We also examined whether the apparent MeV excess could instead be described by a second smooth break in a single continuum component. To test this possibility, we fitted all six time-resolved spectra with a double smoothly broken power-law model (2SBPL), consisting of three power-law segments connected continuously through two smooth transitions \citep{2018A&A...613A..16R}:
\begin{equation}
\begin{aligned}
A(E)={}&K E_{\rm b,1}^{\alpha_1}
\Bigg\{
\left[
\left(\frac{E}{E_{\rm b,1}}\right)^{-\alpha_1 n_1}
+
\left(\frac{E}{E_{\rm b,1}}\right)^{-\alpha_2 n_1}
\right]^{n_2/n_1}
\\
&+
\left(\frac{E}{E_{\rm j}}\right)^{-\beta n_2}
\left[
\left(\frac{E_{\rm j}}{E_{\rm b,1}}\right)^{-\alpha_1 n_1}
+
\left(\frac{E_{\rm j}}{E_{\rm b,1}}\right)^{-\alpha_2 n_1}
\right]^{n_2/n_1}
\Bigg\}^{-1/n_2},
\end{aligned}
\label{eq:2sbpl}
\end{equation}
where $K$ is the normalization, $\alpha_1$, $\alpha_2$, and $\beta$ are the photon indices of the three power-law segments, $E_{\rm b,1}$ is the first break energy, and $E_{\rm j}$ specifies the location of the second smooth transition. We fixed the smoothness parameters at $n_1=5.38$ and $n_2=2.69$, as adopted in the original formulation of the 2SBPL model \citep{2018A&A...613A..16R}. The former is the mean value obtained when $n_1$ was allowed to vary, whereas the latter corresponds to the standard SBPL smoothness adopted in the \textit{Fermi}/GBM catalog. Fixing these parameters avoids introducing two additional, poorly constrained degrees of freedom.

The comparison yields mixed results (Table~\ref{tab:2sbpl_comparison}). In the $T_{\rm ref}+225$--$228$\,s and $230$--$231$\,s intervals, the 2SBPL model gives the lowest AIC and is strongly favored by the AIC over both Band+CPL and Band+BB. In the other four intervals, however, the AIC of the 2SBPL model is higher than those of both Band+CPL and Band+BB. In particular, during $T_{\rm ref}+228$--$230$\,s, when the MeV excess is most prominent, both Band+CPL and Band+BB are strongly favored over 2SBPL by the AIC. Therefore, although the 2SBPL model provides a competitive description of the spectra in some intervals, it is not consistently preferred across the six time-resolved spectra covering $T_{\rm ref}+225$--$235$\,s.

\subsection{Robustness of Sliding-Window Methodology}
\begin{figure*}[t]
\centering
  \includegraphics[width=0.9\textwidth]{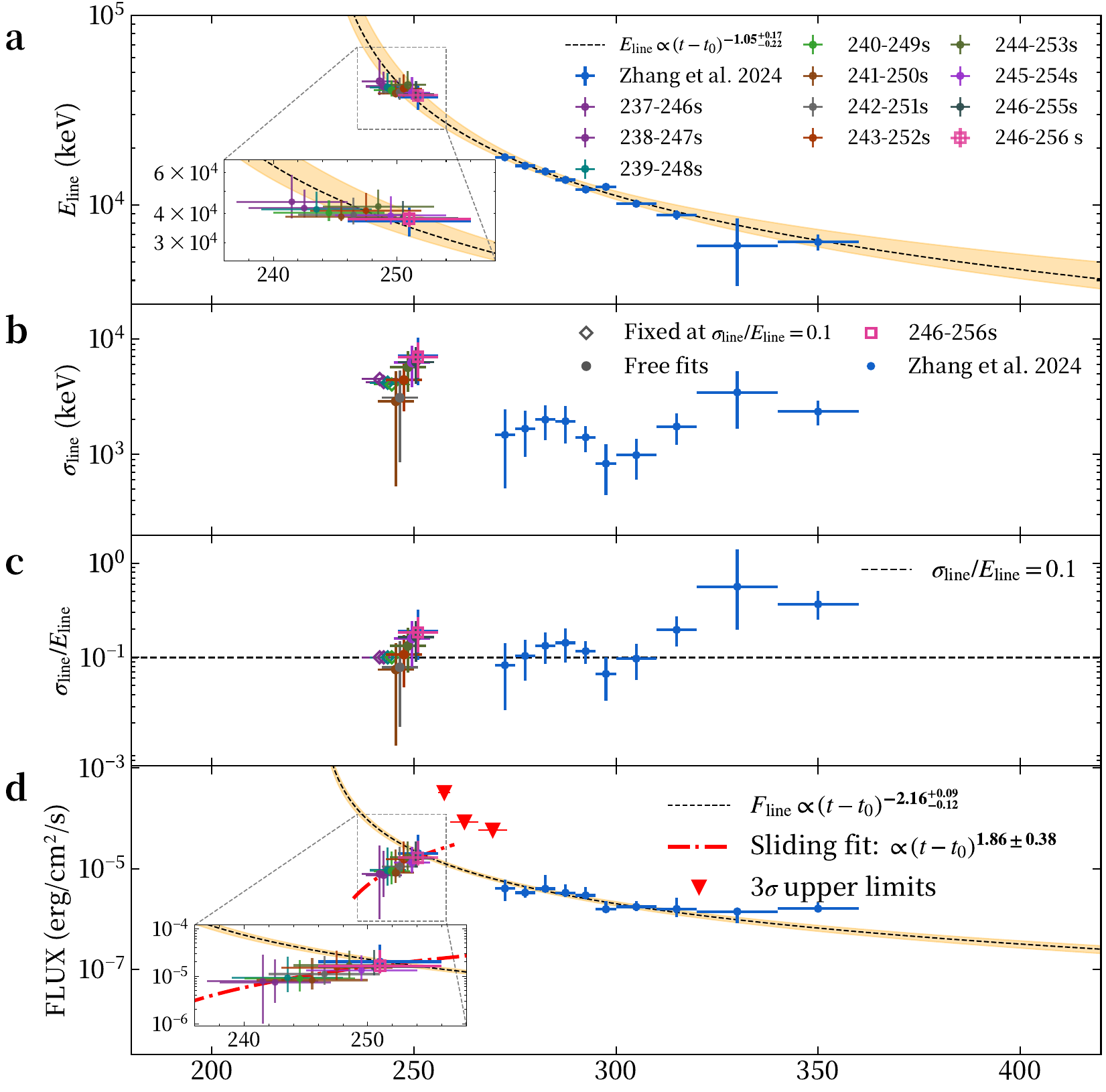}
  \caption{\textbf{Comprehensive sliding-window analysis of the spectral line evolution.} 
  \textbf{(a)} Temporal evolution of the central line energy ($E_{\rm line}$), determined using a 9-s moving window with a 1-s step. Each colored point corresponds to a specific time interval, as defined in the legend. The black dashed line and orange shaded region represent the power-law decay model, $E_{\rm line}\propto(t-t_0)^{-1.05}$, and its associated uncertainty, respectively.
  \textbf{(b)} Temporal evolution of the Gaussian line width ($\sigma_{\rm line}$). The open diamonds denote intervals for which the relative line width was fixed at $\sigma_{\rm line}/E_{\rm line}=0.1$, whereas the filled points represent free fits. The measurements from Zhang et al. (2024) are shown in blue.
  \textbf{(c)} Evolution of the relative line width, $\sigma_{\rm line}/E_{\rm line}$. The horizontal dashed line marks $\sigma_{\rm line}/E_{\rm line}=0.1$.
  \textbf{(d)} Temporal evolution of the emission line flux using the same sliding-window color scheme. The red dash-dotted line denotes the fit to the sliding-window measurements, $F_{\rm line}\propto(t-t_0)^{1.86\pm0.38}$, while the black dashed line and orange shaded region indicate the late-time decay model, $F_{\rm line}\propto(t-t_0)^{-2.16}$, and its associated uncertainty.
  \textbf{Note:} For the intervals from $T_{\rm ref}+237$--$246$\,s through $T_{\rm ref}+240$--$249$\,s, the relative line width was fixed at $\sigma_{\rm line}/E_{\rm line}=0.1$ because of the limited statistics. Beginning with the $T_{\rm ref}+241$--$250$\,s interval, all line parameters were allowed to vary freely.
  }
  
  \label{fig:Sliding_fit}
\end{figure*}

Applying the joint fitting strategy, we performed a sliding-window analysis covering the interval $T_{\rm ref}+(237, 256)$\,s (Fig.~\ref{fig:Sliding_fit}). Starting from the $241$--$250$\,s window, the emission line parameters could be constrained without fixed width priors. However, for the earlier intervals (from $237$--$246$\,s to $240$--$249$\,s), given the limited photon statistics, we employed physically motivated priors by fixing the Gaussian line width ratio to 10\% \citep{2024SCPMA..6789511Z}. The temporal evolution of the freely fitted line width, $\sigma_{\rm line}$, and the relative line width, $\sigma_{\rm line}/E_{\rm line}$, is shown in Fig.~\ref{fig:Sliding_fit}\textbf{b} and \textbf{c}, respectively. Among the free-fitted windows, the $242$--$251$\,s interval was selected for detailed analysis because it exhibited a higher significance than the adjacent $241$--$250$\,s window. A power-law fit to the emission line fluxes obtained during the $237$--$256$\,s interval reveals a rising trend with an index of $1.86 \pm 0.38$ (Fig.~\ref{fig:Sliding_fit}\textbf{d}).

\begin{figure*}[t] 
\centering 
    \includegraphics[width=0.8\textwidth]{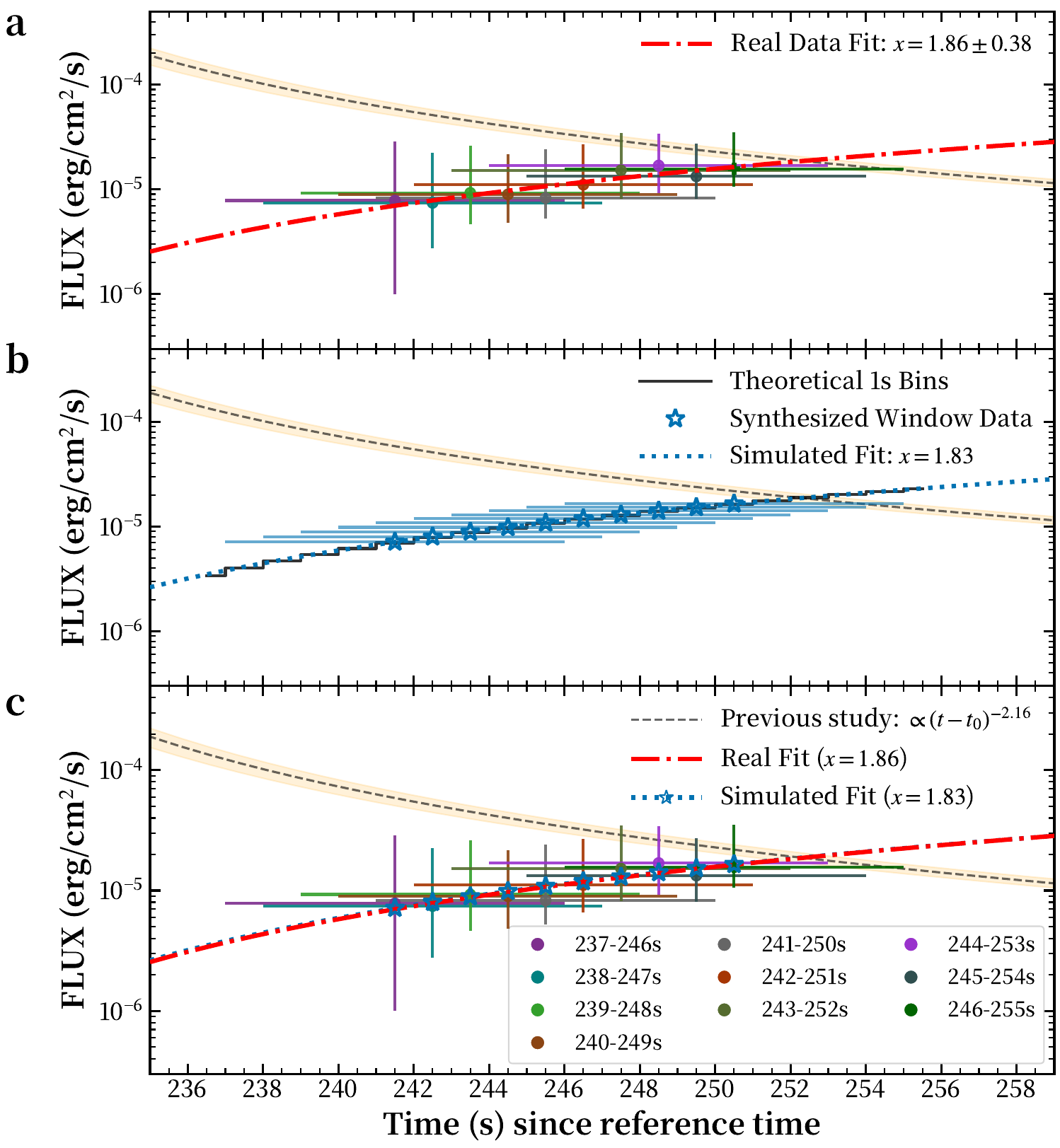}
    
    \caption{\textbf{Simulation-based verification of the sliding-window power-law index.} 
    This simulation validates that the overlapping sliding-window averaging process does not introduce significant systematic bias to the extracted power-law index.
    \textbf{(a)} Power-law fit to the observed emission line flux ($T_{\rm ref} + 237$--$255$~s), where the red dash-dotted line indicates the best-fit model with an index of $1.86 \pm 0.38$.  
    \textbf{(b)} Simulation of the data processing pipeline. A theoretical function $F(t) \propto (t-t_0)^{1.86}$ is sampled at a 1-second resolution (black stepped line). These points are synthesized into overlapping 9-second bins (blue open stars) according to the time intervals used in the real data analysis. The green dotted line represents the power-law fit to these simulated points, yielding an index of $1.83$. 
    \textbf{(c)} Comparison between the real data (red dash-dotted line) and the simulated data (blue dotted line). }
    \label{fig:Sliding_Verification} % note
\end{figure*}

To verify whether the use of overlapping 9-second sliding windows introduce systematic bias to the extracted power-law index, we performed a simulation-based closure test. The procedure strictly mimics our real data analysis pipeline:
\begin{enumerate}
    \item \textbf{Model Assumption:} We assume an underlying physical radiation profile that follows a perfect power-law: $F(t) \propto (t - t_0)^{1.86}$ (Fig.~\ref{fig:Sliding_Verification}\textbf{a}).
    \item \textbf{Dataset Generation:} Based on this above model, we generated a theoretical dataset with a resolution of 1-second (shown as the black stepped line in Fig.~\ref{fig:Sliding_Verification}\textbf{b}).
    \item \textbf{Simulated Sliding-Window Averaging:} These 1-second data points were synthesized into overlapping 9-second bins, exactly matching the time intervals used in the real data analysis (e.g., 237--246 s, 238--247 s, ..., 246--255 s). These synthesized points are shown as blue open stars in Fig.~\ref{fig:Sliding_Verification}\textbf{b}.
    \item \textbf{Comparing power-law indices:} We then performed a power-law fit to these synthesized sliding-window points. As shown in Fig.~\ref{fig:Sliding_Verification}\textbf{c}, the fit to the synthesized dataset recovers a power-law index of 1.83 (blue dotted line), which is in excellent agreement with the input physical index of 1.86 used to generate the data.
\end{enumerate}
This result demonstrates that the sliding-window process does not systematically distort the index of the temporal evolution.

\subsection{Global Statistical Significance of the Line Detection}

\begin{figure*}[t] 
\centering 
    \includegraphics[width=1\textwidth]{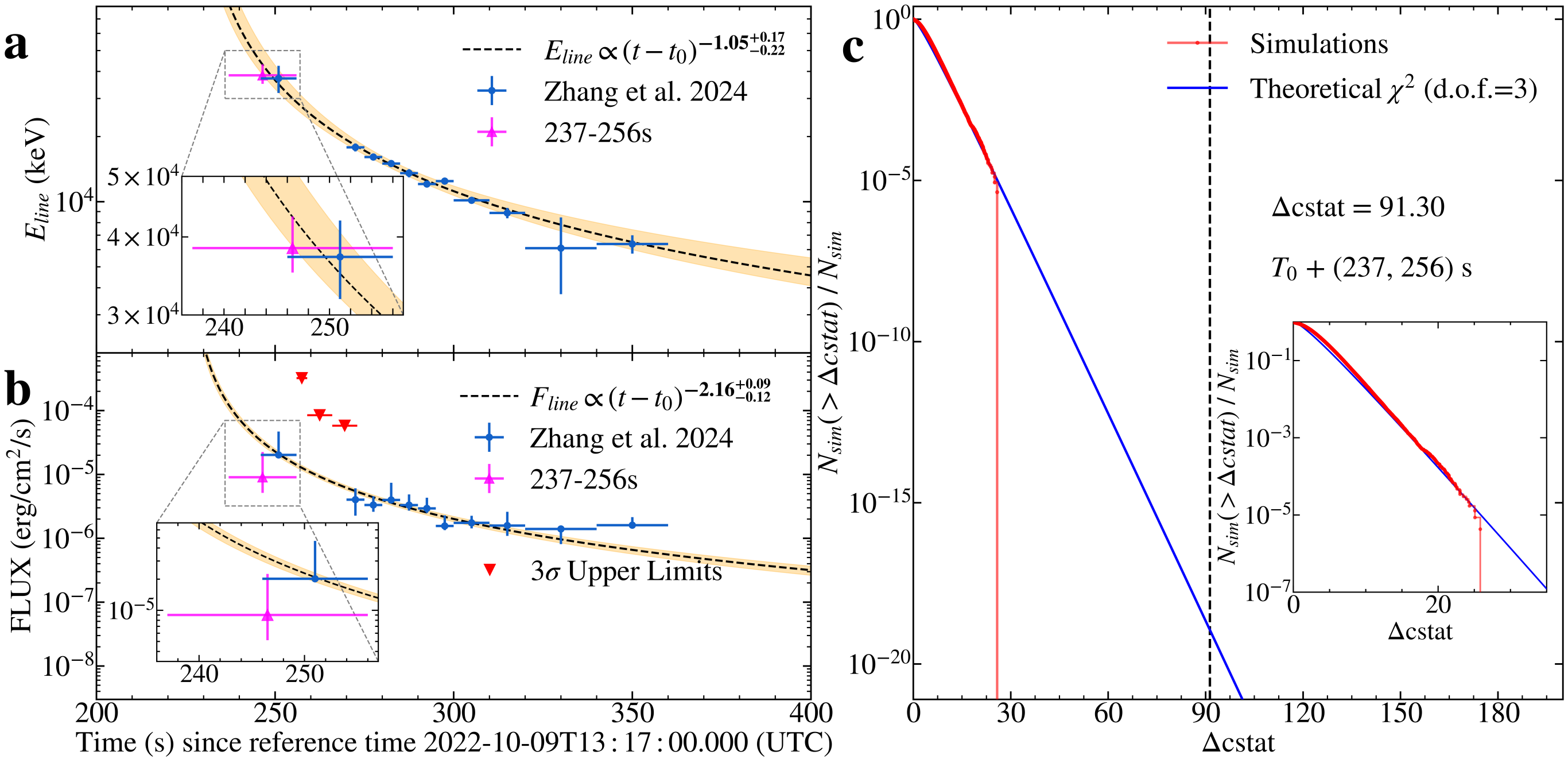}
    
    \caption{\textbf{Evolution and statistical significance of the MeV emission line.} 
    \textbf{(a)} Temporal evolution of the central line energy ($E_{\mathrm{line}}$). The pink marker indicates the measurement for the $T_{\rm ref} + (237, 256)$~s interval; blue crosses represent previous results \citep{2024SCPMA..6789511Z}. The black dashed line and shaded area represent the power-law decay model [$E_{\mathrm{line}} \propto (t-t_0)^{-1.05}$] and its associated uncertainty. 
    \textbf{(b)} Temporal evolution of the emission line flux. The black dashed line indicates the late-time decay trend [$F_{\rm line} \propto (t-t_0)^{-2.16}$], and red triangles denote $3\sigma$ flux upper limits. Markers follow the same color scheme as in \textbf{a}. 
    \textbf{(c)} Verification of the global statistical significance. The red line and markers show the distribution of $\Delta \text{cstat}$ values from $2.3 \times 10^5$ MC simulations of the null hypothesis (the Band function alone). The blue line represents the theoretical $\chi^{2}$ distribution (d.o.f.=3). The observed improvement ($\Delta \text{cstat} = 91.30$) for the $T_{\rm ref} + (237, 256)$~s interval is marked by the vertical black dashed line. }
    \label{fig:Line_all} % note
\end{figure*}

Additionally, we evaluated the global significance of the line detection by performing $2.3 \times 10^5$ MC simulations based on the null model (the Band function alone). As shown in Fig.~\ref{fig:Line_all}\textbf{c}, the observed $\Delta \text{cstat}$ of 91.30 for the $T_{\rm ref} + (237, 256)$ s interval far exceeds the distribution of random fluctuations, confirming a global significance of $> 4.4\sigma$ ($P < 4.3 \times 10^{-6}$).

\section{Detailed Systematic Error Analysis}
\setcounter{equation}{0}

The scientific conclusions presented in this work are underpinned by a rigorous quantification of systematic uncertainties. We dissect the total systematic error budget into its three dominant and independent contributions: instrument response systematics ($\sigma_{\text{rsp}}$), systematics from the correction of instrumental effects ($\sigma_{\text{DCP}}$), and background systematics ($\sigma_{\text{BG}}$). As these error sources are uncorrelated, we combine their contributions in quadrature to determine the total systematic uncertainty, $\sigma_{\text{S}}$:
\begin{equation}
    \sigma_{\text{S}}^2 = \sigma_{\text{rsp}}^2 + \sigma_{\text{DCP}}^2 + \sigma_{\text{BG}}^2.
    \label{eq:sigma_total_app}
\end{equation}
This comprehensive error budget provides a robust and conservative foundation for the scientific results presented in this work.

\subsection{Instrument Response Systematics}

The instrument response must be characterized with the highest possible fidelity. As HEPP-H is a charged-particle detector without pre-launch calibration data for gamma-rays, its response to a GRB can only be determined through simulation. We therefore performed a comprehensive analysis of the systematic uncertainties inherent in the simulation, identifying and rigorously constraining three dominant sources: the satellite mass model, the in-orbit degradation of the detector's energy resolution, and the changing line-of-sight to the burst.

The instrument response was generated via detailed GEANT4 simulations based on CAD models of the HEPP-H detector. While the mass model of the HEPP-H detector itself is highly accurate, the primary source of simulation uncertainty derives from the simplified representation of the surrounding CSES-01 satellite platform and other payloads. To robustly quantify the total systematic uncertainty arising from the simulation, we adopted a data-driven, external validation approach. A joint spectral analysis using data from HEPP-H, GECAM-C, and GBM was performed during a quasi-stable interval ($T_{\rm ref}+246$--$256$\,s). 

The systematic uncertainty $\sigma_{\text{model}}$ is quantified based on the results of the multi-instrument joint spectral analysis. Following the series of response matrix corrections and incorporating the application of the actual geometric factor, the cross-normalization factor for HEPP-H relative to GECAM-C is determined to be $1.05 \pm 0.07$, which is statistically consistent with unity. Accordingly, we adopt the relative fitting uncertainty of 7\% as the estimate for $\sigma_{\text{model}}$.

Second, we accounted for the time-dependent evolution of the detector's energy resolution. To most accurately reproduce the instrument's performance during the GRB, we adopted an in-orbit energy resolution model calibrated directly with cosmic-ray MIP events contemporaneous with the burst (October 2022). We precisely measured the detector's contemporary energy broadening by fitting the MIP spectra with a Landau distribution. Based on the goodness-of-fit and parameter errors, we assess $\sigma_{\text{FWHM}}$ to be a sub-dominant contributor to the total error budget. Finally, we assessed the systematic uncertainty associated with the incident direction of the gamma-rays. Collectively, these factors render the uncertainty introduced by approximating the incident direction with a single vector negligible compared to other systematics. 

These three physically distinct and uncorrelated components are added in quadrature to obtain the systematic uncertainty for the instrument response, $\sigma_{\text{rsp}}$:
\begin{equation}
    \sigma_{\text{rsp}}^2 = \sigma_{\text{model}}^2 + \sigma_{\text{FWHM}}^2 + \sigma_{\text{direction}}^2.
\end{equation}
By accounting for the calculated $\sigma_{\text{model}}$ and allowing a conservative margin, we adopt a total value of 8\% as the systematic uncertainty for the instrument response, $\sigma_{\text{rsp}}$.

\subsection{Correction Systematics}

The unprecedented photon flux from GRB 221009A led to a significant non-linearity in the counting response of the HEPP-H detector, primarily driven by the instrumental effects. Standard analysis techniques are insufficient for such conditions, necessitating the development of a comprehensive, dynamic correction framework based on MC simulations of the full data acquisition process \citep{2026ApJ...996....5D}. 

We quantified the total systematic uncertainty of this pipeline via an extensive validation campaign. First, closed-loop simulations with known input spectra established the framework's internal self-consistency. Second, to establish a conservative upper limit for real-world application, we performed a cross-validation study. The maximum error of each energy channel given by multiple groups of correction data samples is used as the energy spectrum correction error in this study. Based on the reported maximum residual distribution which generally falls within 5$\sigma$ of the statistical error, this campaign obtains a conservative upper bound of 5\% on the total systematic uncertainty of the correction pipeline, denoted $\sigma_{\text{DCP}}$.

\subsection{Background Systematics}
The dominant systematic uncertainty in the background subtraction arises from the intrinsic temporal variability of the in-orbit particle environment, which depends on geographic location, local time, solar activity, and geomagnetic disturbances \citep{2025SpWea..2304197Z, 2025ScChD..68.3776W, wang2026newly}. The model is based on an average of six revisiting orbits with clean backgrounds strictly prior to the burst. We quantify this orbit-to-orbit instability as a systematic uncertainty, $\sigma_{\text{sys}}$, defined by the sample variance of the count rates across the ensemble of $m = 6$ background orbits:
\begin{equation}
    \sigma_{\text{sys}}^2 = \frac{1}{m-1} \sum_{i=1}^{m} \left( \frac{N_i}{t_i} - \bar{R} \right)^2,
\end{equation}
where $N_i$ and $t_i$ are the total counts and exposure time for the $i$-th orbit, and $\bar{R}$ is the average count rate. The total uncertainty of the background model, $\sigma_{\text{BG}}$, is the quadratic sum of this systematic component and the standard statistical uncertainty of the mean, $\sigma_{\text{stat}}$:
\begin{equation}
    \sigma_{\text{BG}}^2 = \sigma_{\text{stat}}^2 + \sigma_{\text{sys}}^2,
\end{equation}
where $\sigma_{\text{stat}}^2 = \frac{1}{m^2} \sum_{i=1}^{m} (N_i / t_i^2)$. Through error propagation across these representative stages, we find that the background systematic uncertainty $\sigma_{\text{BG}}$ remains within a conservative limit of 3\% in our total error budget.

\subsection{Total Systematic}

By combining the three independent and uncorrelated systematic components identified above---instrument response ($\sigma_{\text{rsp}} = 8\%$), correction systematics ($\sigma_{\text{DCP}} = 5\%$), and background modeling ($\sigma_{\text{BG}} = 3\%$), we calculate the total systematic uncertainty $\sigma_{\text{S}}$ using Equation \ref{eq:sigma_total_app}. The resulting combined systematic uncertainty is:
\begin{equation}
    \sigma_{\text{S}} = \sqrt{\sigma_{\text{rsp}}^2 + \sigma_{\text{DCP}}^2 + \sigma_{\text{BG}}^2} \approx 9.9\%
\end{equation}
To ensure the maximum reliability of our physical interpretations, we adopt a conservative total systematic uncertainty of 10\% for all flux measurements and spectral parameters derived from the HEPP-H data.

\section{Theoretical interpretation}
% 黎卓老师
\setcounter{equation}{0}

\begin{figure*}[t]
\centering
  \includegraphics[width=0.9\textwidth]{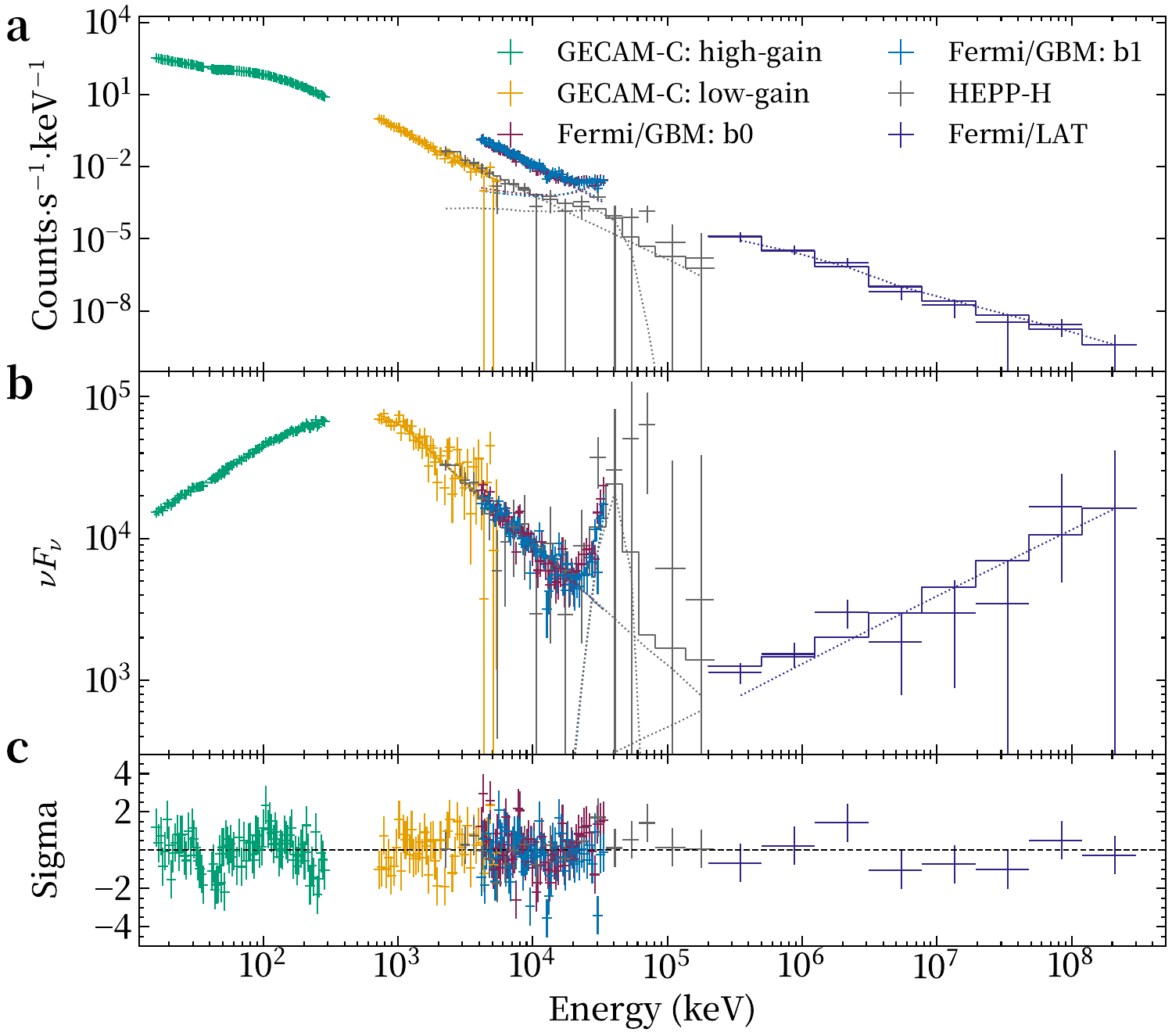}
  \caption{\textbf{Joint spectral fitting results within the time interval $T_{\rm ref} + 246$--$256$\,s.} The energy spectrum is reconstructed using a multicomponent model consisting of Band, Gaussian, and power-law functions. 
  \textbf{(a)} Observed count rate spectrum compared with the folded best-fit model. Data are presented for GECAM-C (high-gain, green; low-gain, orange), Fermi/GBM BGO detectors (b0, purple; b1, blue), HEPP-H (grey), and Fermi/LAT (dark blue). 
  \textbf{(b)} The $\nu F_\nu$ spectral distribution (units: $\mathrm{keV}^2 \cdot \mathrm{Photons} \cdot \mathrm{cm}^{-2} \cdot \mathrm{s}^{-1} \cdot \mathrm{keV}^{-1}$). The total model (solid curve) is the superposition of individual components (dashed curves), which include the low-energy Band, the high-energy power-law, and a Gaussian spectral line feature centered near $37$\,MeV. 
  \textbf{(c)} Residuals in units of standard deviation ($\sigma$) with $1\sigma$ error bars.}
  \label{fig:246_256s}
\end{figure*}

HEPP data fill the gap of $\sim$10--100\,MeV in gamma-ray emission. The broadband spectra of GRB 221009A from combining HEPP data with the other instruments in general show, as in Fig.~\ref{fig:246_256s}, two big spectral humps, where the low energy one peaking at a few hundred keV is explained as the prompt emission, believed to be produced by synchrotron radiation from electrons, whereas the second spectral hump dominating beyond few hundreds MeV could be consistent with the inverse Compton radiation generated by electrons in the afterglow shock \citep{2023Sci...380.1390L}.

The extension of the HEPP data up to 100 MeV without obviously steepening break may imply that the prompt emission does not suffer gamma-gamma annihilation up to at least 100 MeV, which can put a constraint on the GRB Lorentz factor. 
For a power-law photon number spectrum $dN/dE\propto E^{-(a+1)}$, the observed photon energy where the optical depth for $\gamma\gamma\rightarrow e^\pm$ absorption is $\tau_{\gamma\gamma}=1$ can be expressed as \citep{2024ApJ...973L..17Z, 1987MNRAS.227..403S}
\begin{equation}
    E_{\gamma\gamma}=\Gamma^2
\left[
\frac{4\pi r^2 m_e c^2}{(a \zeta(a) / 2) \sigma_T \delta E_\gamma}
\right]^{1/a}, 
\end{equation}
where $r$ is the radius of the prompt emission region, $\Gamma$ is the bulk Lorentz factor of the GRB jet, $\delta E_{\gamma}$ is the observed gamma-ray energy released in one light-curve variability, and $\zeta(a)= (7/6)(2+a)^{-1}(1+a)^{-5/3}$. Taking $E_{\gamma\gamma}> 200 \ {\rm MeV}$, and $a\simeq 1.6$ for GRB 220109A, we obtain a lower limit,
$\Gamma > 24~(r/10^{16} {\rm cm})^{-5/8} (\delta E_{\gamma}/10^{54}  {\rm erg})^{5/16}$.

Beyond the spectral continuum fitted with Band function, the data show some excess features, narrow line emission, and small spectral bump. We discuss their origins in the following.

\subsection{MeV spectral bump}
\label{sec:theory-bump}

HEPP detects additional spectral bumps at a few MeV beyond the Band function at the first gamma-ray light-curve peak of GRB 221009A, which can be fitted comparably well by either a blackbody or a CPL model (Table~\ref{tab:CPL_vs_BB_model}). In observations such a MeV spectral component is discovered for the first time in GRBs, and in theory there are rare predictions of MeV components in prompt emission, except for blackbody-like thermal emission from the fireball \citep{1986ApJ...308L..43P,2000ApJ...530..292M}.

Consider the thermal emission in the fireball model. With the total luminosity, $L_{w}=10^{53}L_{w,53}\rm erg\,s^{-1}$, and the radius of the initial fireball, $r_0=10^7r_{0,7}$cm, the temperature of the fireball at rest is 
\begin{equation}
    kT_{\rm 0}=(L_w/4\pi r_0^2ca)^{1/4}= 2.10  \ L_{w,53}^{1/4} r_{0,7}^{-1/2}\ {\rm MeV}
\end{equation}
The fireball expands and accelerates due to the radiation pressure, as $\Gamma(r)=r/r_0$. When the ejecta expands to the so-called photosphere radius where it becomes optically thin, a thermal emission is released, apparently as a blackbody spectrum. This may occur in either the accelerating phase or the coasting phase after the acceleration, depending on the initial dimensionless entropy $\eta=L_w/\dot{M}c^2$ which represents the mass loading rate $\dot{M}$. Define $\eta_*$ the critical entropy where the fireball acceleration saturates at the photosphere radius. The thermal emission occurs in the acceleration phase if $\eta>\eta_*$, or in the coasting phase if $\eta<\eta_*$.

According to previous studies, the apparent blackbody temperature and luminosity are 
\begin{equation}
    kT_{\rm ph}=\left\{\begin{array}{ll}
     \frac{kT_0}{(1+z_g)(1+z)}= 2.10  \ L_{w,53}^{1/4} r_{0,7}^{-1/2} (1+z_g)^{-1}(1+z)^{-1}\ {\rm MeV} & \eta>\eta_*,\\
      \frac{kT_0}{(1+z_g)(1+z)} (\frac{\eta}{\eta_*})^{\frac{8}{3}} = 0.102  \ L_{w,53}^{-5/12} r_{0,7}^{1/6} (\frac{\eta}{500})^{8/3} (1+z_g)^{-1}(1+z)^{-1}\ {\rm MeV} & \eta<\eta_*,
    \end{array}\right.
\end{equation}
and
\begin{equation}
    L_{\rm ph}=\left\{\begin{array}{ll}
    L_w=10^{53}L_{w,53}{\rm erg\,s^{-1}} & \eta>\eta_*, \\
     L_w (\frac{\eta}{\eta_*})^{\frac{8}{3}}= 4.84\times 10^{51} L_{w,53}^{1/3} r_{0,7}^{2/3} (\frac{\eta}{500})^{8/3} {\rm erg\,s^{-1}} & \eta<\eta_*,
     \end{array}\right.
\end{equation}
where $1+z_g=1/\sqrt{1-r_s/r_0}$ with $r_s$ the Schwarzschild radius of the central object, and
\begin{equation}
    \eta_*=(L_w\sigma_TY/8\pi m_p c^3 r_0)^{1/4}= 1.56\times 10^3 (L_{w,53}Y/r_{0,7})^{1/4},
\end{equation}
with  $Y$ being the electron number per baryon \citep{2000ApJ...530..292M,2019pgrb.book.....Z}.

For GRB 221009A, the blackbody fit gives a temperature of $kT_{\rm ph}\simeq 2$ MeV and a flux of $F_{\rm bump}\sim2\times10^{-3}\rm erg\,s^{-1}cm^{-2}$, thus the thermal emission luminosity is $L_{\rm ph}\simeq10^{53}\rm erg\,s^{-1}$. %If the thermal emission occurs in the coasting phase, we obtain $L{w}=L_{\rm ph}\simeq 10^{53} \ {\rm erg \ s^{-1}}$ and $r_0\simeq 8.314 \times 10^{6}\ {\rm cm}$. 
If the thermal emission occurs in the coasting phase, i.e., $\eta<\eta_*$ case, plugging these results into the above equations gives  
\begin{equation}
    L_w=8.84\times 10^{52} L_{\rm ph,53}^{4/3} (kT_{\rm ph}/2 \ {\rm MeV})^{-4/3} r_{0,7}^{-2/3} (1+z_g)^{-4/3} \ {\rm erg \ s^{-1}}
\end{equation}
\begin{equation}
    \eta=1.58\times 10^3 L_{\rm ph,53}^{5/24}(kT_{\rm ph}/2 \ {\rm MeV})^{1/6} r_{0,7}^{-1/6} (1+z_g)^{1/6}
\end{equation}
As the above $\eta$ value is close to the $\eta_*$ value when $L_w=10^{53}\rm erg\ s^{-1}$, this case is close to the transition point where the fireball evolves from acceleration to the coasting phase, i.e., it is not in the deep coasting phase. Furthermore, the final ejecta Lorentz factor in this case, $\Gamma=\eta\simeq1500$, is inconsistent with the constraint imposed by fast cooling requirement for annihilating pairs discussed in the previous study \citep{2024ApJ...973L..17Z}.

On the other hand, the observations are more consistent with the $\eta>\eta_*$ case, where $L_w=L_{\rm ph}\simeq10^{53}\rm erg\ s^{-1}$. In this case $\eta$ is constrained only by a lower limit, $\eta>\eta_*(L_w=10^{53}{\rm erg\ s^{-1}})$. However, if the Lorentz factor $\Gamma$ is known after acceleration ends, it is constrained to \citep{2000ApJ...530..292M,2019pgrb.book.....Z}
\begin{equation}
    \eta= \frac{L_w\sigma_T Y}{8\pi m_p c^3 r_0 \Gamma^3}= 4.7\times 10^4 L_{w,53} Y r_{0,7}^{-1} (\Gamma/500)^{-3}.
\end{equation}
Such a high value of $\eta$ indicates that the initial fireball is extremely pure with few baryons.
In this case the initial fireball radius is constrained by the observed luminosity and temperature of the thermal emission, %$r_0\simeq 8.314 \times 10^{6}\ {\rm cm}$ cm, 
$r_0\sim10^7$cm, similar to the radius of the last stable orbit around a $10M_\odot$ Schwarzschild black hole. 

By time-integration, the observed isotropic-equivalent energy of the thermal emission from GRB 221009A is derived to be $E_{\rm therm,iso}=9\times 10^{53}$erg, which is only a fraction $6\%$ of the total gamma-ray energy release. As the fireball becomes optically thin before saturation and transitions into the coasting phase, the emitted thermal energy is comparable to the total fireball energy. Thus we conclude that the thermal fireball energy is only a fraction of the total energy, and the observed non-thermal emission should be produced by additional components, most likely the energy dissipation of a Poynting flux energy dominated jet. This is consistent with the argument for the narrow MeV line width that in order for a narrow line width, $\sigma/E\sim10\%$, the annihilating pairs should radiatively cool rapidly to non-relativistic stage, which requires radiative cooling processes in addition to inverse-Compton scatterings off the prompt GRB photons, most likely the synchrotron cooling in a magnetic field with the energy density much larger than the prompt radiation, implying a magnetically dominated GRB jet in GRB 221009A \citep{2024ApJ...973L..17Z}.

\subsection{MeV line emission}
\label{sec:theory-line}

The observed line energy evolution can be well fit with scaling $E_{\text{line}}\propto(t-t_0)^{-k}$, giving $k=-1.05_{-0.17}^{+0.22}$ and $t_0=T_{\rm ref}+226^{+8}_{-10}$s \citep{2024SCPMA..6789511Z}. The constraint of the temporal index is well consistent with the value $k=-1$ predicted by the curvature effect (see below). The best-fit value of $t_0\simeq T_{\rm ref}+226$s, consistent with the peak time of the second gamma-ray light-curve peak, further supports the curvature-effect explanation, where the line emission is produced by the ejecta that also accounts for the second gamma-ray light-curve peak. Moreover, as pointed out by the study \citep{2024ApJ...973L..17Z}, the other explanation of the line emission face many problems, thus here we will explain the line emission as pair annihilation and the temporal evolution of the line energy by the curvature effect \citep{2024ApJ...973L..17Z}. 

Consider that a spherical surface expands with a Lorentz factor $\Gamma$ ($\beta\equiv\sqrt{1-\Gamma^{-2}}$). Prompt emission, as well as the line emission, is produced abruptly at radius $r$ from the central engine. The time for photons emitted from a certain latitude $\theta$ ($\mu \equiv \cos \theta$) to arrive at the observer is expressed as 
\begin{equation}
    t=t_0+(1+z)(1-\beta \mu)\frac{r}{\beta c}, 
    \label{eq:high-latitude effect}
\end{equation}
where $t_0$ should be considered as the arrival time of photons emitted from the center $r=0$ at the point that the ejecta is released \citep{2024ApJ...973L..17Z}.
%解释线心能量
The narrow line width $\sigma/E_{\text{line}}\sim10\%$ requires that the pair should be non-relativistic in the comoving frame of the ejecta. The pair annihilation produce a central line energy in the comoving frame of $m_ec^2$. The observed line central energy for emission from a latitude $\theta$ should be $E_{\text{line}}=m_ec^2/\Gamma(1-\beta\mu)(1+z)$.
With Eq.~\ref{eq:high-latitude effect}, we obtain 
\begin{equation}\label{eq:line energy}
    E_{\text{line}}=m_e c^2\frac{r }{\Gamma \beta c} (t-t_0)^{-1},
\end{equation}
the standard value $k=-1$ from curvature effect. 
%Previous study \citep{2024SCPMA..6789511Z} finds that the observed line energy well follows $E_{line}\propto (t-t_0)^{-1}$, with which as assumption the best fit gives $t_0=T_{\rm ref}+226^{+9}_{-10} \ {\rm s}$. 
The new results obtained from HEPP-H observations (Fig. \ref{fig:line_evolution}) show that the the line energy at $T_{\rm ref}+(236-246)$s and $T_{\rm ref}+(246-256)$s are consistent with the $-1$ index, and then the curvature effect explanation. By fitting line energy evolution with Eq. \ref{eq:line energy}, we have radius of the emission region, $r=2.45\times10^{16}(\Gamma/500)$cm \citep{2024ApJ...973L..17Z}.

%解释流量

In the $\delta$-function approximation for the temporal behavior of the line emission in the ejecta comoving frame, and if the line emission is only $\theta$-dependent, the observed flux of the line emission as function of time $t$ may be given by
\begin{equation}\label{eq:flux_line}
    F_{\text{line}}=\frac{2\pi c(1+z)^4}{r D_L^2} \epsilon(\theta) f_{\Omega}(\theta), 
\end{equation}
where $\epsilon(\theta)$ is the line emission energy per unit solid angle of the emitting spherical surface, $f_{\Omega}(\theta)=\frac{dW}{Wd\Omega}$ is the fractional, angular distribution of the line emission energy, and $D_L$ is the luminosity distance. Note $t$ and $\theta$ relation is given by Eq. \ref{eq:high-latitude effect}. With the Lorentz transformation $dW/d\Omega=(dW'/d\Omega')/\gamma^3(1-\beta \mu)^3$ and $W=\int (dW/d\Omega) d\Omega$, we can obtain $f_{\Omega}$ if the angular distribution of emission in the comoving frame $dW'/d\Omega'$ is given. For the simple case of $\epsilon=$constant and $dW'/d\Omega'=$constant, we obtain the known relation $F\propto1/(1-\beta \mu)^{3}\propto(t-t_0)^{-3}$. 

%第一点：通过设定fOmega或epsilon的角分布得到流量光变曲线
The observed flux of line emission shown in Fig.~\ref{fig:line_evolution} behaves in a pattern different from the simple case. The flux first rises and starts to decrease at about $\sim t_0+20$s roughly as $F_{\text{line}}\propto(t-t_0)^{-2}$. 
The deviation of the behavior of flux evolution from the simple case indicates that the line emission is angle-dependent, i.e., either $\epsilon$ or $f_{\Omega}$, or both, should be angle dependent. 

Consider isotropic emission in the comoving frame, $dW'/d\Omega'=$const, then we have, from Eq. \ref{eq:flux_line}, $F_{\text{line}}\propto(t-t_0)^{-3}\epsilon(\theta)$. 
As the flux of line emission at $T_{\rm ref}+(237-246)$s is lower than the extrapolation back from the late-time power-law decay, consider that the peak time of the flux is about $t_{p}\sim t_0+20$\,s. This, using Eq. \ref{eq:high-latitude effect} and small angle approximation ($\theta\ll1$) in ultra-relativistic limit ($\Gamma\gg1$), corresponds to a latitude of $\theta_p\sim0.01[(t_{p}-t_0)/20\,\text{s}]^{1/2}(r/10^{16}\text{cm})^{-1/2}$ (The uncertainty of $\theta_p$ is a factor of 25\% given the uncertainty of $t_0$ is about 10 s).  In order for small line emission flux at $t<t_p$, $\epsilon(\theta)$ should be very small at $\theta<\theta_p\simeq5/\Gamma(\Gamma/500)$.

Assume $\epsilon\propto\theta^\delta$ at $\theta>\theta_p$. In ultra-relativistic and small angle limit, $t-t_0\propto\theta^2$ for $\theta\gg1/\Gamma$, thus $F_{\text{line}}\propto (t-t_0)^{-3} \theta^{\delta}\propto (t-t_0)^{-3+\frac{\delta}{2}}$ at $t>t_p$. The observed decay is $F_{\text{line}}\propto (t-t_0)^{-2.13}$, thus $\delta\simeq1.7$, i.e., $\epsilon\propto\theta^{1.7}$, increasing toward larger latitude.

The facts that the line emission is weak in the central jet but increases with latitude may be consistent with a Poynting jet released from the central engine, where the magnetic field is small in the central part but larger at high latitude, so that pairs could be produced, cooling to non-relativistic phase, and annihilate more efficiently at higher latitude, as shown in Fig. \ref{fig:physical_process_picture}.

\clearpage
\section{Additional Figures and Tables}

This appendix presents additional figures and tables supporting the instrumental characterization, data processing, spectral analysis, and robustness tests. Figures F1–F4 describe the instrument configuration, observational data, calibration, and instrumental-effect corrections. Figures F5–F8 present the time-resolved spectral analyses. Figures F9–F15 show the observational geometry, detector configuration, instrumental responses, and cross-validation results. Figures F16–F18 present the sliding-window and emission-line evolution analyses. Tables F1–F4 summarize the instrument performance, gain evolution, spectral fitting, and alternative-model comparisons.

%Fig S1
\begin{figure*}[htbp]
\centering
  \includegraphics[width=0.7\textwidth]{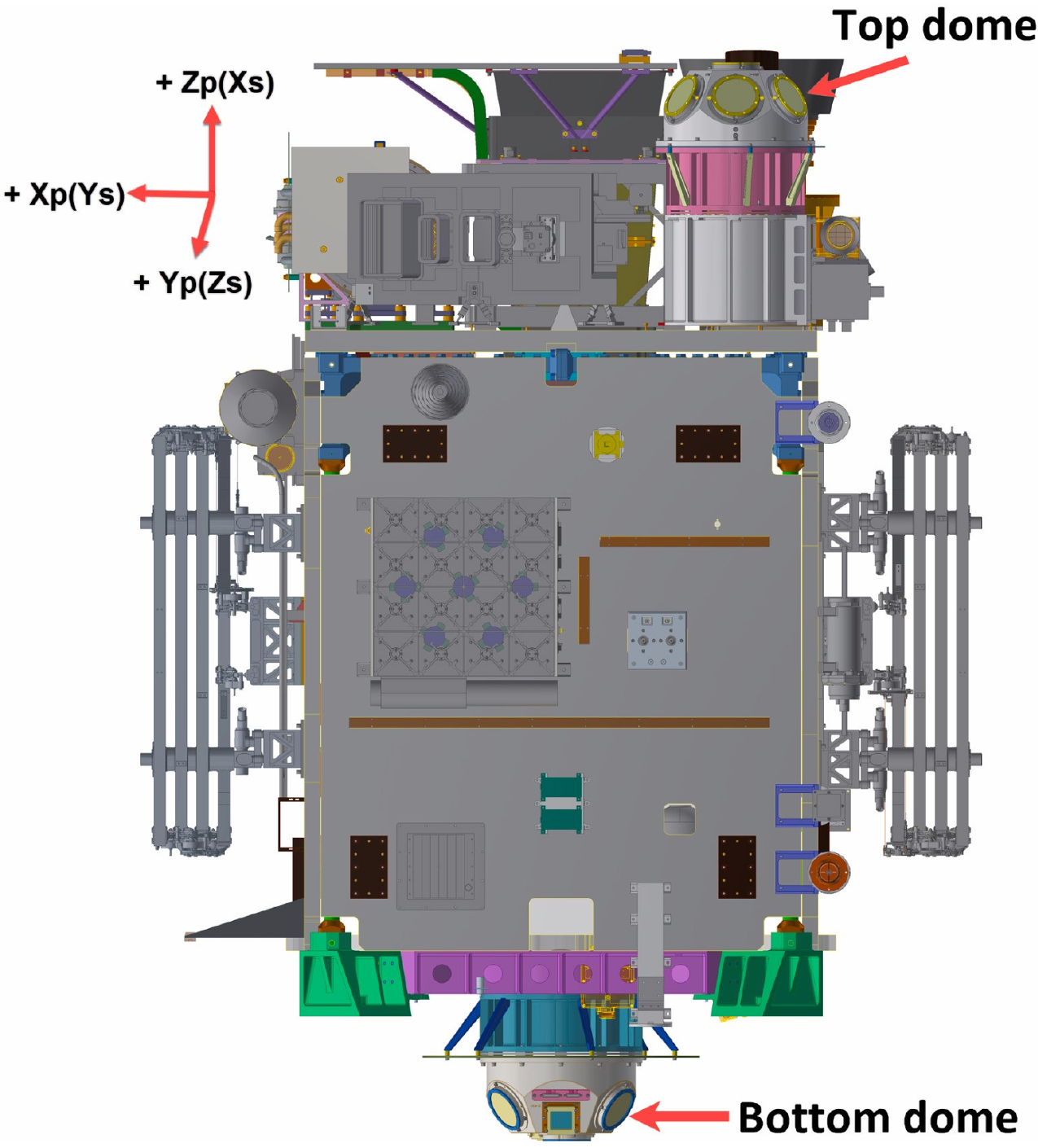}
  \caption{\textbf{Schematic layout of the GECAM-C (HEBS) payload on the SATech-01 satellite.} The HEBS instrument is designed with a two-compartment structure to maximize its field of view. The red arrows indicate the two main components: the top and bottom domes, which collectively form the primary structure of the HEBS payload and house its detector units. The payload coordinate system (X$_{\text{p}}$, Y$_{\text{p}}$, Z$_{\text{p}}$) is also shown for reference.}
  \label{fig:GECAM_C}
\end{figure*}

%Fig S2
\begin{figure*}[t] 
    \centering 
    \begin{subfigure}[t]{0.9\textwidth}
    \caption{}
        \centering
        \includegraphics[width=\linewidth]{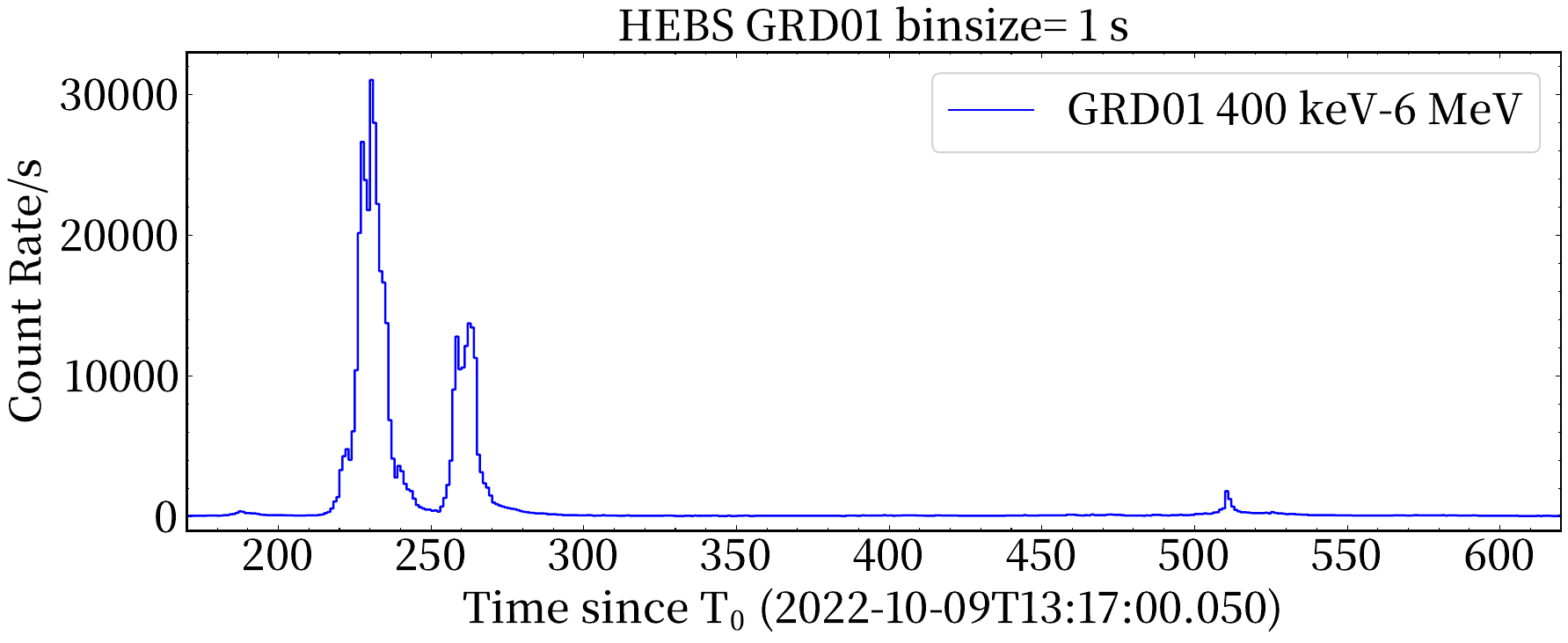}
        \label{subfig:lc_1s}
    \end{subfigure}
        
    \vspace{0.15cm} 

    \begin{subfigure}[t]{0.9\textwidth}
    \caption{}
        \centering
        \includegraphics[width=\linewidth]{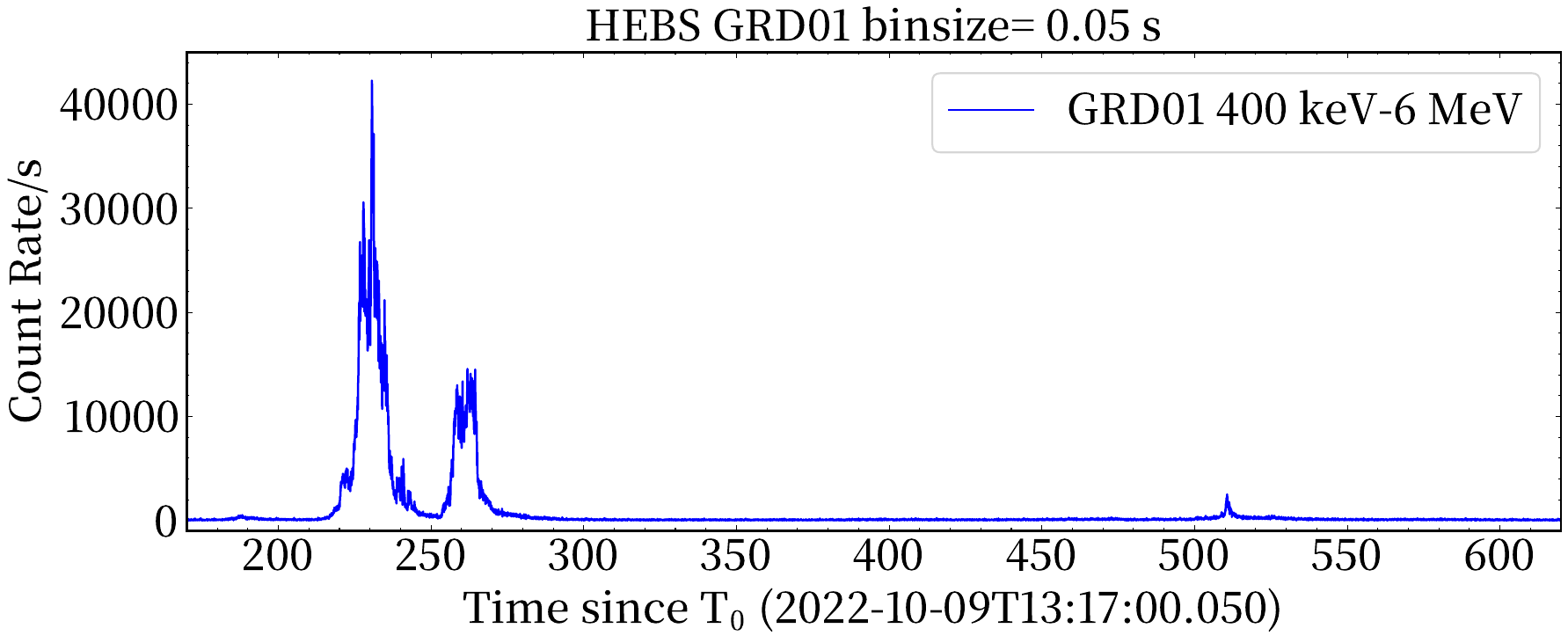}
        \label{subfig:lc_005s}
    \end{subfigure}
    
  \caption{\textbf{Multi-resolution light curves of GRB 221009A observed by GECAM-C (HEBS).} 
  The figure shows the light curve in the 400\,keV--6\,MeV range from a single Gamma-Ray Detector (GRD01) unit. 
  \textbf{(a)} The light curve binned at 1-second resolution, showing the overall morphology of the main emission episode, which consists of two prominent peaks. \citep{2024SCPMA..6789511Z, 2023arXiv230301203A}. 
  \textbf{(b)} The same data binned at a high resolution of 0.05 seconds, revealing the rapid and complex variability within these peaks.}
  \label{fig:GRB221009A_LC}
\end{figure*}

%Fig S3
\begin{figure*}[t]
    \centering
    \begin{subfigure}[t]{0.9\textwidth}
        \includegraphics[width=\linewidth]{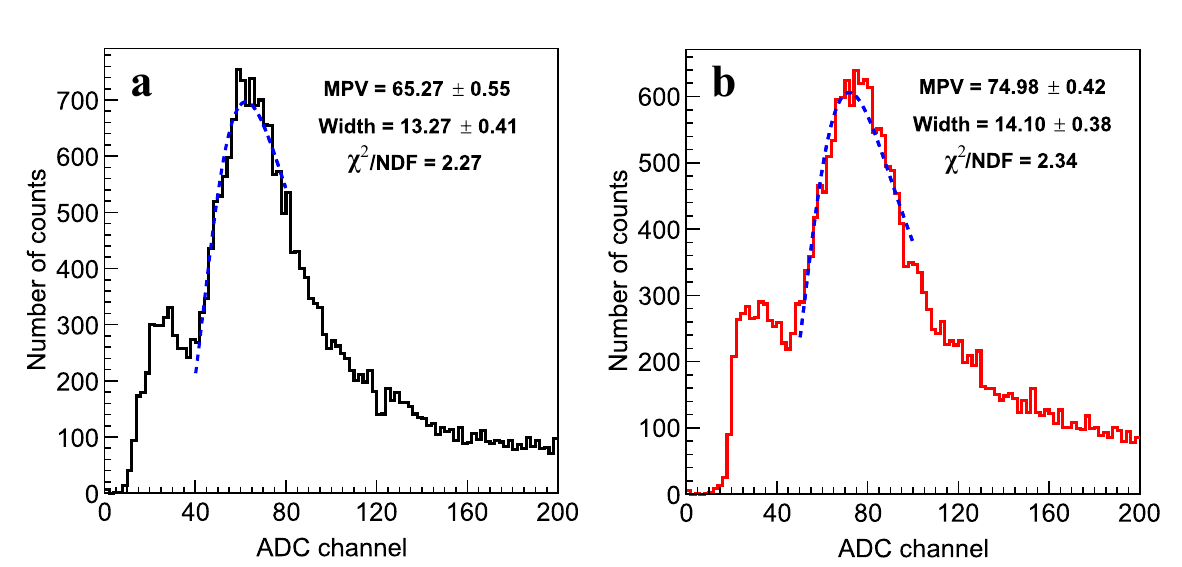}
    \end{subfigure}

    \vspace{0.05cm} % note
    
    \begin{subfigure}[t]{0.95\textwidth}
        \includegraphics[width=\linewidth]{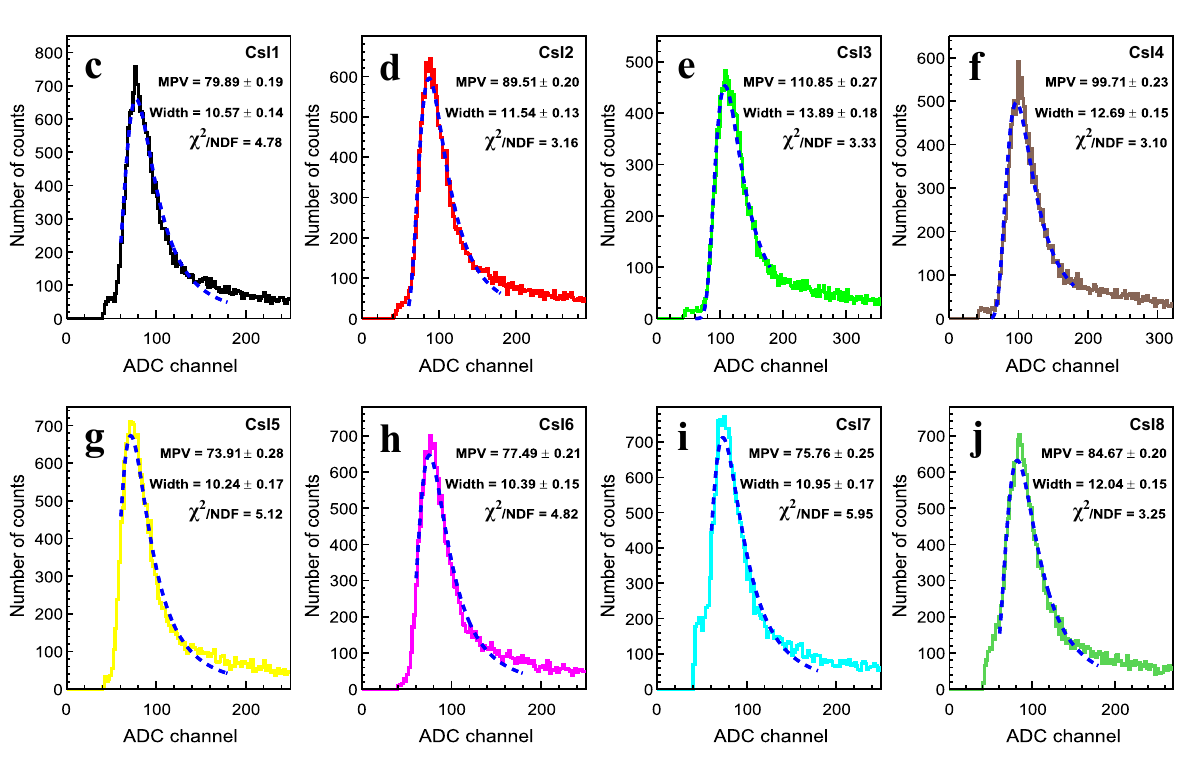}
    \end{subfigure}\hfill

    \caption{\textbf{ In-orbit MIP calibration spectra for the HEPP-H PSD and CsI calorimeter.}
    \textbf{(a, b)} ADC spectra from the two PSD layer PMTs. 
    \textbf{(c--j)} Spectra from the eight PMTs coupled to the four CsI calorimeter layers. 
    All data were recorded on 14 October 2022 (contemporary with the burst time period).}
    \label{fig:HEPP_CAL_PSD}
\end{figure*}

%Fig S4
\begin{figure*}[t]
  \centering

  \begin{minipage}{0.96\textwidth}
      \centering
      \includegraphics[width=\linewidth]{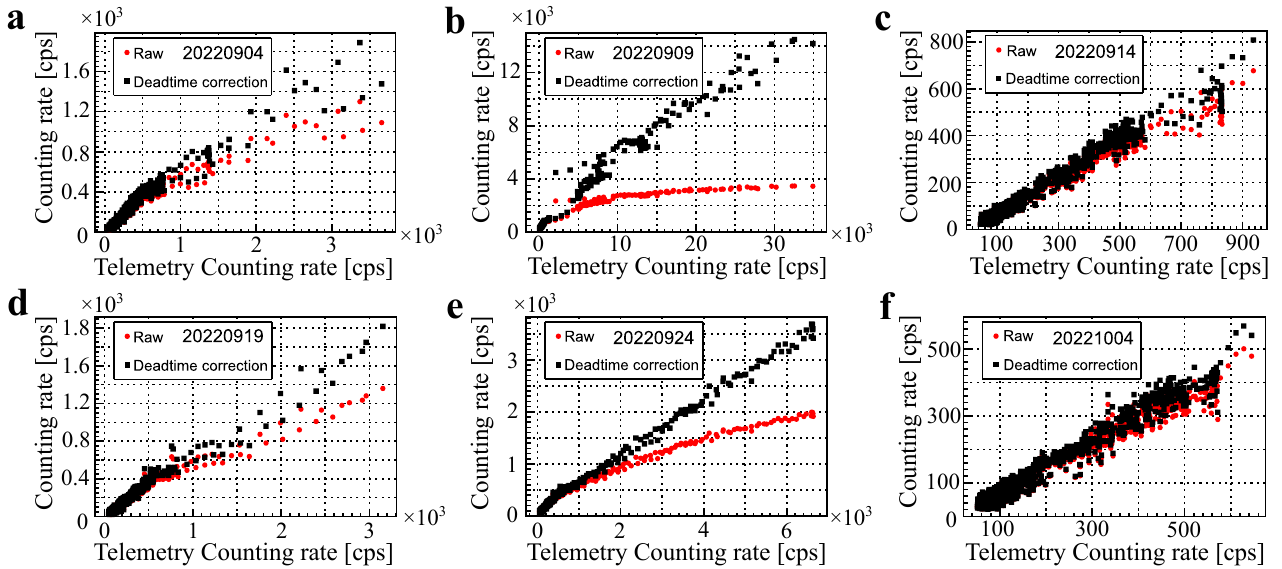}
  \end{minipage}

  \vspace{0.2cm} 

  \begin{subfigure}[t]{0.5\textwidth}
      \centering
      \includegraphics[width=\linewidth]{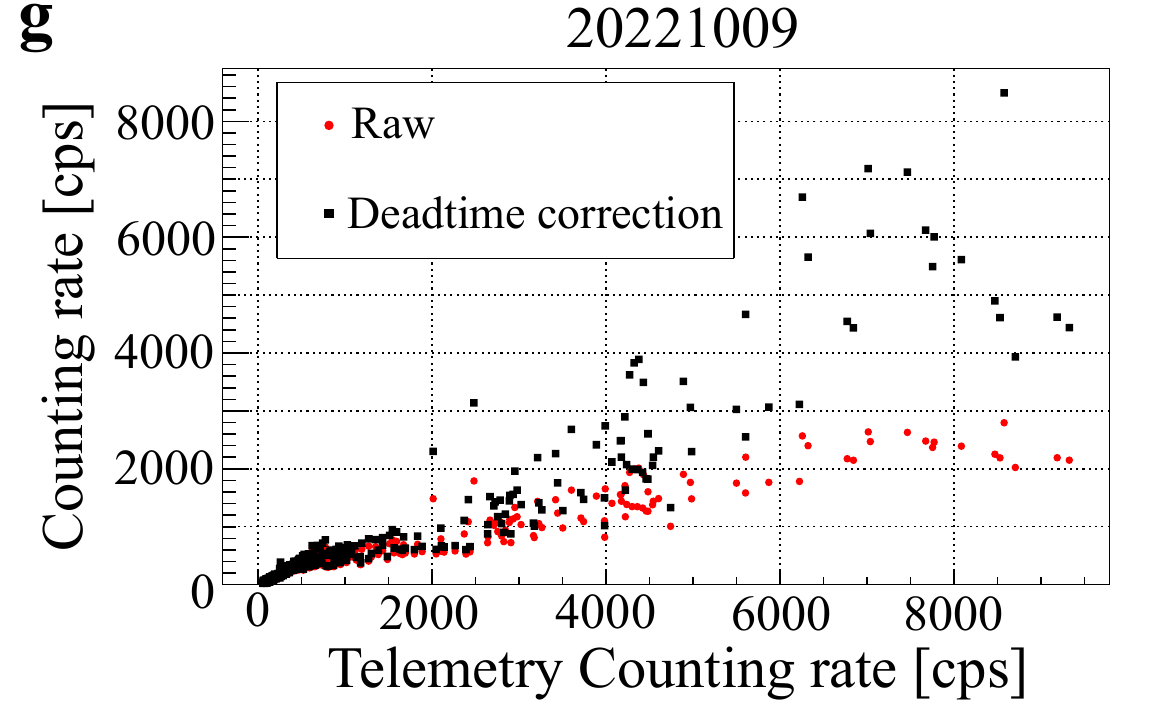}
      \label{subfig:de_09a}
  \end{subfigure}\hfill
  \begin{subfigure}[t]{0.5\textwidth}
      \centering
      \includegraphics[width=\linewidth]{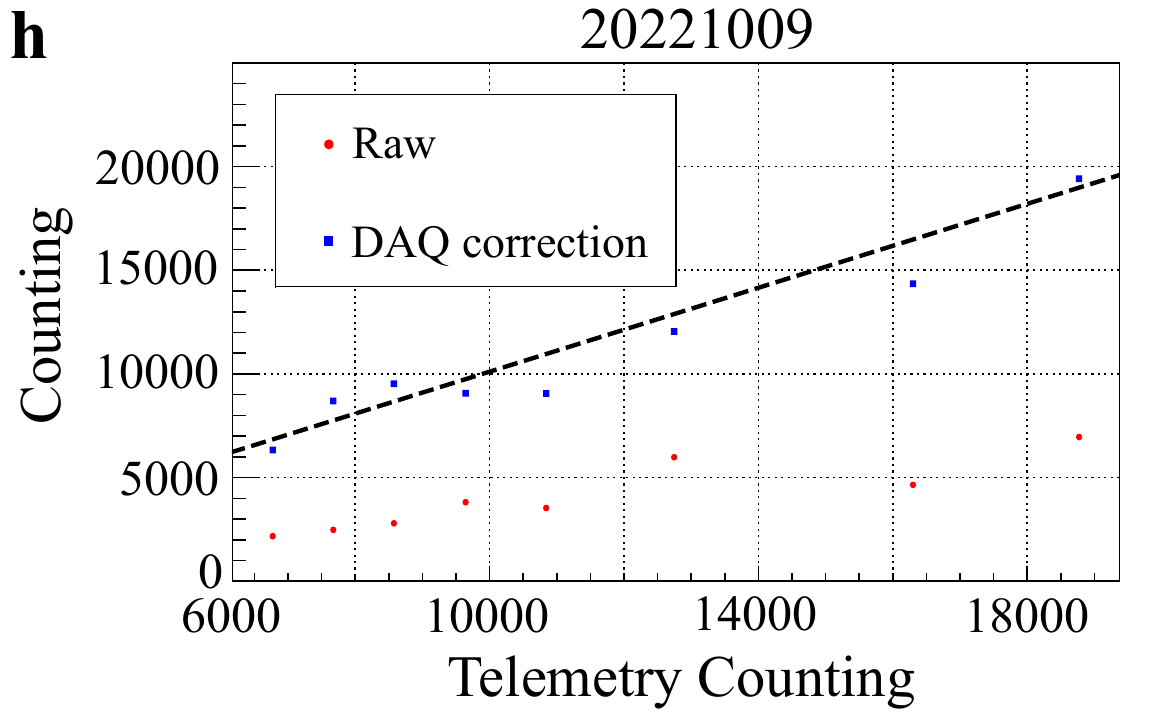}
      \label{subfig:c3_correction}
  \end{subfigure}

  \caption{
  \textbf{Correction for instrumental effects across varying flux levels.} 
  \textbf{(a--f)} Verification of the baseline correction under low-flux conditions using six revisiting orbits preceding the burst. Compared to the raw data (red data points), the dead time corrected the energy spectrum counting rate (black data points) show a linear relationship with the total telemetry counting rate, where $y=x$.
  \textbf{(g)} Saturation analysis during the main burst phase of GRB~221009A. The conventional dead-time correction (black points) deviates from linearity as telemetry rates exceed $\sim$4,000\,cps. 
  \textbf{(h)} High-flux signal recovery via the comprehensive DAQ correction (blue points). The black dashed line represents the ideal linear identity ($y=x$).}
  \label{fig:Dead_Time_Correction}
\end{figure*}

%Fig S6
\begin{figure*}[t] % note
    \centering
    % note
    \begin{subfigure}[b]{0.32\textwidth}
        \centering
        \includegraphics[width=\textwidth]{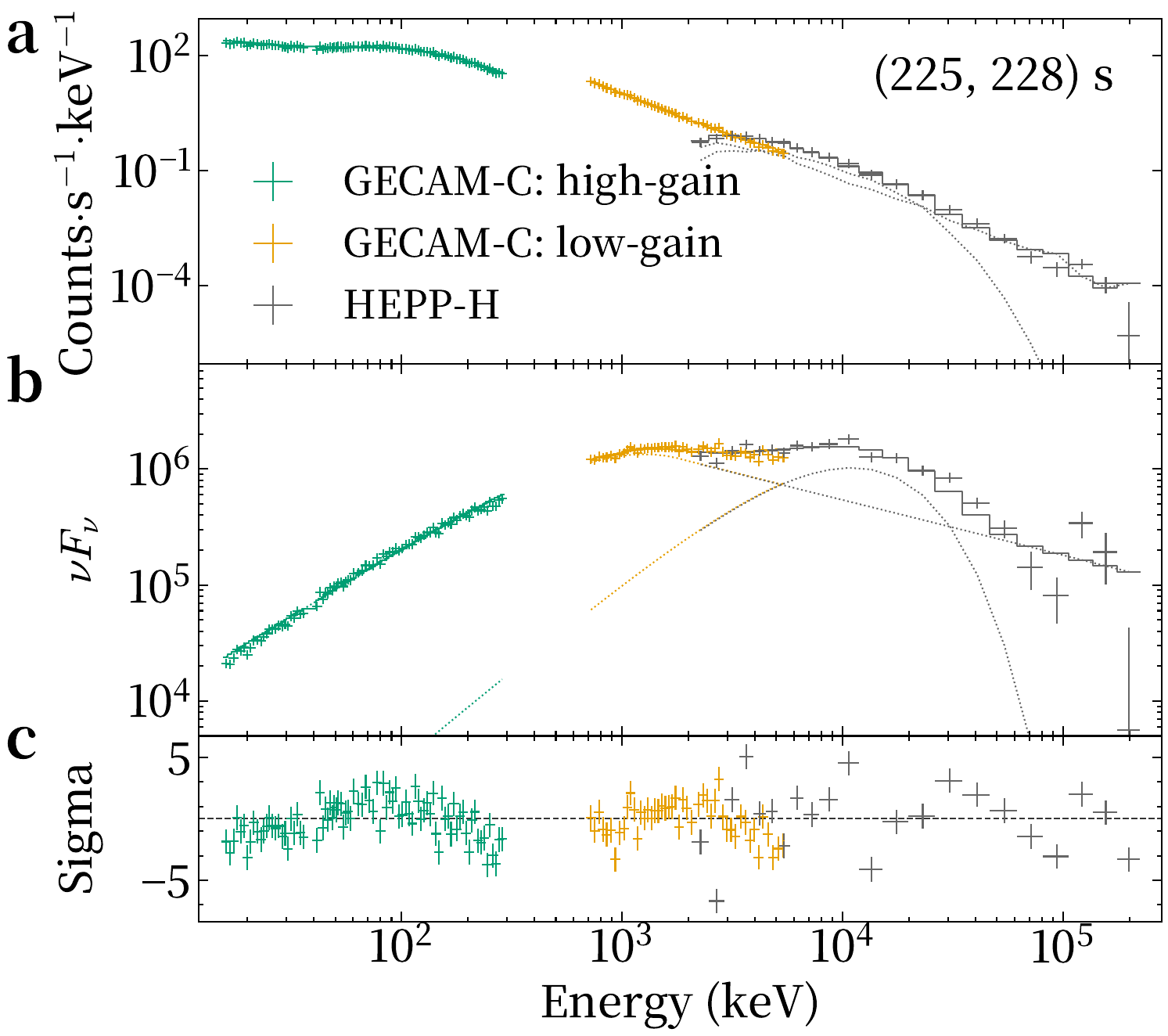} % note
    \end{subfigure}
    \hfill
    \begin{subfigure}[b]{0.32\textwidth}
        \centering
        \includegraphics[width=\textwidth]{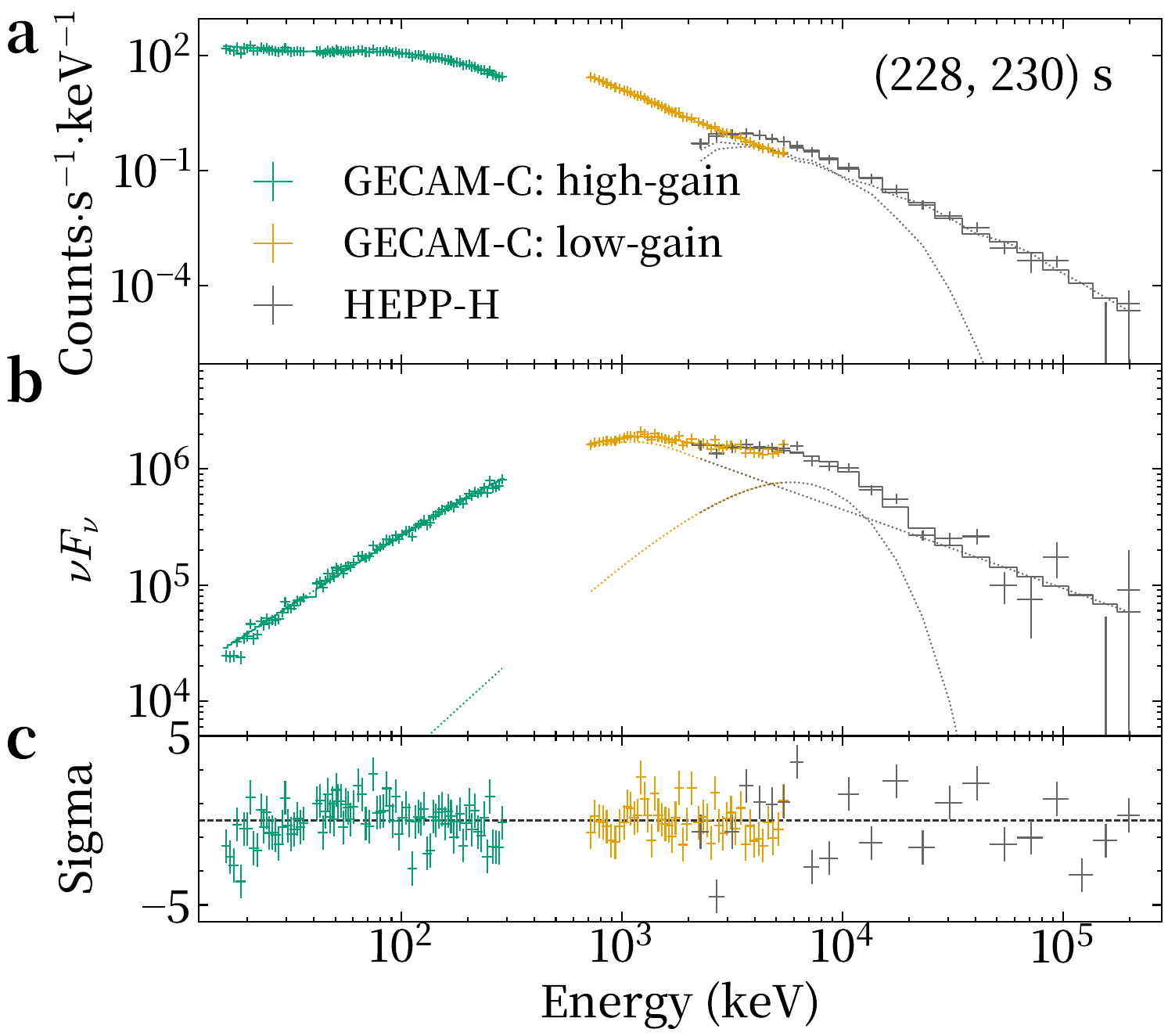}
    \end{subfigure}
    \hfill
    \begin{subfigure}[b]{0.32\textwidth}
        \centering
        \includegraphics[width=\textwidth]{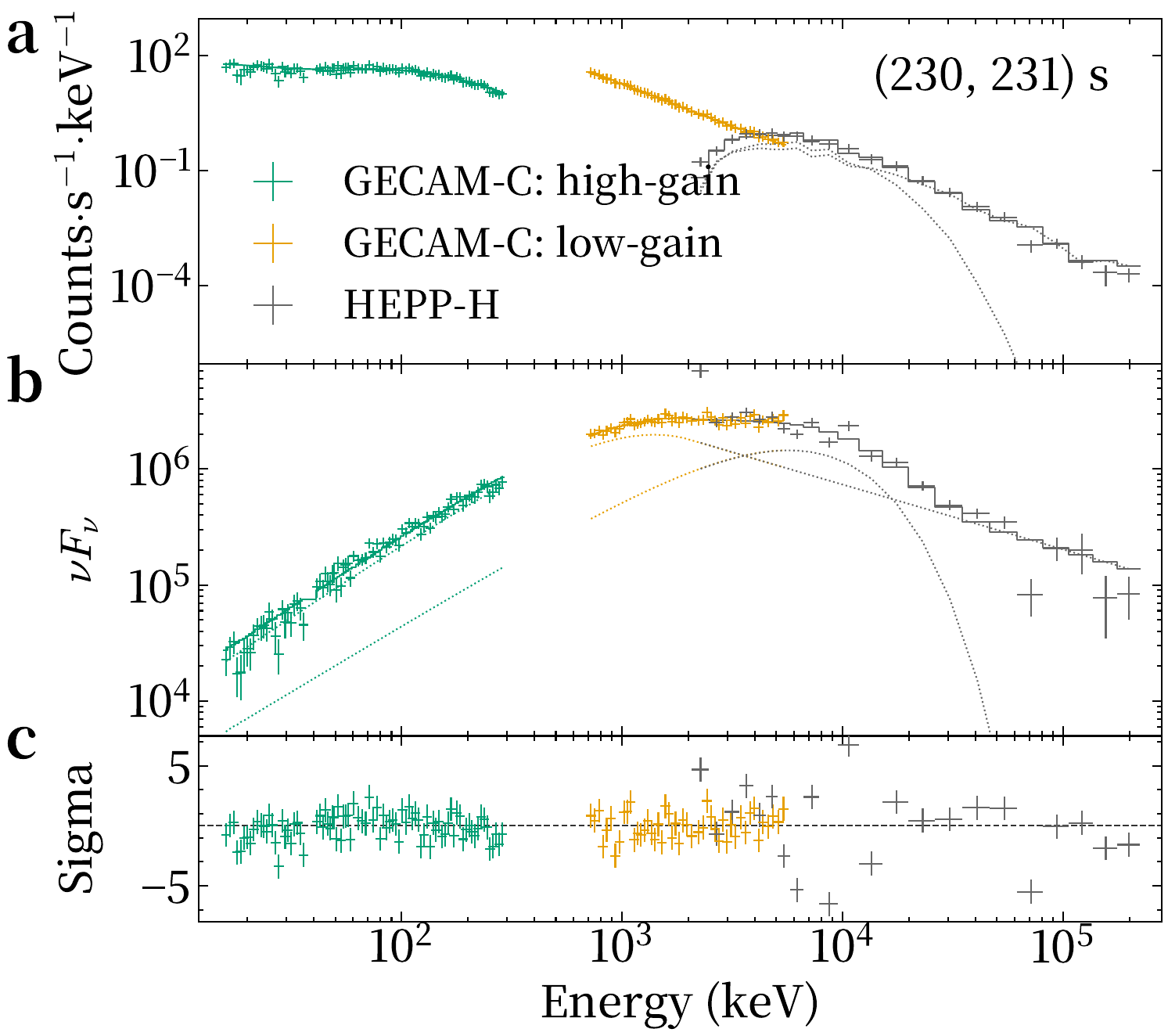}
    \end{subfigure}
    
    \vspace{0.2cm} % note

    % note
    \begin{subfigure}[b]{0.32\textwidth}
        \centering
        \includegraphics[width=\textwidth]{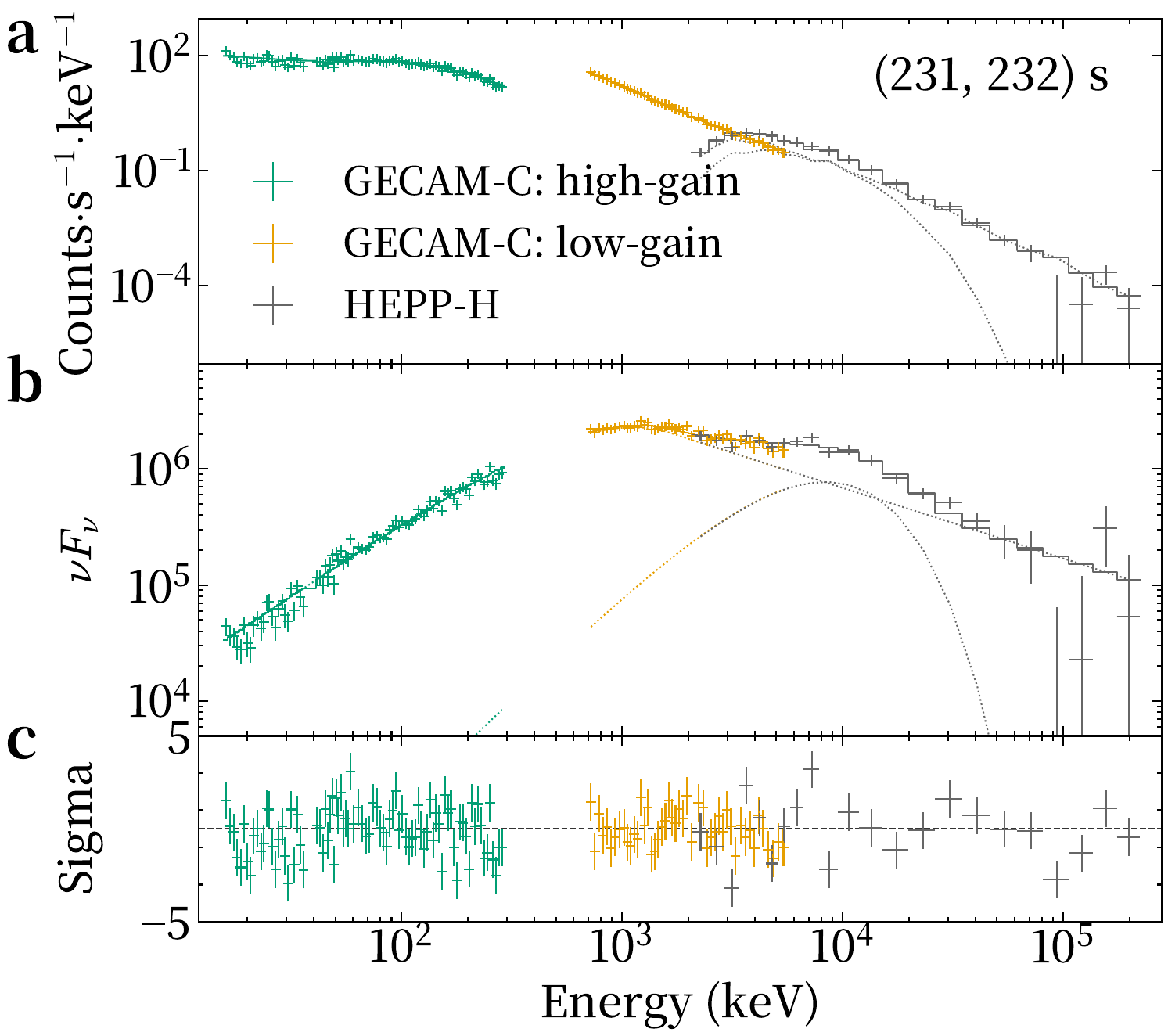}
    \end{subfigure}
    \hfill
    \begin{subfigure}[b]{0.32\textwidth}
        \centering
        \includegraphics[width=\textwidth]{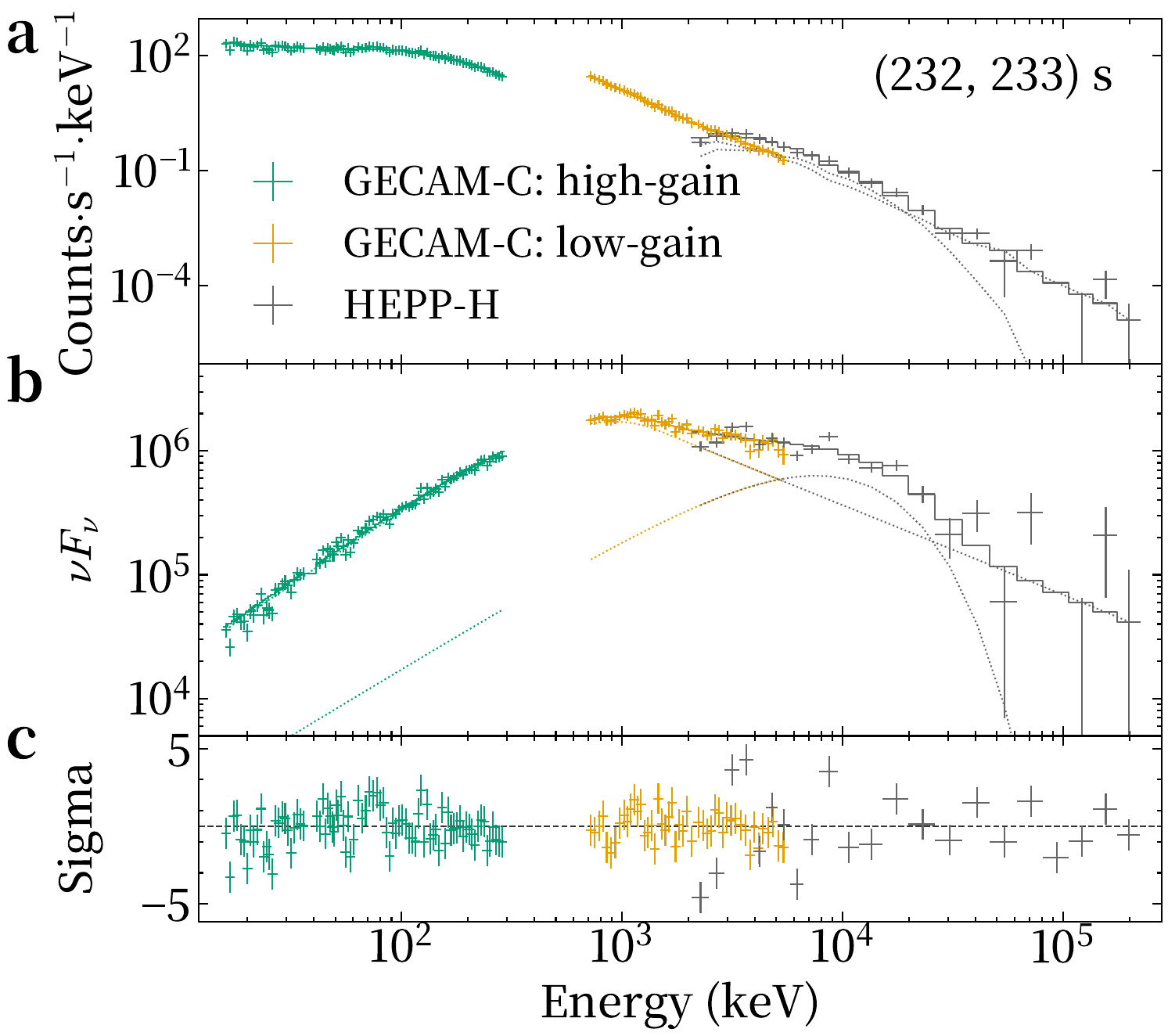}
    \end{subfigure}
    \hfill
    \begin{subfigure}[b]{0.32\textwidth}
        \centering
        \includegraphics[width=\textwidth]{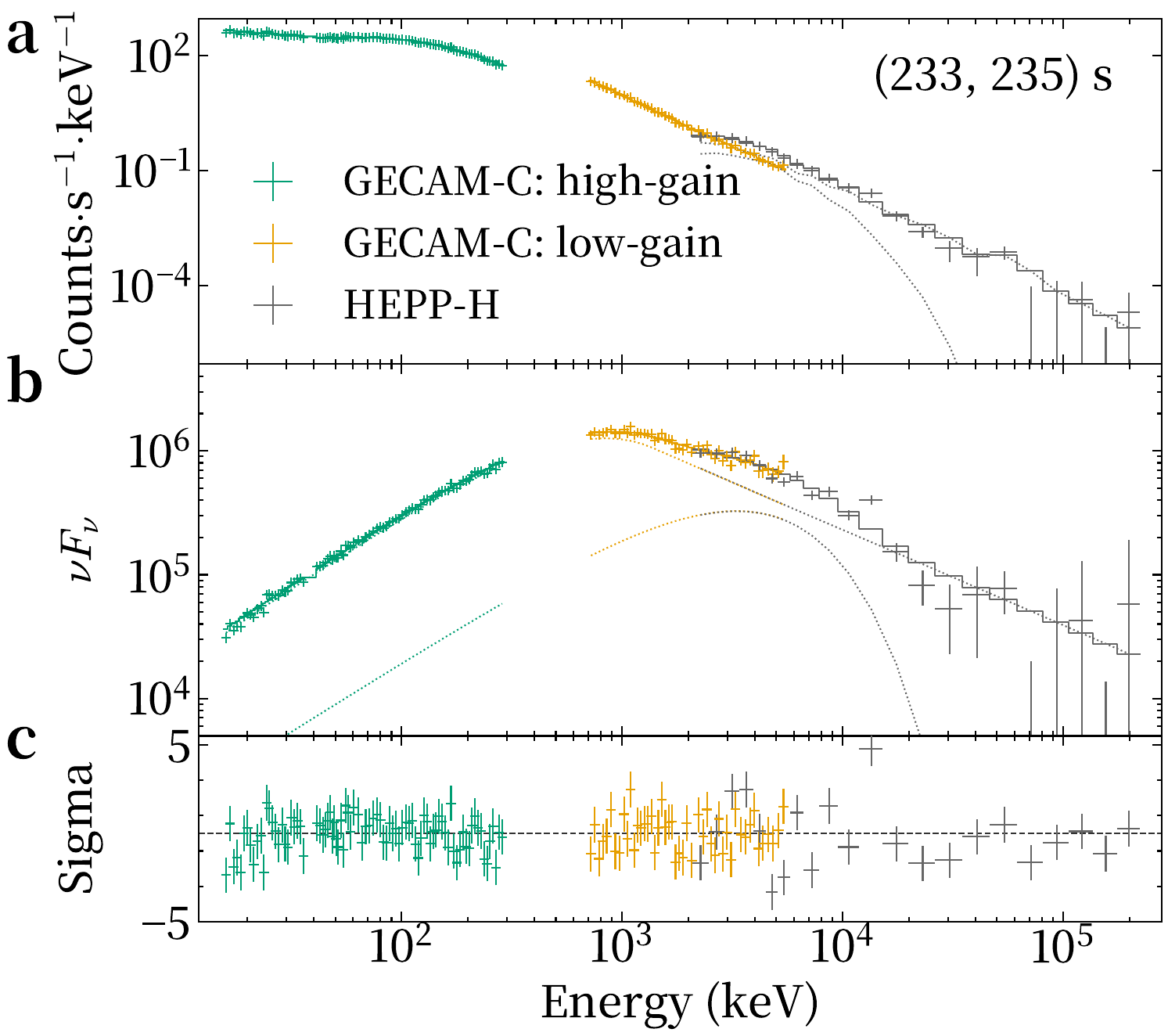}
    \end{subfigure}

    \vspace{0.2cm}

    % note
    \begin{subfigure}[b]{0.32\textwidth}
        \centering
        \includegraphics[width=\textwidth]{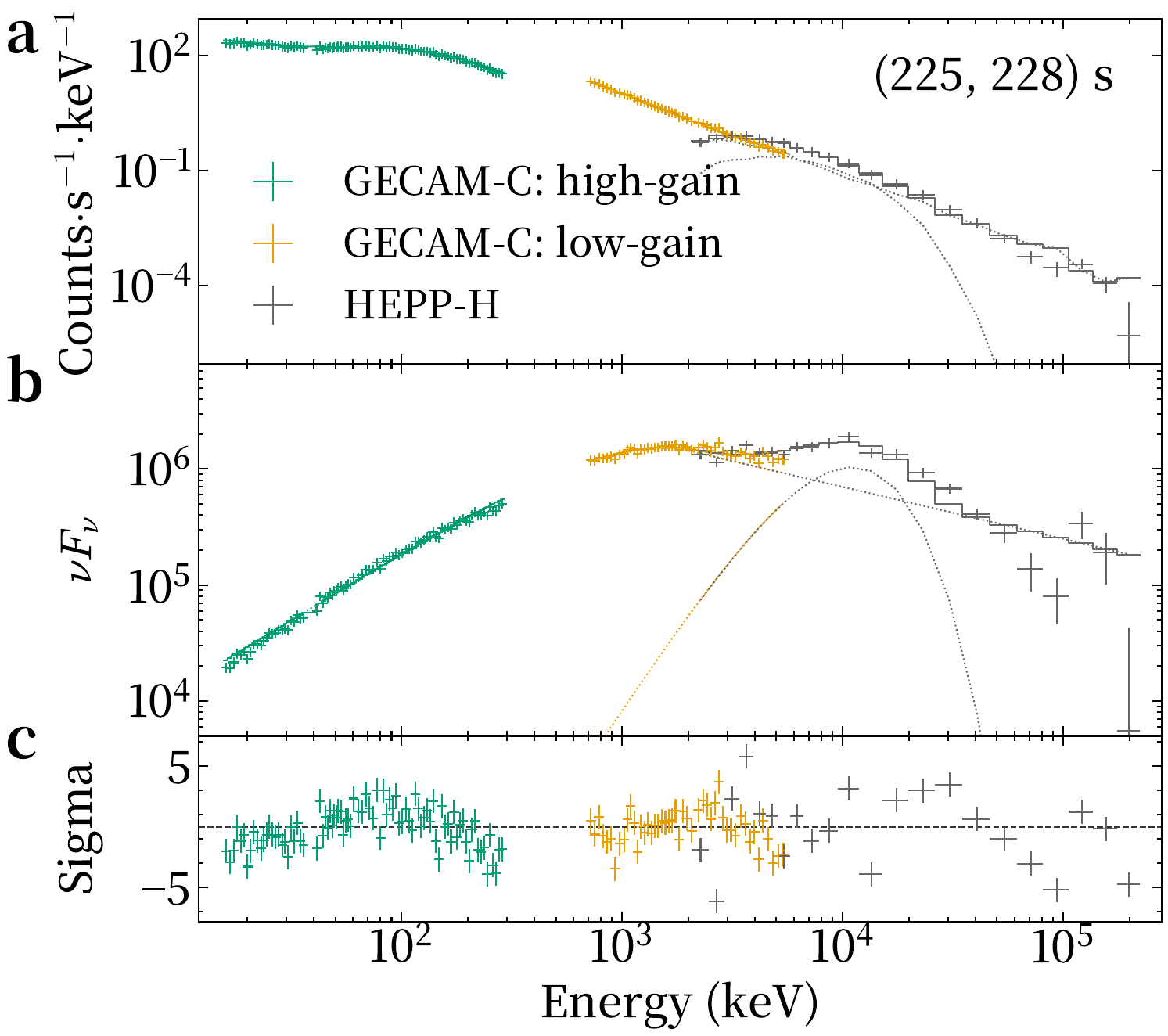}
    \end{subfigure}
    \hfill
    \begin{subfigure}[b]{0.32\textwidth}
        \centering
        \includegraphics[width=\textwidth]{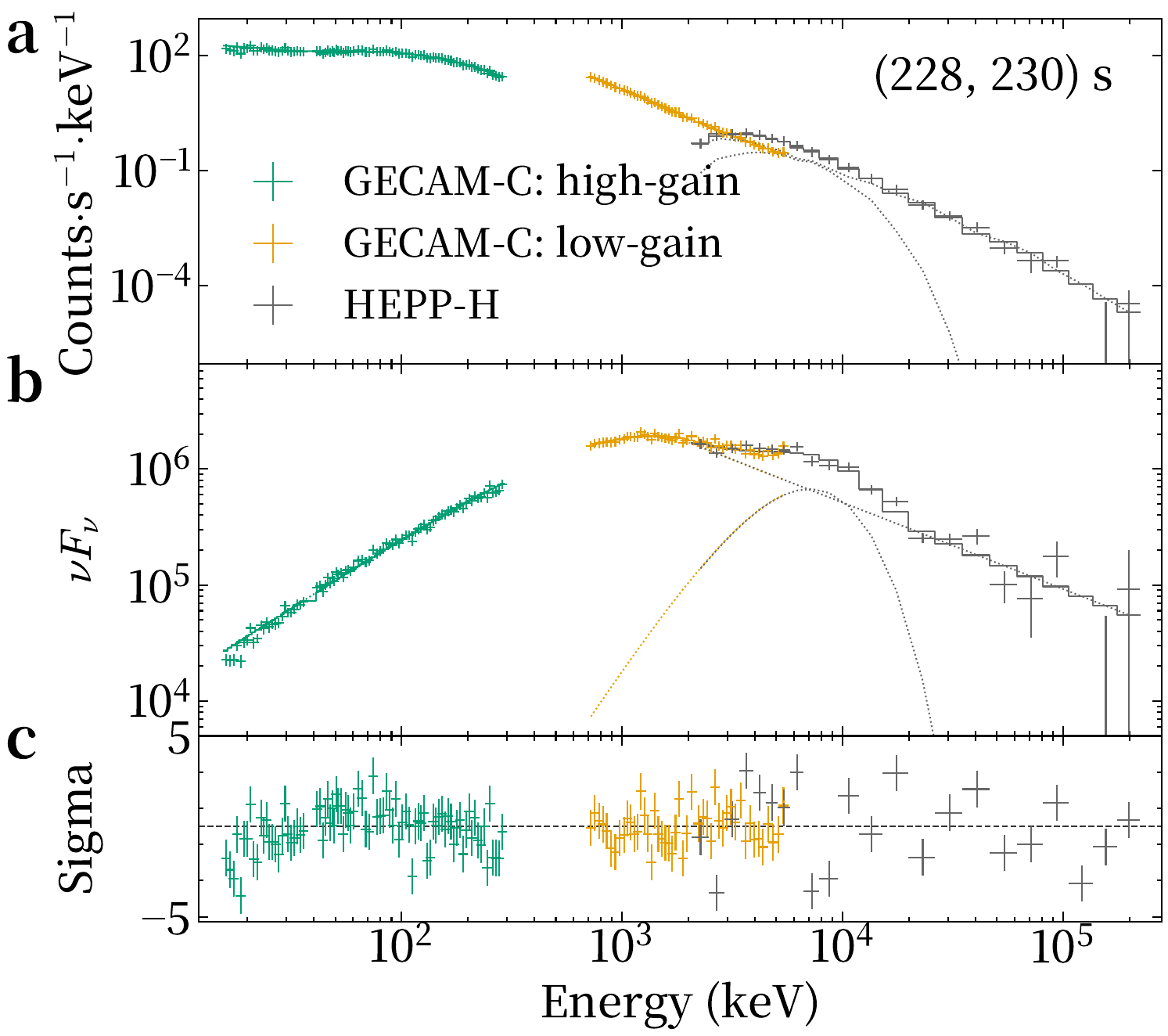}
    \end{subfigure}
    \hfill
    \begin{subfigure}[b]{0.32\textwidth}
        \centering
        \includegraphics[width=\textwidth]{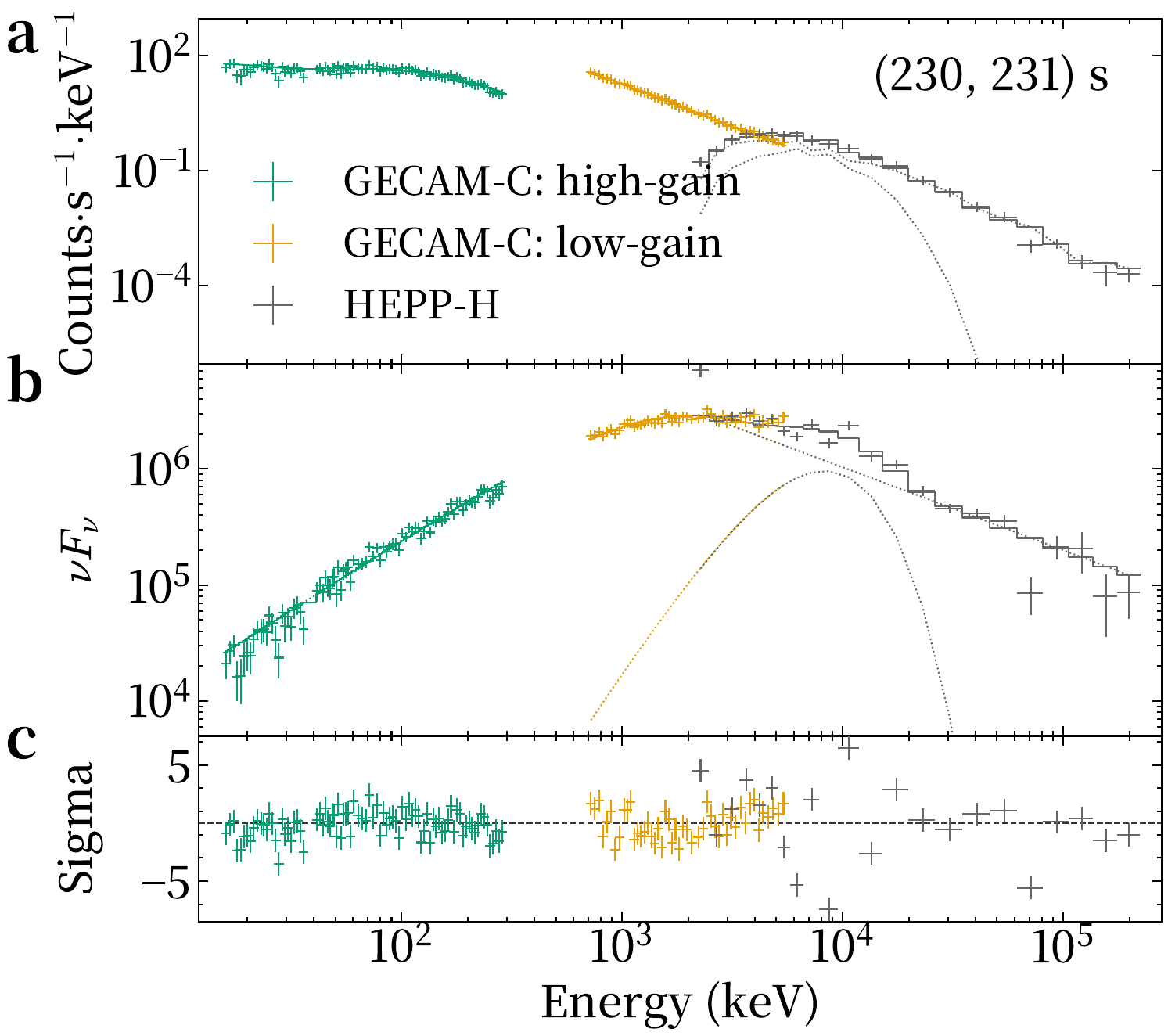}
    \end{subfigure}

    \vspace{0.2cm}

    % note
    \begin{subfigure}[b]{0.32\textwidth}
        \centering
        \includegraphics[width=\textwidth]{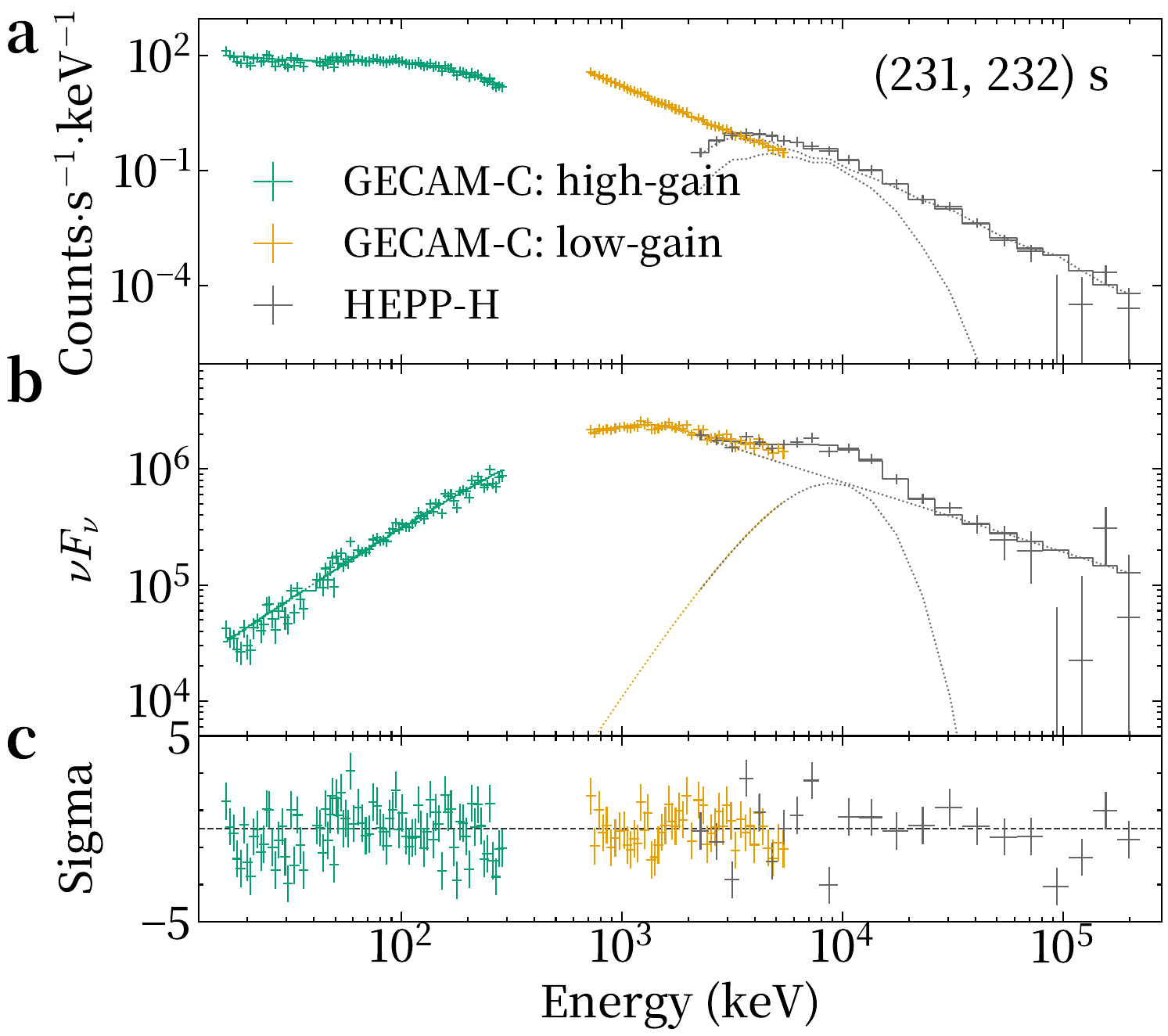}
    \end{subfigure}
    \hfill
    \begin{subfigure}[b]{0.32\textwidth}
        \centering
        \includegraphics[width=\textwidth]{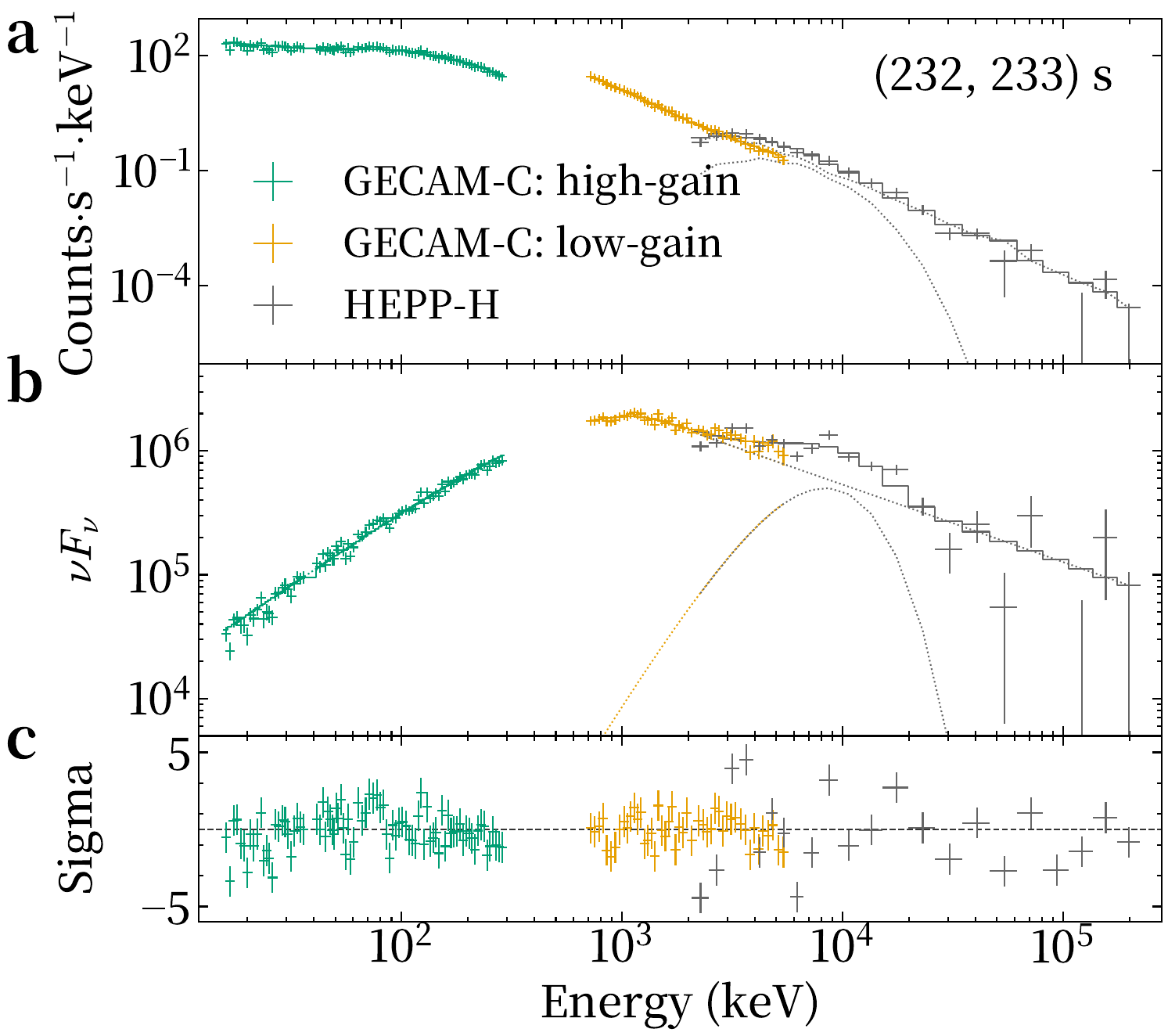}
    \end{subfigure}
    \hfill
    \begin{subfigure}[b]{0.32\textwidth}
        \centering
        \includegraphics[width=\textwidth]{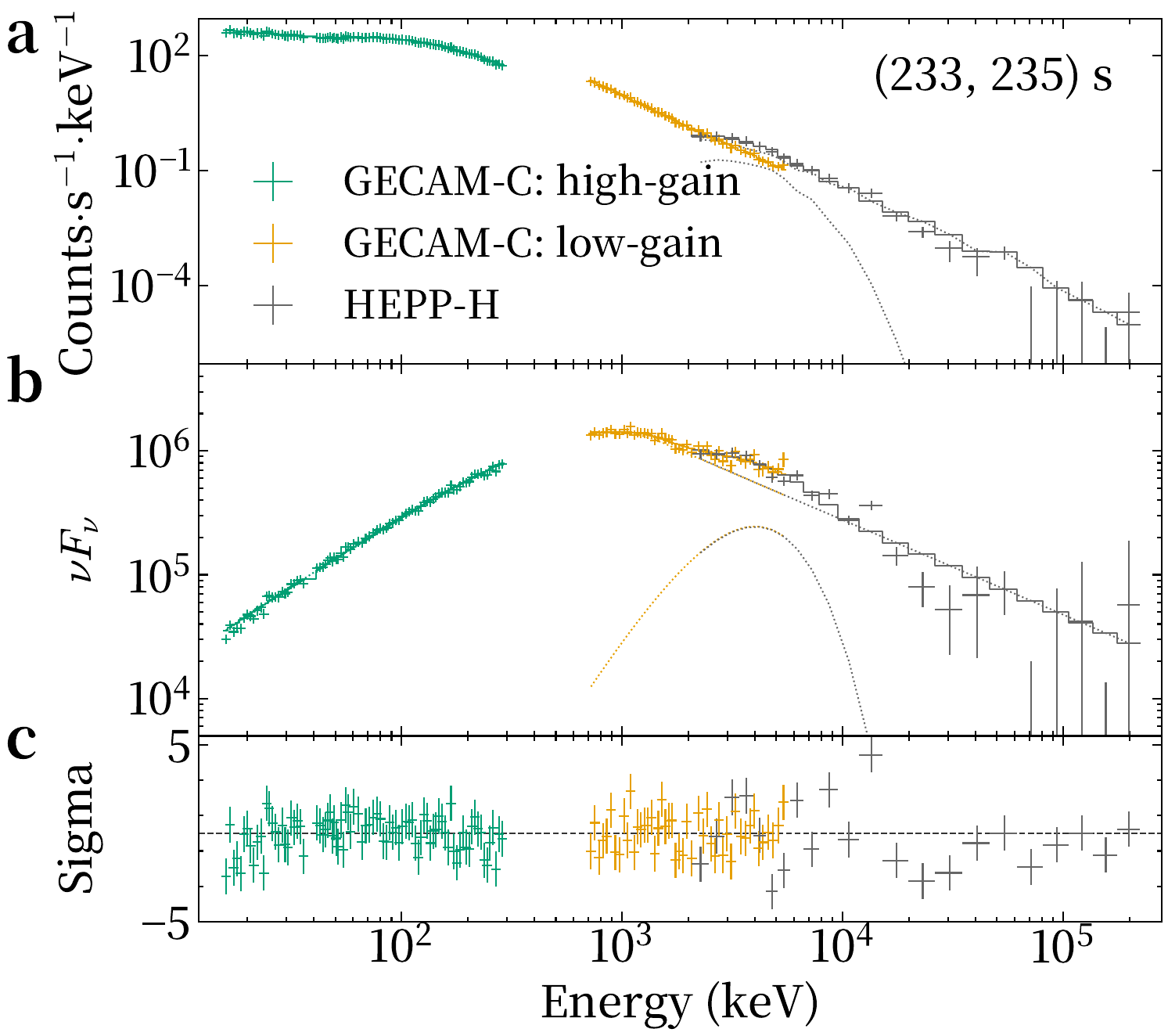}
    \end{subfigure}

    \caption{\textbf{Spectral fitting of the peak emission phase.} Each subplot displays: 
    \textbf{(a)} the count spectrum; \textbf{(b)} the $\nu F_\nu$ spectral energy distribution (units: $\mathrm{keV}^2 \cdot \mathrm{Photons} \cdot \mathrm{cm}^{-2} \cdot \mathrm{s}^{-1} \cdot \mathrm{keV}^{-1}$); and \textbf{(c)} the residuals. Time intervals indicated in parentheses are relative to $T_{\rm ref}$. The top two rows and bottom two rows correspond to fits using the Band + Cutoff Power-Law and Band + Blackbody models, respectively.}
    \label{fig:225_235_all_spectra}
\end{figure*}

%Fig S7
\begin{figure*}[t] 
    \centering 
    \begin{subfigure}[t]{0.49\textwidth}
    \caption{}
        \centering
        \includegraphics[width=\linewidth]{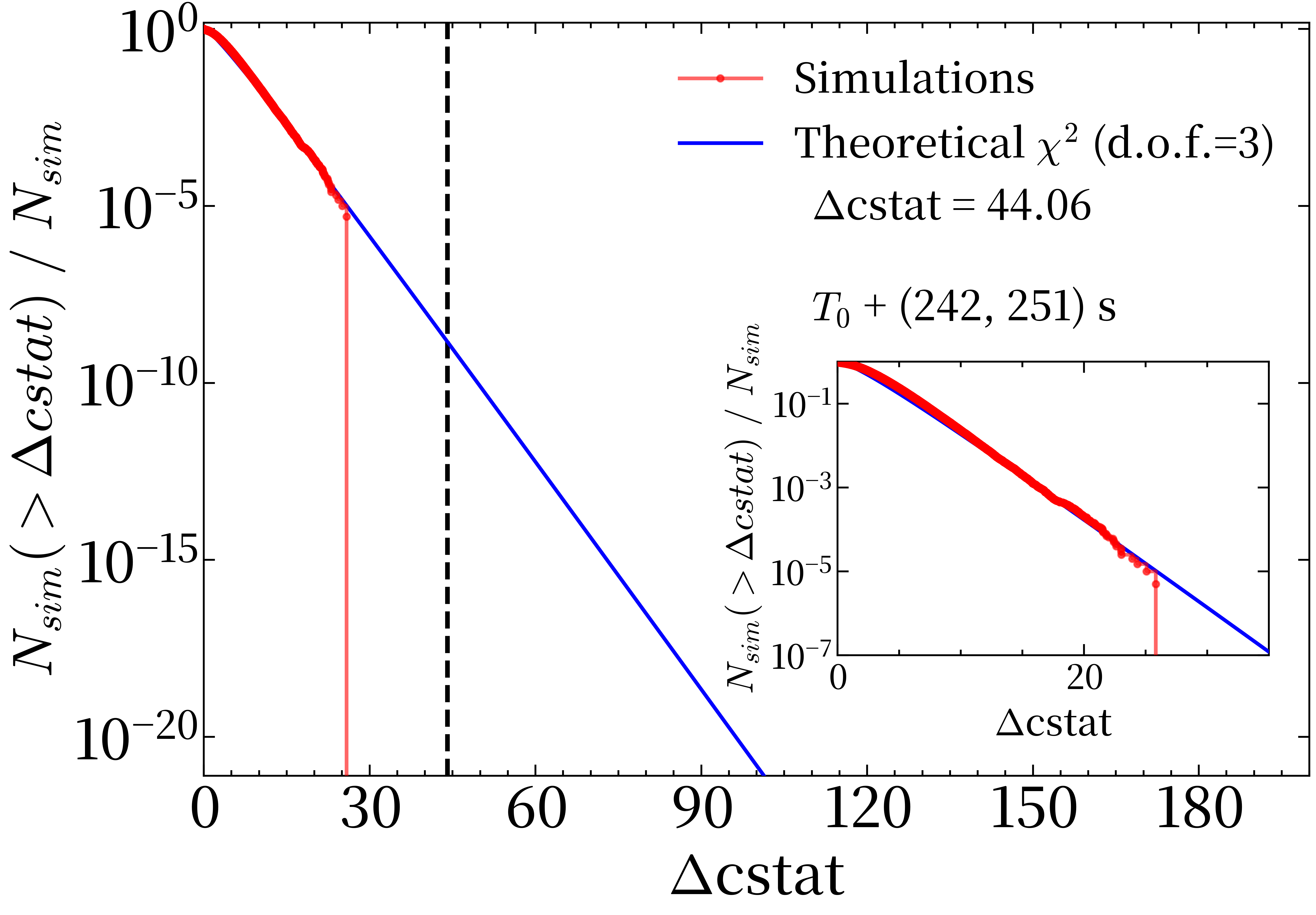}
        \label{subfig:242-251}
    \end{subfigure}%
    \hfill 
    \begin{subfigure}[t]{0.49\textwidth}
    \caption{}
        \centering
        \includegraphics[width=\linewidth]{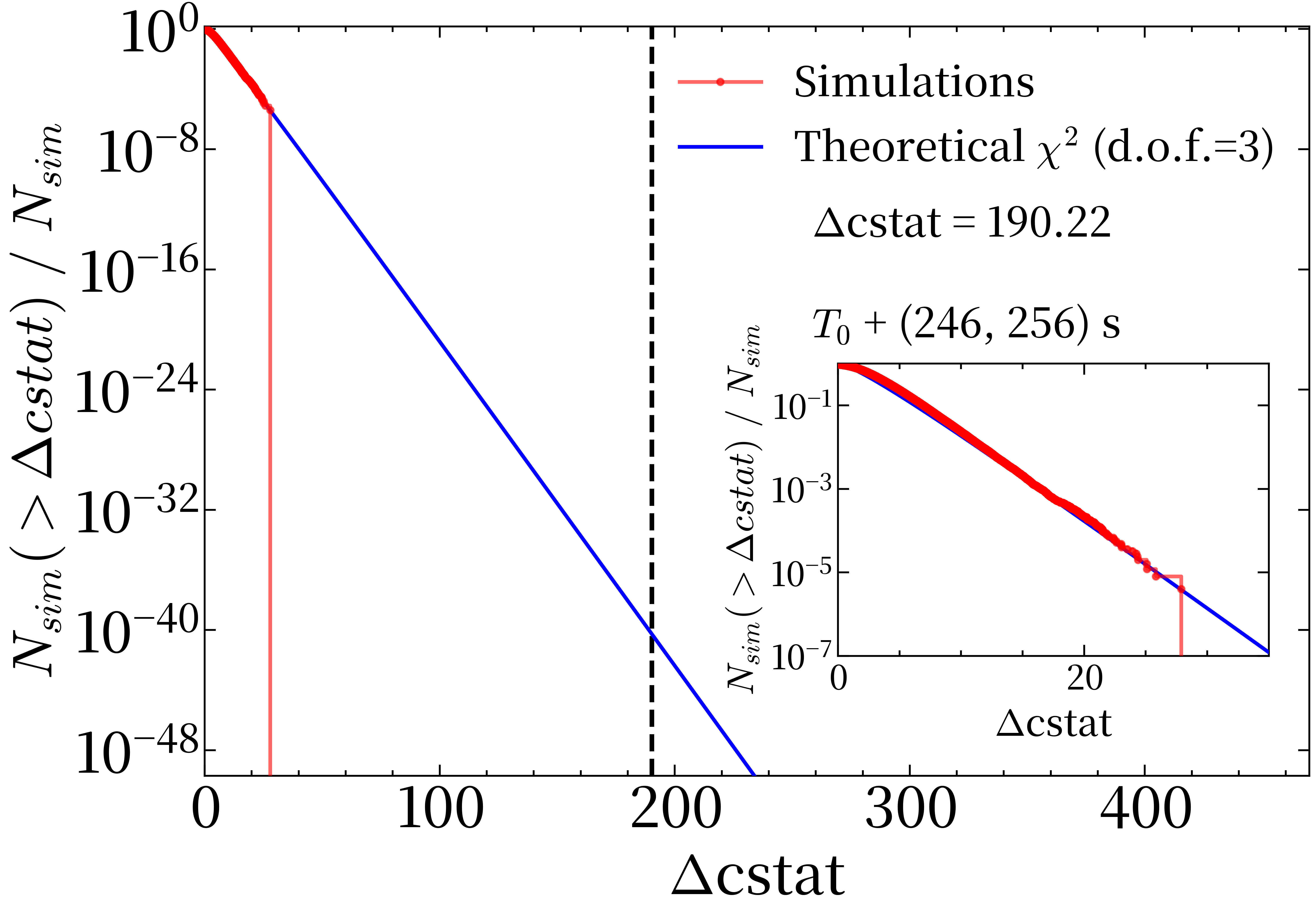}
        \label{subfig:246-256}
    \end{subfigure}%
    \caption{\textbf{Statistical significance verification via Monte Carlo simulations.}
    Red points represent the distribution of $\Delta\text{cstat}$ values derived from synthetic datasets, while blue line correspond to the theoretical $\chi^2$ distribution with 3 degrees of freedom (d.o.f.). The vertical black dashed line indicates the observed $\Delta\text{cstat}$ improvement obtained by adding the Gaussian component to the fit. Empirical $p$-values are calculated directly where the observed statistic falls within the simulated range. For observed values exceeding the simulated distribution, where direct computation is computationally prohibitive, an upper limit of $p < 1/N$ is adopted, where $N$ represents the number of simulations, alongside the theoretical significance.
    \textbf{(a)} Results for the interval $T_{\rm ref}+(242, 251)$\,s.
    \textbf{(b)} Results for the interval $T_{\rm ref}+(246, 256)$\,s.}
    \label{fig:line_significance} % note
\end{figure*}

%Fig S8
\begin{figure*}[t] 
\centering 
    \includegraphics[width=0.9\textwidth]{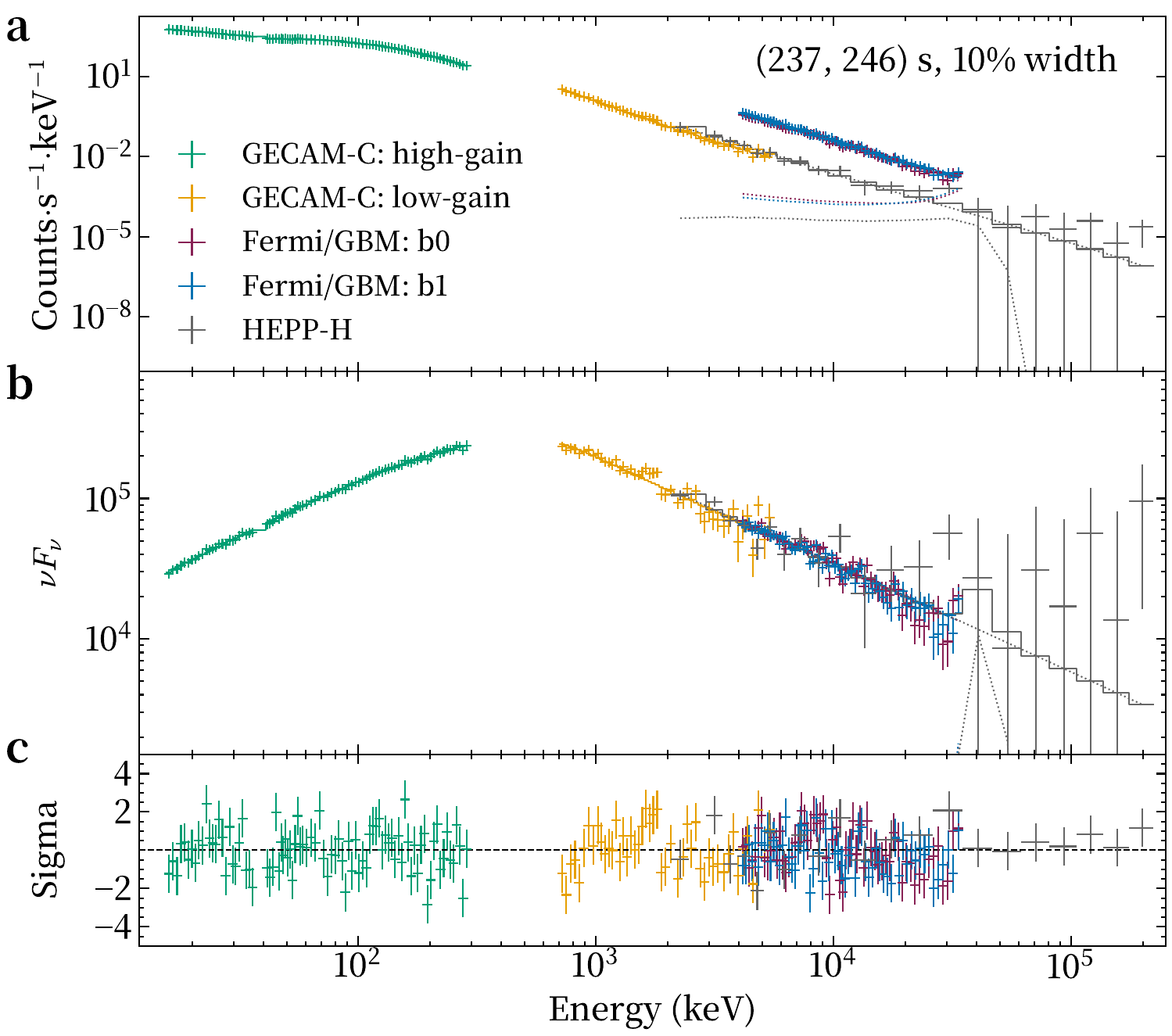}
        
    \caption{\textbf{Spectral identification of an emission line feature ($T_{\rm ref} + 237$--$246$\,s).} The analysis incorporates unsaturated observations from HEPP-H, GECAM-C, and Fermi/GBM. \textbf{(a)} the count spectrum. \textbf{(b)} The $\nu F_\nu$ spectral distribution (units: $\mathrm{keV}^2 \cdot \mathrm{Photons} \cdot \mathrm{cm}^{-2} \cdot \mathrm{s}^{-1} \cdot \mathrm{keV}^{-1}$). 
    \textbf{(c)} Residuals in units of standard deviation ($\sigma$) with $1\sigma$ error bars. Solid lines indicate the best-fitting model (Band function plus Gaussian line), while dashed lines denote individual components. Time intervals indicated in parentheses are relative to $T_{\rm ref}$.}
    \label{fig:line_fit} % note
\end{figure*}

%Fig S9

%Fig S10

%Fig S11

%Fig S12

%Fig S13

%Fig S14

%Fig S15

%Fig S16

%Fig S17

%Fig S18

%Fig S19

\clearpage

%\subsection{Table}

\clearpage

%Table S1
\begin{table}
  \centering % 核心修复：添加居
  \rmfamily 
  \caption{Main performance parameters of HEPP for electron detection}
  \label{tab:hepp_params}
  \renewcommand{\arraystretch}{1.5}
  \setlength{\tabcolsep}{18pt}
  \begin{tabular}{@{} ll cc @{}}
    \toprule[1.3pt] 
    \multirow{2}{*}{No.} & \multirow{2}{*}{Item} & \multicolumn{2}{c}{Index (measured value)} \\
    \cmidrule(lr){3-4} 
     & & HEPP-H & HEPP-L \\
    \midrule[0.5pt] 
    1 & Energy range & 1.5--255 MeV$^{*}$ & 0.1--3 MeV \\
    2 & Energy resolution & $\le 8.3\%$ @ $\ge 10$ MeV & $\le 8.9\%$ @ 1 MeV \\
    3 & Identification efficiency & $>$ 98.9\% & $>$ 96.68\% \\
    4 & Geometric Factor$^{a}$ & 102.5--184.3 cm$^2\cdot$sr & -- \\
    \bottomrule[1.3pt] 
  \end{tabular}
  \begin{flushleft}
  \small
  *The electron calibration of HEPP-H was performed up to 40~MeV, and the proton calibration was performed up to 180~MeV. The detection energy range for secondary electrons produced by gamma rays is extended up to 255~MeV, mainly through extrapolation of electron calibration data and simulations. \\
  $^{a}$The geometric factor for HEPP-H is energy-dependent, with the quoted range corresponding to energies of 2--15 MeV.
  \end{flushleft}
\end{table}

%Table S2
\begin{table}
  \centering
  \rmfamily 
  \renewcommand{\arraystretch}{1.5}
  \setlength{\tabcolsep}{16pt}
  \caption{Gain evolution of the HEPP-H calorimeter channels}
  \label{tab:gain_degradation}
  \begin{tabular}{@{} c l c c c @{}}
    \toprule[1.3pt]
    \multirow{2}{*}{\textbf{Layer}} & \multirow{2}{*}{\textbf{Detector Unit}} & \textbf{Peak (May 2018)} & \textbf{Peak (Oct 2022)} & \textbf{Relative Gain} \\
     & & (ADC Channels) & (ADC Channels) & (\%) \\
    \midrule[0.5pt]
    \multirow{2}{*}{PS Layer} & PS1 & 68.00 & 64.50 & 94.9 \\
     & PS2 & 76.00 & 75.55 & 99.4 \\
    \midrule
    \multirow{2}{*}{CsI Layer 1} & CsI1 & 95.00 & 79.10 & 83.3 \\
     & CsI2 & 102.00 & 88.20 & 86.5 \\
    \midrule
    \multirow{2}{*}{CsI Layer 2} & CsI3 & 127.00 & 112.95 & 88.9 \\
     & CsI4 & 115.00 & 101.20 & 88.0 \\
    \midrule
    \multirow{2}{*}{CsI Layer 3} & CsI5 & 88.00 & 75.45 & 85.7 \\
     & CsI6 & 90.00 & 77.60 & 86.2 \\
    \midrule
    \multirow{2}{*}{CsI Layer 4} & CsI7 & 85.00 & 75.95 & 89.4 \\
     & CsI8 & 94.00 & 85.25 & 90.7 \\
    \bottomrule[1.3pt]
  \end{tabular}
  
  \begin{flushleft}
  \small
  \textbf{Note:} The table compares the Minimum Ionizing Particle (MIP) spectral peak positions measured shortly after launch (May 2018) and contemporaneous with GRB~221009A (October 2022). The relative gain is calculated as the ratio of the peak position in 2022 to that in 2018, quantifying the gain attenuation mentioned in the text.
  \end{flushleft}
\end{table}

\clearpage

% Table E3
\begin{table*}[t]
  \centering
  \caption{Comparison of the Band, Band+CutoffPL, and Band+Blackbody spectral fits}
  \label{tab:CPL_vs_BB_model}

  \scriptsize
  \renewcommand{\arraystretch}{1.5}
  \setlength{\tabcolsep}{0pt}

  \begin{tabular*}{\textwidth}
    {@{\extracolsep{\fill}} ll ccc cc cccc @{}}
    \toprule
    \multirow{2}{*}{Time range (s)}
    & \multirow{2}{*}{Model}
    & \multicolumn{3}{c}{Band Component}
    & \multicolumn{2}{c}{Added Component}
    & \multicolumn{4}{c}{Statistics} \\
    \cmidrule(lr){3-5}
    \cmidrule(lr){6-7}
    \cmidrule(lr){8-11}
    & & $\alpha$ & $\beta$ & $E_{\rm c}$ (keV)
    & $\alpha_{\rm cpl}$ & $E_{\rm c,cpl}$ or $kT$ (keV)
    & C-stat/dof & $\chi^2$/dof & AIC & $\Delta$AIC \\
    \midrule
    
    \multirow{3}{*}{(225, 228)}
    & Band
    & $-0.85\pm0.01$ & $-2.38\pm0.01$ & $1807\pm40$
    & -- & -- & 1296/152 & 1408/152 & 1308 & -- \\
    & Band+CPL
    & $-0.79\pm0.02$ & $-2.49\pm0.03$ & $959\pm69$
    & $-0.86\pm0.17$ & $9403\pm1163$
    & 451/149 & 463/149 & 469 & 839 \\
    & Band+BB
    & $-0.82\pm0.01$ & $-2.45\pm0.02$ & $1281\pm39$
    & -- & $2801\pm59$
    & 509/150 & 526/150 & 525 & 783 \\
    \cmidrule{1-11}
    
    \multirow{3}{*}{(228, 230)}
    & Band
    & $-0.80\pm0.01$ & $-2.70\pm0.02$ & $1426\pm30$
    & -- & -- & 693/152 & 724/152 & 705 & -- \\
    & Band+CPL
    & $-0.74\pm0.01$ & $-2.68\pm0.03$ & $864\pm56$
    & $-0.20\pm0.43$ & $3234\pm588$
    & 244/149 & 249/149 & 262 & 443 \\
    & Band+BB
    & $-0.76\pm0.01$ & $-2.75\pm0.02$ & $1073\pm33$
    & -- & $1803\pm51$
    & 261/150 & 267/150 & 277 & 428 \\
    \cmidrule{1-11}
    
    \multirow{3}{*}{(230, 231)}
    & Band
    & $-0.85\pm0.02$ & $-2.86\pm0.02$ & $2288\pm68$
    & -- & -- & 549/152 & 539/152 & 561 & -- \\
    & Band+CPL
    & $-0.71\pm0.14$ & $-2.56\pm0.06$ & $1088\pm186$
    & $-0.86\pm0.44$ & $4963\pm1107$
    & 372/149 & 379/149 & 390 & 171 \\
    & Band+BB
    & $-0.77\pm0.02$ & $-2.72\pm0.02$ & $1560\pm63$
    & -- & $2105\pm71$
    & 401/150 & 410/150 & 417 & 144 \\
    \cmidrule{1-11}
    
    \multirow{3}{*}{(231, 232)}
    & Band
    & $-0.75\pm0.02$ & $-2.48\pm0.02$ & $1132\pm42$
    & -- & -- & 438/152 & 466/152 & 450 & -- \\
    & Band+CPL
    & $-0.71\pm0.02$ & $-2.61\pm0.03$ & $870\pm60$
    & $-0.12\pm0.56$ & $4441\pm1160$
    & 229/149 & 235/149 & 247 & 203 \\
    & Band+BB
    & $-0.72\pm0.02$ & $-2.60\pm0.03$ & $970\pm42$
    & -- & $2269\pm89$
    & 231/150 & 239/150 & 247 & 203 \\
    \cmidrule{1-11}
    
    \multirow{3}{*}{(232, 233)}
    & Band
    & $-0.80\pm0.02$ & $-2.56\pm0.02$ & $1058\pm42$
    & -- & -- & 392/152 & 416/152 & 404 & -- \\
    & Band+CPL
    & $-0.74\pm0.05$ & $-2.73\pm0.08$ & $726\pm81$
    & $-0.93\pm0.48$ & $7041\pm2430$
    & 264/149 & 270/149 & 282 & 122 \\
    & Band+BB
    & $-0.77\pm0.02$ & $-2.63\pm0.04$ & $863\pm44$
    & -- & $2123\pm112$
    & 285/150 & 291/150 & 301 & 103 \\
    \cmidrule{1-11}
    
    \multirow{3}{*}{(233, 235)}
    & Band
    & $-0.85\pm0.01$ & $-2.82\pm0.02$ & $1018\pm27$
    & -- & -- & 293/152 & 294/152 & 305 & -- \\
    & Band+CPL
    & $-0.79\pm0.09$ & $-2.77\pm0.06$ & $690\pm97$
    & $-0.88\pm0.99$ & $2934\pm1511$
    & 202/149 & 201/149 & 220 & 85 \\
    & Band+BB
    & $-0.80\pm0.01$ & $-2.77\pm0.03$ & $761\pm38$
    & -- & $1019\pm62$
    & 207/150 & 207/150 & 223 & 82 \\
    \bottomrule
  \end{tabular*}

  \vspace{10pt}
  \begin{minipage}{\textwidth}
  \scriptsize
  \textbf{Note:}
    For the Band+CPL model, the added-component parameters are $\alpha_{\rm cpl}$ and $E_{\rm c,cpl}$; for the Band+BB model, the added-component parameter is $kT$. We define $\Delta\mathrm{AIC} =\mathrm{AIC}_{\rm Band}-\mathrm{AIC}_{\rm model}$, where ``model'' denotes Band+CPL or Band+BB.
    
  \end{minipage}
\end{table*}

% Table E4
\begin{table*}[!t]
\centering
\caption{AIC comparison among alternative models for the MeV spectral structure}
\label{tab:2sbpl_comparison}

\small
\renewcommand{\arraystretch}{1.25}
\setlength{\tabcolsep}{0pt}

\begin{tabular*}{0.72\textwidth}
{@{\extracolsep{\fill}} lrrrr @{}}
\toprule
Time range (s)
& Band+CPL
& Band+BB
& 2SBPL
& $\Delta{\rm AIC}_{\rm 2SBPL}$ \\
\midrule
$(225,228)$ & 469 & 525 & \textbf{456} & 0 \\
$(228,230)$ & \textbf{262} & 277 & 293 & 31 \\
$(230,231)$ & 390 & 417 & \textbf{382} & 0 \\
$(231,232)$ & \textbf{247} & \textbf{247} & 252 & 5 \\
$(232,233)$ & \textbf{282} & 301 & 306 & 24 \\
$(233,235)$ & \textbf{220} & 223 & 230 & 10 \\
\bottomrule
\end{tabular*}

\vspace{5pt}
\begin{minipage}{0.72\textwidth}
\small
\textbf{Note.}
The lowest AIC value in each interval is shown in bold.
We define
$\Delta{\rm AIC}_{\rm 2SBPL}
={\rm AIC}_{\rm 2SBPL}-{\rm AIC}_{\rm min}$,
where ${\rm AIC}_{\rm min}$ is the minimum AIC among the three models
in the corresponding interval.
\end{minipage}
\end{table*}

\bibliographystyle{aasjournalv7}
\bibliography{maintex}

%\noindent\textbf{References}

%\vspace{5mm}
%\input{./ref2.tex}

%\clearpage
% note
%\input{supplementary.tex} 

\end{document}